\documentclass[a4paper,11pt]{article}

\usepackage{hyperref}
\usepackage{cite}
\usepackage{amsmath}
\usepackage{amssymb}
\usepackage{bm}
\usepackage{color}
\usepackage{colortbl}
\usepackage{url}
\usepackage{graphicx}
\usepackage{tikz}
\numberwithin{equation}{section}
\usepackage{geometry}
\numberwithin{equation}{section}
\allowdisplaybreaks

\usepackage{braketmod}
\usepackage{esint}

\global\long\def\ket#1{\Ket{#1}}%

\global\long\def\kket#1{\Kket{#1}}%

\global\long\def\braket#1{\Braket{#1}}%

\global\long\def\brakket#1{\Brakket{#1}}%

\global\long\def\bbrakket#1{\Bbrakket{#1}}%

\begin{document}

\newcommand{\aff}[1]{${}^{#1}$}
\renewcommand{\thefootnote}{\fnsymbol{footnote}}

\begin{titlepage}
\begin{flushright}
USTC-ICTS/PCFT-25-15
\end{flushright}
\begin{center}
{\Large\bf
Exact large $N$ expansion of $\mathcal{N}=4$ circular quiver Chern-Simons theories and squashing
}\\
\bigskip\bigskip
\bigskip\bigskip
{\large Naotaka Kubo,\footnote{\tt naotaka.kubo@yukawa.kyoto.u-ac.jp}}\aff{1}
{\large Tomoki Nosaka\footnote{\tt nosaka@simis.cn}}\aff{2,3}
{\large and Yi Pang\footnote{\tt pangyi1@tju.edu.cn}}\aff{1,4}

\bigskip\bigskip

\aff{1} {\small
\it Center for Joint Quantum Studies and Department of Physics, School of Science,
Tianjin University, 135 Yaguan Road, Tianjin 300350, China
}\\
\aff{2} {\small
\it Fudan Center for Mathematics and Interdisciplinary Study, Fudan University, Shanghai 200433, China
}\\
\aff{3} {\small
\it Shanghai Institute for Mathematics and Interdisciplinary Sciences,\\
Block A, International Innovation Plaza, No.~657 Songhu Road, Yangpu District, Shanghai, China
}\\
\aff{4} {\small
\it Peng Huanwu Center for Fundamental Theory, Hefei, Anhui 230026, China
}

\bigskip
\end{center}
\bigskip
\bigskip
\begin{abstract}
In this work, we revisit the exact computation of the round sphere partition function of 3d $\mathcal{N}=4$ circular quiver Chern-Simons theories with mass and Fayet-Iliopoulos (FI) deformations.
Utilizing the Fermi gas formalism, we derive the large $N$ expansion of the partition function and determine the Airy function structure, parameterized by three functions $C$, $B$ and $A$.
We propose a novel closed-form expression for $A$ that incorporates the effects of FI parameters and satisfies various consistency constraints from quiver reductions. 
As an application, by using an accidental coincidence of the Fermi gas density matrices we extend our results to the squashed sphere partition function of $\mathcal{N}=4$ super Yang-Mills theories with an adjoint hypermultiplet and multiple fundamental hypermultiplets. 
Our findings provide further evidence for the universality of the Airy function structure in supersymmetric gauge theories of multiple M2-branes.
\end{abstract}

\bigskip\bigskip\bigskip

\end{titlepage}

\setcounter{footnote}{0}
\renewcommand{\thefootnote}{$\dagger$\arabic{footnote}}

\tableofcontents

\section{Introduction and Summary\label{sec:Introduction}}

The AdS/CFT correspondence enables us to study the quantum effect of gravity through the $1/N$ expansion in the quantum field theory living on the boundary.
In order to understand the $1/N$ corrections in the language of the gravity side, it will be useful to focus on their universal structures for a wide class of quantum field theories which are dual to the same type of background geometry.
For this purpose, the sphere partition function of the theories of M2-branes provides a useful playground.
In a large class of the theories of $N$ M2-branes realized by the type IIB brane construction, we can calculate 
the three sphere partition function $Z_{S^3}$
by the supersymmetric localization \cite{Kapustin:2009kz}, and study its large $N$ limit by the large $N$ saddle point approximation.
The AdS/CFT correspondence suggests that the free energy $F=-\log Z_{S^3}$ coincides in the large $N$ limit with the action of eleven-dimensional supergravity in $\text{AdS}_4\times Y_7$ background, which scales as $\sim N^{3/2}$.
The large $N$ saddle point approximation indeed reproduces this $N^{3/2}$ scaling, including its prefactor \cite{Herzog:2010hf}.
Furthermore in these theories, through the rewriting of the Fermi gas formalism \cite{Marino:2011eh}\footnote{
In the ABJM theory \cite{Hosomichi:2008jd,Aharony:2008ug}, the Airy form was originally obtained by resumming the genus expansion in the 't Hooft limit \cite{Fuji:2011km}.
}
one can further show that the large $N$ partition function including all order $1/N$ perturbative corrections has the following universal structure
\begin{equation}
Z^{\mathcal{T}}\left(N;\boldsymbol{\xi}\right)
=e^{A^{\mathcal{T}}\left(\boldsymbol{\xi}\right)}C^{\mathcal{T}}\left(\boldsymbol{\xi}\right)^{-\frac{1}{3}}
\text{Ai}\left[C^{\mathcal{T}}\left(\boldsymbol{\xi}\right)^{-\frac{1}{3}}\left(N-B^{\mathcal{T}}\left(\boldsymbol{\xi}\right)\right)\right]\left(1+{\cal O}\left(e^{-\#\sqrt{N}}\right)\right).
\label{eq:LargeN-Gen}
\end{equation}
Here $\boldsymbol{\xi}$ is a set of parameters of the theory $\mathcal{T}$ except $N$. 
An important point of the universality is that the perturbative part is parameterized by only three functions $C^{\mathcal{T}}\left(\boldsymbol{\xi}\right)$, $B^{\mathcal{T}}\left(\boldsymbol{\xi}\right)$ and $A^{\mathcal{T}}\left(\boldsymbol{\xi}\right)$, and thus the attempt to determine the perturbative part exactly reduces to the task of obtaining these three functions.

When $\boldsymbol{\xi}$ contains a continuous deformation parameter of the theory such as the mass parameters of the matter fields, the Fayet-Illipoulos parameter or the squashing parameter, the correlation function of the operators associated with the deformation can be calculated as the derivatives of
the partition function
\cite{Closset:2012ru, Nishioka:2013gza, Binder:2018yvd, Binder:2019mpb, Chester:2021gdw}.
This method has been applied in various contexts, such as in the conformal bootstrap analysis \cite{Agmon:2017xes} and in the calculation of the entanglement entropy \cite{Hirano:2019szi}.
The Airy form \eqref{eq:LargeN-Gen} provides a powerful tool to calculate the large $N$ expansion of the correlation function \cite{Chester:2020jay}.

Extensions of the Fermi gas formalism to more general setups were carried out, including the theories with mass deformations \cite{Nosaka:2015iiw}, non-uniform ranks of the gauge groups \cite{Awata:2012jb,Honda:2013pea,Matsumoto:2013nya,Kubo:2020qed}, non-circular quiver diagrams \cite{Assel:2015hsa,Moriyama:2015jsa,Kubo:2024raz} and/or $O/USp$-type gauge groups \cite{Mezei:2013gqa,Honda:2015rbb,Okuyama:2016xke,Moriyama:2016xin,Moriyama:2016kqi}, whose large $N$ partition function also turned out to takes the Airy form. 
More recently, motivated by these results, along with various nontrivial checks, such as higher order corrections in the 't Hooft limit \cite{Geukens:2024zmt} and exact or numerical evaluations of the partition function in finite $N$ 
\cite{Gaiotto:2019mmf,Kubo:2024qhq,Bobev:2025ltz},
it has been proposed that the Airy form exhibits universality, even for the partition functions without the Fermi gas formalism \cite{Bobev:2022jte,Hristov:2022lcw,Bobev:2022eus,Bobev:2023lkx,Bobev:2025ltz}. The universality and simplicity of the Airy form certainly facilicate the precision test of AdS/CFT correspondence. In the dual gravity side, the rather restrictive structure of the supergravity invariants implies that the partition function of the gravity theories is also characterized by a few parameters. Indeed, assuming the Airy form of the CFT partition function, one can determine the ${\cal B}$ and ${\cal C}$ coefficients from the perturbative calculations in ${\cal N}=2, D=4$ supergravity extended by four-derivative terms \cite{Bobev:2020egg,Bobev:2021oku, Hristov:2022lcw}. Attempts to reproduce the exact Airy form of the gravitational partition have been made by applying the localization technique to the pure AdS$_4$ vacuum solution in ${\cal N}=2, D=4$ minimal supergravity \cite{Dabholkar:2014wpa} and the mini-superspace approximation to the gravitational path integral \cite{Caputa:2018asc}.
It is conceivable that the gravitational partition functions for more general asymptotically AdS$_4$ solutions may also take the Airy form, as suggested by holography. This, however, necessitates a deeper understanding of the gravitational path integral.


Among the three parameteres entering the Airy form \eqref{eq:LargeN-Gen}, the coefficients $C^{\mathcal{T}}(\boldsymbol{\xi})$ and $B^{\mathcal{T}}(\boldsymbol{\xi})$ are relatively well understood compared with $A^{\mathcal{T}}(\boldsymbol{\xi})$.
For those partition functions written in the Fermi gas formalism, these coefficients can be easily determined by the semiclassical expansion.
On the other hand, the coefficient $A^{\mathcal{T}}(\boldsymbol{\xi})$ is an infinite series of the Planck constant $\hbar$ which is hard to determine even in the Fermi gas formalism.
Note that although $A^{\mathcal{T}}(\boldsymbol{\xi})$ is simply an overall factor in the Airy form \eqref{eq:LargeN-Gen}, it plays a crucial role for example in the application for the calculation of the correlation function mentioned above, where the derivatives with respect to the deformation parameters also act on $A^{\mathcal{T}}(\boldsymbol{\xi})$.
In the ABJM theory with the Chern-Simons level $k$, the coefficient $A^{\mathcal{T}}(\boldsymbol{\xi})=A^{\text{ABJM}}(k)$ was determined by combining the data of the small $\hbar=2\pi k$ expansion and the result of the 't Hooft expansion \cite{Hanada:2012si}.
Curiously, it turned out that in many setups the coefficient $A^{\mathcal{T}}(\boldsymbol{\xi})$ is written as a simple linear combination of $A^{\text{ABJM}}(k)$ with rescaled arguments.
However, some results suggest that in more general setups $A^{\mathcal{T}}(\boldsymbol{\xi})$ cannot be expressed in terms of $A^{\text{ABJM}}(k)$ \cite{Chester:2023qwo,Kubo:2024qhq}.

In this paper,
to provide more evidence for the conjectured uiversal Airy form \eqref{eq:LargeN-Gen},
we revisit the computation of the exact partition function of the circular quiver Chern-Simons theory with mass and FI deformations.
The most difficult part of our work resides in finding a closed form expression of $A^{\mathcal{T}}(\boldsymbol{\xi})$.
While $A^{\mathcal{T}}(\boldsymbol{\xi})$ of this model without FI deformation was already proposed in \cite{Nosaka:2015iiw} as a linear combination of $A^{\text{ABJM}}(k)$, the expression for $A^{\mathcal{T}}(\boldsymbol{\xi})$ with the FI parameteres turned on was not studied.
In this paper, based on the data obtained by the $\hbar$ expansion from the Fermi gas formalism we propose a new function ${\cal A}\left(\kappa\right)$ \eqref{eq:calA-Def}, in terms of which $A^{\mathcal{T}}(\boldsymbol{\xi})$ of the circular quiver Chern-Simons theories can be written in a unified manner \eqref{Aofpqmodel}.
When the values of the FI parameters are tuned appropriately, the partition function of one theory reduces to the partition function of another theory with a smaller quiver and the rescaled Chern-Simons levels.
Our proposal for $A^{\mathcal{T}}(\boldsymbol{\xi})$ satisfies all non-trivial constraints arising from these quiver reductions.

As a non-trivial application of the above result, we also study the $\mathcal{N}=4$ $\text{U}(N)$ super Yang-Mills (SYM) theory with an adjoint hypermultiplet and $N_\text{f}$ fundamental hypermultiplets on the squashed sphere $S^3_b$.
\footnote{
The same setup was also studied through the expansion around $b=1$ with vanishing mass deformations and FI parameters in \cite{Chester:2020jay}.
}
When $N_\text{f}=1$, the Fermi gas formalism has been applied to the SYM theory by tuning the mass parameter for the adjoint hypermultiplet \cite{Hatsuda:2016uqa,Kubo:2024qhq}, where it was also shown that the inverse of the resulting density matrix can be written as the sum of three operators, and $A^{\mathcal{T}}(\boldsymbol{\xi})$ can be obtained by employing the result in \cite{Hatsuda:2015oaa}.
The extension of the Fermi gas approach to general $N_\text{f}$ does not require any technical advance.
(The Fermi gas formalism for the general $N_\text{f}$ case was also discussed in \cite{Bobev:2025ltz}.)
However, in the next step, namely when we compute $A^{\mathcal{T}}(\boldsymbol{\xi})$ in the Fermi gas formalism, one can no longer use the same technique for the $N_\text{f} \geq 2$ case as in the $N_\text{f}=1$ case.
We illustrate that the density matrix of the SYM theory with general $N_\text{f}$ (including the $N_\text{f}=1$ case) accidentally coincides with the one of the circular quiver Chern-Simons theory with specific mass deformations and FI parameters.
More explicitly, we show that the former is a special case of the latter one.
This allows us to determine $C^{\mathcal{T}}(\boldsymbol{\xi})$, $B^{\mathcal{T}}(\boldsymbol{\xi})$ and especially $A^{\mathcal{T}}(\boldsymbol{\xi})$ for the $S^3_b$ partition function of the same Yang-Mills theory with more than one fundamental hypermultiplets.
Moreover, by comparing our result with the result for $N_\text{f}=1$ in \cite{Kubo:2024qhq}, we further provide a non-trivial check of our proposal \eqref{Aofpqmodel}.

The rest of this paper is organized as follows.
In section \ref{sec:PartitionFunc}, we first explain the circular quiver Chern-Simons theory we focus on in this paper, which we call the $\left(q,\tilde{q}\right)$ model.
We then show the matrix model for the $S^3$ partition function of the $\left(q,\tilde{q}\right)$ model and review how to apply the Fermi gas formalism.
In section \ref{sec:LargeN}, we study the large $N$ expansion of the matrix model by using the density matrix obtained in the previous section.
This is our main result in this paper.
In section \ref{sec:Consistency}, we perform various consistency checks of our results.
In section \ref{sec:SYM-Squash}, we study the SYM theory on the squashed three sphere as an application of our result.
Finally, in section \ref{sec:Conclusion}, we summarise this paper and discuss various future directions.
In appendix \ref{sec:SpecialFunc}, we list special functions and their properties which we use in the main text.
In appendix \ref{app_WignerKirkwood}, we explain an analytic treatment of the spectral zeta function through the Wigner-Kirkwood expansion, which also proves an assumption underlying the more efficient numerical approach to the spectral zeta function adopted in the main text.
In appendix \ref{app_listofD}, we list the analytic expression for the coefficients of the spectral zeta function in the $\hbar$ expansion.

\section{\texorpdfstring{$S^3$}{S3} partition function with mass and FI deformations\label{sec:PartitionFunc}}

In this section we review the matrix model of the $\left(q,\tilde{q}\right)$ model and the Fermi gas formalism.

\begin{figure}
\begin{center}
\includegraphics[width=7cm]{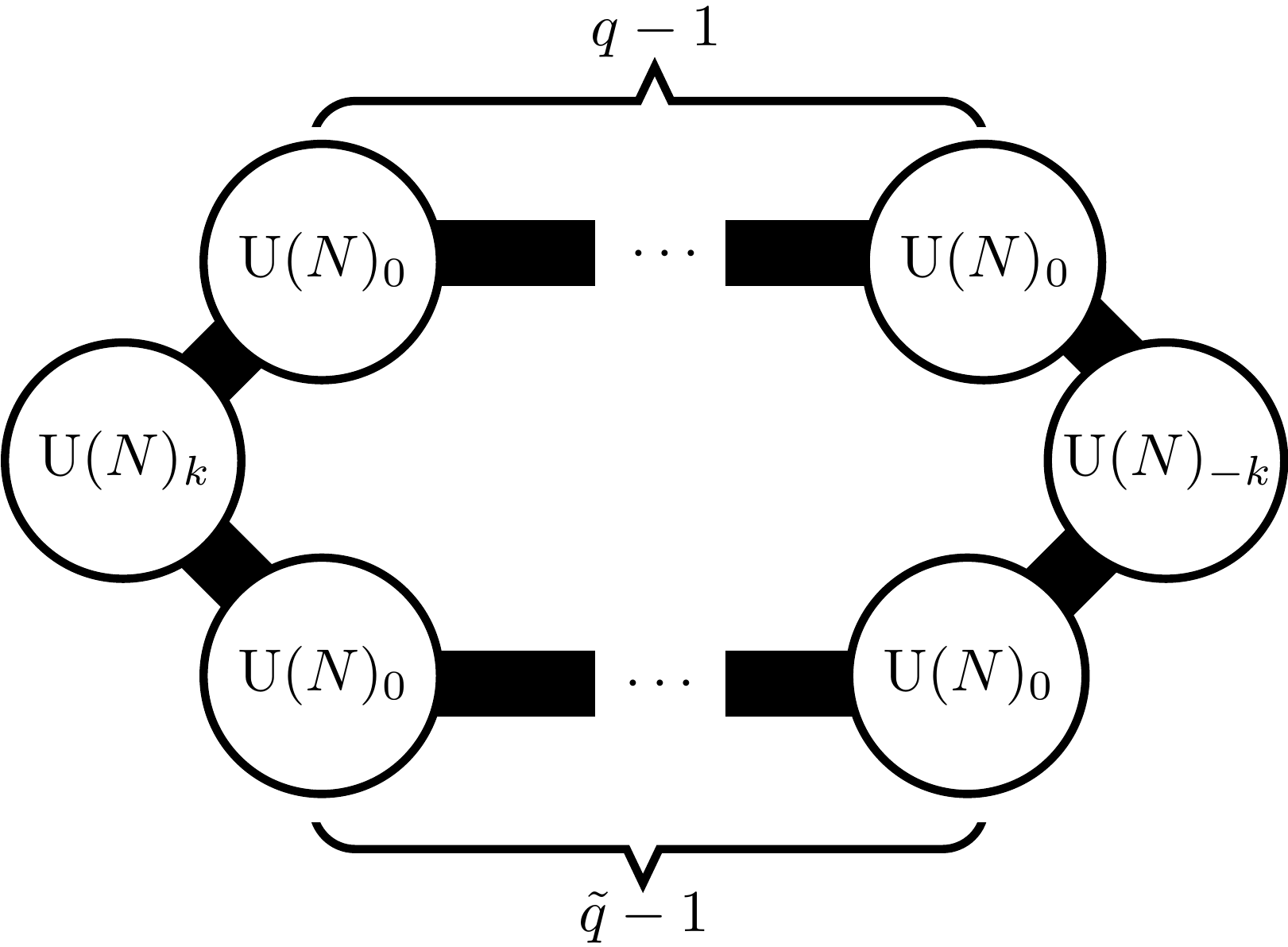}
\caption{
Quiver diagram of the $\left(q,\tilde{q}\right)$ model.
Each node which has non-vanishing Chern-Simons level denotes an ${\cal N}=2$ vector multiplet, each node with vanishing Chern-Simons level denotes an ${\cal N}=4$ (twisted) vector multiplet and each edge denotes the $\mathcal{N}=4$ bi-fundamental (twisted) hypermultiplet.
}
\label{quiver}
\end{center}
\end{figure}

The $\left(q,\tilde{q}\right)$ model is an $\mathcal{N}=4$ supersymmetric Chern-Simons theory described by a circular quiver diagram in figure \ref{quiver} \cite{Hosomichi:2008jd}.\footnote{
The name $\left(q, \tilde{q}\right)$ model was introduced in \cite{Moriyama:2014gxa}.
}
Each node of the quiver diagram is $\mathrm{U}\left(N\right)$, and the number of the nodes is $q+\tilde{q}$. 
Two $\mathrm{U}\left(N\right)$ gauge factors have Chern-Simons terms with Chern-Simons levels $\pm k$, and the numbers of nodes between two $\mathrm{U}\left(N\right)_{\pm k}$ are $q-1$ and $\tilde{q}-1$.
Namely, the gauge group is
\begin{equation}
\text{\ensuremath{\mathrm{U}\left(N\right)}}_{k}\times\underset{q-1}{\underbrace{\mathrm{U}\left(N\right)_{0}\times\cdots\times\mathrm{U}\left(N\right)_{0}}}\times\mathrm{U}\left(N\right)_{-k}\times\underset{\tilde{q}-1}{\underbrace{\mathrm{U}\left(N\right)_{0}\times\cdots\times\mathrm{U}\left(N\right)_{0}}}.
\end{equation}
Here the subscripts denote the Chern-Simons levels. 
Each $\mathrm{U}\left(N\right)$ gauge factor is associated with an ${\cal N}=2$ vector multiplet (${\cal N}=4$ (twisted) vector multiplet for those with vanishing Chern-Simons level).
In addition to the vector multiplets, there are $\mathcal{N}=4$ bi-fundamental (twisted) hypermultiplets connecting adjacent nodes.
The $\left(q,\tilde{q}\right)$ model admits FI deformations for the $\mathrm{U}(N)$ nodes and mass deformations for the bi-fundamental matters.

\subsection{Matrix model\label{sec:MatrixModel}}

Thanks to the supersymmetric localization \cite{Pestun:2007rz,Kapustin:2009kz}, the $S^{3}$ partition functions reduce to matrix models.
The matrix model for the $\left(q,\tilde{q}\right)$ model is given by
\begin{align}
 & Z_{k}^{\left(q,\tilde{q}\right)}\left(N;\eta_{\alpha},M;\tilde{\eta}_{\alpha},\tilde{M}\right)\nonumber \\
 & =\frac{1}{\left(N!\right)^{q+\tilde{q}}}\int_{-\infty}^{\infty}\left(\prod_{\alpha=1}^{q}\prod_{i=1}^{N}\frac{d\lambda_{i}^{\left(\alpha\right)}}{2\pi}\right)\left(\prod_{\alpha=1}^{\tilde{q}}\prod_{i=1}^{N}\frac{d\tilde{\lambda}_{i}^{\left(\alpha\right)}}{2\pi}\right)\nonumber \\
 & \quad\quad\times e^{-\frac{ik}{4\pi}\sum_{i}^{N}\left(\left(\tilde{\lambda}_{i}^{\left(\tilde{q}\right)}\right)^{2}-\left(\lambda_{i}^{\left(q\right)}\right)^{2}\right)}e^{-\frac{i}{2}\sum_{\alpha=1}^{q}\eta_{\alpha}\sum_{i}^{N}\left(\lambda_{i}^{\left(\alpha-1\right)}-\lambda_{i}^{\left(\alpha\right)}\right)}e^{-\frac{i}{2}\sum_{\alpha=1}^{\tilde{q}}\tilde{\eta}_{\alpha}\sum_{i}^{N}\left(\tilde{\lambda}_{i}^{\left(\alpha-1\right)}-\tilde{\lambda}_{i}^{\left(\alpha\right)}\right)}\nonumber \\
 & \quad\quad\times\prod_{\alpha=1}^{q}\frac{\prod_{i<j}^{N}2\sinh\frac{\lambda_{ij}^{\left(\alpha-1\right)}}{2}\prod_{i<j}^{N}2\sinh\frac{\lambda_{ij}^{\left(\alpha\right)}}{2}}{\prod_{i,j}^{N}2\cosh\frac{\lambda_{i}^{\left(\alpha-1\right)}-\lambda_{j}^{\left(\alpha\right)}+\pi M}{2}}\prod_{\alpha=1}^{\tilde{q}}\frac{\prod_{i<j}^{N}2\sinh\frac{\tilde{\lambda}_{ij}^{\left(\alpha-1\right)}}{2}\prod_{i<j}^{N}2\sinh\frac{\tilde{\lambda}_{ij}^{\left(\alpha\right)}}{2}}{\prod_{i,j}^{N}2\cosh\frac{\tilde{\lambda}_{i}^{\left(\alpha-1\right)}-\tilde{\lambda}_{j}^{\left(\alpha\right)}+\pi\tilde{M}}{2}},\label{eq:MM-pq-Def}
\end{align}
where $\lambda_{i}^{\left(0\right)}=\tilde{\lambda}_{i}^{\left(\tilde{q}\right)}$, $\tilde{\lambda}_{i}^{\left(0\right)}=\lambda_{i}^{\left(q\right)}$ and $\lambda_{ij}=\lambda_{i}-\lambda_{j}$.
The integration variables correspond to the Cartans of the gauge fields, the factors in the second line come from the Chern-Simons terms and the FI terms, and the factors in the third line come from the 1-loop determinants of the gauge fields and the bi-fundamental matter fields.
$\eta_{\alpha}$ and $\tilde{\eta}_{\alpha}$ denote the FI deformations, and $M$ and $\tilde{M}$ denote the mass deformations.
Here by shifting the integration variables $\lambda^{(q)}_i$ and $\tilde{\lambda}^{(\tilde{q})}_i$ (which have non-zero Chern-Simons levels in the corresponding nodes) we have chosen
\begin{equation}
\sum_{\alpha=1}^{q}\eta_{\alpha}=\sum_{\alpha=1}^{\tilde{q}}\tilde{\eta}_{\alpha}=0.\label{eq:eta-Sum}
\end{equation}
(We have ignored additional factors appearing in this step since they are independent of the integration variables and thus they are just overall factors.)
We have also chosen all the mass parameters of $q$ nodes ($\tilde{q}$ nodes) to be equal $M$ ($\tilde{M}$) by shifting the remaining integration variables $\lambda^{(\alpha)}_i$ ($\alpha \neq q$) ($\tilde{\lambda}^{(\alpha)}_i$ ($\alpha \neq \tilde{q}$)) appropriately.
Consequently, $q+\tilde{q}-2$ FI parameters and two mass parameters remain.
We restrict the range of these parameters in the strip
\begin{equation}
\left|\mathrm{Im}\left(\eta_{\alpha}\right)\right|<1,\quad\left|\mathrm{Im}\left(\tilde{\eta}_{\alpha}\right)\right|<1,\label{eq:eta-Range}
\end{equation}
and 
\begin{equation}
\left|\mathrm{Im}\left(M\right)\right|<1,\quad\left|\mathrm{Im}\left(\tilde{M}\right)\right|<1.\label{eq:M-Range}
\end{equation}
\eqref{eq:eta-Range} and \eqref{eq:M-Range} are demanded to avoid the divergence of the matrix model.

\subsection{Fermi gas formalism\label{sec:FGF}}

In this section we see that the matrix model \eqref{eq:MM-pq-Def} can be written as a partition function of an ideal Fermi gas system as \cite{Marino:2011eh}. We first use the Cauchy determinant formula
\begin{equation}
\frac{\prod_{i<j}^{N}2\sinh\frac{\mu_{i}-\mu_{j}}{2}\prod_{i<j}^{N}2\sinh\frac{\nu_{i}-\nu_{j}}{2}}{\prod_{i,j}^{N}2\cosh\frac{\mu_{i}-\nu_{j}+c}{2}}=\det\left(\left[\frac{1}{2\cosh\frac{\mu_{i}-\nu_{j}+c}{2}}\right]_{i,j}^{N\times N}\right).\label{eq:CauchyDet}
\end{equation}
Here $\left(\left[a_{i,j}\right]_{i,j}^{N\times N}\right)$ denotes an $N\times N$ matrix whose $\left(i,j\right)$ element is $a_{i,j}$.
This formula allows us to rewrite the third line of \eqref{eq:MM-pq-Def} to the $q+\tilde{q}$ determinants.
After rescaling the integration variables as $\lambda_{i}^{\left(\alpha\right)}\rightarrow\lambda_{i}^{\left(\alpha\right)}/k$ and $\tilde{\lambda}_{i}^{\left(\alpha\right)}\rightarrow\tilde{\lambda}_{i}^{\left(\alpha\right)}/k$, we rewrite the elements of the matrices in terms of a quantum mechanical system by using a formula
\begin{equation}
\frac{1}{2\cosh\left(\frac{\mu-\nu}{2k}+c\right)}=\frac{1}{2\pi}\int_{-\infty}^{\infty}dp\frac{e^{\frac{i}{2\pi k}p\left(\mu-\nu+2kc\right)}}{2\cosh\frac{p}{2}}=k\braket{\mu|\frac{e^{\frac{ic}{\pi}\hat{p}}}{2\cosh\left(\frac{\pi k}{\hbar}\hat{p}\right)}|\nu}.\label{eq:Cosh-op}
\end{equation}
Here the commutation relation of the canonical position operator $\hat{x}$ and the momentum operator $\hat{p}$ is
\begin{equation}
\left[\hat{x},\hat{p}\right]=i\hbar,\quad\hbar=2\pi k,
\end{equation}
and the normalization of the position eigenstate $\ket x$ and momentum eigenstate $\kket p$ is
\begin{equation}
\braket{x|y}=2\pi\delta\left(x-y\right),\quad\bbrakket{p|p'}=2\pi\delta\left(p-p'\right),\quad\brakket{x|p}=\frac{1}{\sqrt{k}}e^{\frac{ixp}{\hbar}}.
\end{equation}
We then appropriately put the remaining factors into the determinants and change them to position operators as
\begin{align}
 & Z_{k}^{\left(q,\tilde{q}\right)}\left(N;\eta_{\alpha},M;\tilde{\eta}_{\alpha},\tilde{M}\right)\nonumber \\
 & =\frac{1}{\left(N!\right)^{q+\tilde{q}}}\int_{-\infty}^{\infty}\left(\prod_{\alpha=1}^{q}\prod_{i=1}^{N}\frac{d\lambda_{i}^{\left(\alpha\right)}}{2\pi}\right)\left(\prod_{\alpha=1}^{\tilde{q}}\prod_{i=1}^{N}\frac{d\tilde{\lambda}_{i}^{\left(\alpha\right)}}{2\pi}\right)\nonumber \\
 & \quad\quad\times\prod_{\alpha=1}^{q}\det\left(\left[\braket{\lambda_{i}^{\left(\alpha-1\right)}|e^{-\frac{i}{4\pi k}\delta_{\alpha,1}\hat{x}^{2}}e^{-\frac{i}{2k}\eta_{\alpha}\hat{x}}\frac{e^{\frac{i}{2}M\hat{p}}}{2\cosh\frac{\hat{p}}{2}}e^{\frac{i}{2k}\eta_{\alpha}\hat{x}}e^{\frac{i}{4\pi k}\delta_{\alpha,q}\hat{x}^{2}}|\lambda_{j}^{\left(\alpha\right)}}\right]_{i,j}^{N\times N}\right)\nonumber \\
 & \quad\quad\times\prod_{\alpha=1}^{\tilde{q}}\det\left(\left[\braket{\tilde{\lambda}_{i}^{\left(\alpha-1\right)}|e^{-\frac{i}{2k}\tilde{\eta}_{\alpha}\hat{x}}\frac{e^{\frac{i}{2}\tilde{M}\hat{p}}}{2\cosh\frac{\hat{p}}{2}}e^{\frac{i}{2k}\tilde{\eta}_{\alpha}\hat{x}}|\tilde{\lambda}_{j}^{\left(\alpha\right)}}\right]_{i,j}^{N\times N}\right).
\end{align}
Now we can perform the integrations by using the Andr\'eief identity
\begin{align}
 & \frac{1}{N!}\int_{-\infty}^{\infty}\left(\prod_{i=1}^{N}\frac{d\nu_{i}}{2\pi}\right)\det\left(\left[\braket{\mu_{i}|\hat{A}|\nu_{j}}\right]_{i,j}^{N\times N}\right)\det\left(\left[\braket{\nu_{i}|\hat{B}|\sigma_{j}}\right]_{i,j}^{N\times N}\right)\nonumber \\
 & =\det\left(\left[\braket{\mu_{i}|\hat{A}\hat{B}|\sigma_{j}}\right]_{i,j}^{N\times N}\right).\label{eq:GlueForm}
\end{align}
Then, the integrand is written by a single determinant of a matrix
\begin{align}
 & Z_{k}^{\left(q,\tilde{q}\right)}\left(N;\eta_{\alpha},M;\tilde{\eta}_{\alpha},\tilde{M}\right)\nonumber \\
 & =\frac{1}{N!}\int_{-\infty}^{\infty}\left(\prod_{i=1}^{N}\frac{d\lambda_{i}}{2\pi}\right)\det\left(\left[\braket{\lambda_{i}|\prod_{\alpha=1}^{q}\frac{e^{\frac{i}{2}M\left(\hat{x}+\hat{p}\right)}}{2\cosh\frac{\hat{x}+\hat{p}+\pi\eta_{\alpha}}{2}}\prod_{\alpha=1}^{\tilde{q}}\frac{e^{\frac{i}{2}\tilde{M}\hat{p}}}{2\cosh\frac{\hat{p}+\pi\tilde{\eta}_{\alpha}}{2}}|\lambda_{j}}\right]_{i,j}^{N\times N}\right).
\end{align}
Here we have used formulas of similarity transformations
\begin{equation}
e^{-\frac{ic}{\hbar}\hat{x}}f\left(\hat{p}\right)e^{\frac{ic}{\hbar}\hat{x}}=f\left(\hat{p}+c\right),\quad e^{-\frac{ic}{2\hbar}\hat{x}^{2}}f\left(\hat{p}\right)e^{\frac{ic}{2\hbar}\hat{x}^{2}}=f\left(\hat{p}+c\hat{x}\right).
\end{equation}
In this expression, it is clear that the value of the partition function is invariant under similarity transformations.
We perform the similarity transformation with $e^{-\frac{i}{2\hbar}\hat{p}^{2}}$, which shift the position operator as $\hat{x}\rightarrow\hat{x}-\hat{p}$ through a formula
\begin{equation}
e^{-\frac{ic}{2\hbar}\hat{p}^{2}}f\left(\hat{x}\right)e^{\frac{ic}{2\hbar}\hat{p}^{2}}=f\left(\hat{x}-c\hat{p}\right).
\end{equation}
Then, we finally obtain
\begin{align}
 & Z_{k}^{\left(q,\tilde{q}\right)}\left(N;\eta_{\alpha},M;\tilde{\eta}_{\alpha},\tilde{M}\right)\nonumber \\
 & =\frac{1}{N!}\int_{-\infty}^{\infty}\left(\prod_{i=1}^{N}\frac{d\lambda_{i}}{2\pi}\right)\det\left(\left[\braket{\lambda_{i}|\hat{\rho}_{k}^{(q,\tilde{q})}\left(\hat{x},\hat{p};\eta_{\alpha},M,\tilde{\eta}_{\alpha},\tilde{M}\right)|\lambda_{j}}\right]_{i,j}^{N\times N}\right).\label{Fermigas}
\end{align}
This expression can be regarded as a partition function of an ideal Fermi gas system, and the operator $\hat{\rho}_{k}^{(q,\tilde{q})}$ can be regarded as a one-particle density matrix. The density matrix in this case is
\begin{equation}
\hat{\rho}_{k}^{\left(q,\tilde{q}\right)}\left(\hat{x},\hat{p};\eta_{\alpha},M;\tilde{\eta}_{\alpha},\tilde{M}\right)=\prod_{\alpha=1}^{q}\frac{e^{\frac{i}{2}M\hat{x}}}{2\cosh\frac{\hat{x}+\pi\eta_{\alpha}}{2}}\prod_{\alpha=1}^{\tilde{q}}\frac{e^{\frac{i}{2}\tilde{M}\hat{p}}}{2\cosh\frac{\hat{p}+\pi\tilde{\eta}_{\alpha}}{2}}.\label{eq:DM-qq}
\end{equation}

From the Fermi gas formalism \eqref{Fermigas} it follows that the grand partition function $\Xi_k^{\left(q,{\tilde q}\right)}\left(\mu;\eta_\alpha,M;{\tilde\eta}_\alpha,{\tilde M}\right)$ defined by
\begin{align}
\Xi_k^{\left(q,{\tilde q}\right)}\left(\mu;\eta_\alpha,M;{\tilde\eta}_\alpha,{\tilde M}\right)
=1+\sum_{N=1}^\infty e^{\mu N} Z^{\left(q,{\tilde q}\right)}_k\left(N;\eta_\alpha,M;{\tilde\eta}_\alpha,{\tilde M}\right),
\end{align}
is given by the Fredholm determinant of the density matrix
\begin{align}
\Xi_k^{\left(q,{\tilde q}\right)}\left(\mu;\eta_\alpha,M;{\tilde\eta}_\alpha,{\tilde M}\right)=\text{Det}\left(1+e^{\mu}{\hat\rho}_k^{\left(q,{\tilde q}\right)}\left({\hat x},{\hat p};\eta_\alpha,M;{\tilde \eta}_\alpha,{\tilde M}\right)\right).
\label{Fredholmdet}
\end{align}
By defining the modified grand potential $J^{(q,{\tilde q})}_k\left(\mu;\eta_\alpha,M;{\tilde\eta}_\alpha,{\tilde M}\right)$ as
\begin{align}
\Xi_k^{\left(q,{\tilde q}\right)}\left(\mu;\eta_\alpha,M;{\tilde\eta}_\alpha,{\tilde M}\right)=\sum_{n=-\infty}^\infty e^{J^{\left(q,{\tilde q}\right)}_k\left(\mu+2\pi in;\eta_\alpha,M;{\tilde\eta}_\alpha,{\tilde M}\right)},
\label{XivsJ}
\end{align}
we can express the partition function as
\begin{align}
Z^{\left(q,{\tilde q}\right)}_k\left(N;\eta_\alpha,M;{\tilde\eta}_\alpha,{\tilde M}\right)
=\int_{-i\infty}^{i\infty}\frac{d\mu}{2\pi i}e^{
J^{\left(q,{\tilde q}\right)}_k\left(\mu;\eta_\alpha,M;{\tilde\eta}_\alpha,{\tilde M}\right)
-\mu N}.
\label{JtoZ}
\end{align}
In the next section we study the large $N$ expansion of the matrix model by using $J^{\left(q,{\tilde q}\right)}_k$ instead of directly studying $Z^{\left(q,{\tilde q}\right)}_k$.
Especially, the large $N$ in $Z^{\left(q,{\tilde q}\right)}_k$ corresponds to large $\mu$ in $J^{\left(q,{\tilde q}\right)}_k$.

\section{Large \texorpdfstring{$N$}{N} expansion from Fermi gas formalism\label{sec:LargeN}}

In this section we show our main result.
Namely, we find that the modified grand potential $J^{\left(q,{\tilde q}\right)}_k\left(\mu;\eta_\alpha,M;{\tilde \eta}_\alpha,{\tilde M}\right)$ is given in the large $\mu$ expansion as
\begin{align}
&J_k^{\left(q,{\tilde q}\right)}\left(\mu;\eta_\alpha,M,{\tilde \eta}_\alpha;{\tilde M}\right)\nonumber \\
&=\frac{C_k^{\left(q,{\tilde q}\right)}\left(M;{\tilde M}\right)}{3}\mu^3+
B_k^{\left(q,{\tilde q}\right)}\left(\eta_\alpha,M;{\tilde \eta}_\alpha,{\tilde M}\right)\mu
+ A_k^{\left(q,{\tilde q}\right)}\left(\eta_\alpha,M;{\tilde \eta}_\alpha,{\tilde M}\right)+{\cal O}\left(e^{-\# \mu}\right),
\label{Jpert}
\end{align}
with
\begin{subequations}
\label{CBAofpqmodel}
\begin{align}
C_{k}^{\left(q,\tilde{q}\right)}\left(M;\tilde{M}\right) & =\frac{2}{\pi^{2}kq\tilde{q}\left(1+M^{2}\right)\left(1+\tilde{M}^{2}\right)},\label{Cofpqmodel}\\
B_{k}^{\left(q,\tilde{q}\right)}\left(\eta_{\alpha},M;\tilde{\eta}_{\alpha},\tilde{M}\right) & =-\frac{1}{2kq\left(1+M^{2}\right)}\left(\sum_{\alpha=1}^{\tilde{q}}\tilde{\eta}_{\alpha}^{2}+\frac{\tilde{q}}{3}\right)-\frac{1}{2k\tilde{q}\left(1+\tilde{M}^{2}\right)}\left(\sum_{\alpha=1}^{q}\eta_{\alpha}^{2}+\frac{q}{3}\right)\nonumber \\
 & \quad
 + \frac{2}{3kq\tilde{q}\left(1+M^{2}\right)\left(1+\tilde{M}^{2}\right)}
 +\frac{kq\tilde{q}}{24},\label{Bofpqmodel}\\
A_{k}^{\left(q,\tilde{q}\right)}\left(\eta_{\alpha},M;\tilde{\eta}_{\alpha},\tilde{M}\right) & =\frac{1}{4}\sum_{\pm}\left[\sum_{\alpha,\beta=1}^{\tilde{q}}{\cal A}\left(\left(1\pm iM\right)qk,\tilde{\eta}_{\alpha\beta}\right)+\sum_{\alpha,\beta=1}^{q}{\cal A}\left(\left(1\pm i\tilde{M}\right)\tilde{q}k,\eta_{\alpha\beta}\right)\right],\label{Aofpqmodel}
\end{align}
\end{subequations}
where $\eta_{\alpha\beta}=\eta_{\alpha}-\eta_{\beta}$, $\tilde{\eta}_{\alpha\beta}=\tilde{\eta}_{\alpha}-\tilde{\eta}_{\beta}$, and ${\cal A}\left(\kappa,\chi\right)$ is given by
\begin{equation}
{\cal A}\left(\kappa,\chi\right)=\frac{2\zeta\left(3\right)}{\pi^{2}\kappa}+\frac{\chi^{2}}{2\kappa}-\frac{\kappa}{12}+\frac{1}{\pi}\int_{0}^{\infty}dy\frac{1}{e^{2\pi y}-1}\frac{d}{dy}\left[\frac{\cos\left(\pi\chi y\right)}{y\tanh\frac{\pi\kappa y}{2}}-\frac{2}{\pi\kappa y^{2}}\right].\label{eq:calA-Def}
\end{equation}
Plugging \eqref{Jpert} into the inverse transformation \eqref{JtoZ}, we find that the partition function is completely determined up to the non-perturbative corrections in $1/N$ as \eqref{eq:LargeN-Gen} with
$C^{\mathcal{T}}\left(\boldsymbol{\xi}\right)=C^{\left(q,{\tilde q}\right)}_k\left(M;{\tilde M}\right)$, $B^{\mathcal{T}}\left(\boldsymbol{\xi}\right)=B^{\left(q,{\tilde q}\right)}_k\left(\eta_\alpha,M;{\tilde\eta}_\alpha,{\tilde M}\right)$ and $A^{\mathcal{T}}\left(\boldsymbol{\xi}\right)=A^{\left(q,{\tilde q}\right)}_k\left(\eta_\alpha,M;{\tilde\eta}_\alpha,{\tilde M}\right)$.
In the following two subsections, we illustrate how we have obtained these coefficients.

\subsection{\texorpdfstring{$C^{\left(q,{\tilde q}\right)}_k\left(M;{\tilde M}\right)$}{C} and \texorpdfstring{$B^{\left(q,{\tilde q}\right)}_k\left(\eta_\alpha,M;{\tilde\eta}_\alpha,{\tilde M}\right)$}{B} from semi-classical Fermi surface\label{subsec:LargeN-CB}}

First let us calculate the coefficients $C^{\left(q,{\tilde q}\right)}_k\left(M;{\tilde M}\right)$ and $B^{\left(q,{\tilde q}\right)}_k\left(\eta_\alpha,M;{\tilde\eta}_\alpha,{\tilde M}\right)$.

In order to obtain the large $\mu$ expansion of the modified grand potential \eqref{Jpert}, first we write the relation between the grand partition function and the modified grand potential \eqref{XivsJ} as
\begin{align}
&J_k^{\left(q,{\tilde q}\right)}\left(\mu;\eta_\alpha,M;{\tilde \eta}_\alpha,{\tilde M}\right)\nonumber \\
&=
\text{Tr}\log
\left(1+e^{\mu}{\hat\rho}_k^{\left(q,{\tilde q}\right)}\left({\hat x},{\hat p};\eta_\alpha,M;{\tilde \eta}_\alpha,{\tilde M}\right)\right)\nonumber \\
&\quad -\log\left[
1+
\sum_{n\neq 0}
e^{J_k^{\left(q,{\tilde q}\right)}\left(\mu+2\pi in;\eta_\alpha,M;{\tilde \eta}_\alpha,{\tilde M}\right)
-J_k^{\left(q,{\tilde q}\right)}\left(\mu;\eta_\alpha,M;{\tilde \eta}_\alpha,{\tilde M}\right)
}
\right],
\label{decomposeXiosc}
\end{align}
where we have also used \eqref{Fredholmdet}.
Now if we assume that the large $J_k^{\left(q,{\tilde q}\right)}\left(\mu;\eta_\alpha,M;{\tilde\eta}_\alpha,{\tilde M}\right)$ has the structure \eqref{Jpert}, namely
$J_k^{\left(q,{\tilde q}\right)}\left(\mu;\eta_\alpha,M;{\tilde\eta}_\alpha,{\tilde M}\right)=\frac{C}{3}\mu^3+\cdots$
with some real positive constant $C$, we find that 
the second term in the right-hand side of \eqref{decomposeXiosc} gives only non-perturbatively small corrections in $1/\mu$, ${\cal O}\left(e^{-4\pi^2 C\mu}\right)$.
Hence we can ignore these corrections for our purpose of obtaining the perturbative part of the large $\mu$ expansion of $J_k^{\left(q,{\tilde q}\right)}\left(\mu;\eta_\alpha,M;{\tilde\eta}_\alpha,{\tilde M}\right)$,
\begin{align}
&J_k^{\left(q,{\tilde q}\right)}\left(\mu;\eta_\alpha,M;{\tilde \eta}_\alpha,{\tilde M}\right)
\approx 
\text{Tr}\log
\left(1+e^{\mu}{\hat\rho}_k^{\left(q,{\tilde q}\right)}\left({\hat x},{\hat p};\eta_\alpha,M;{\tilde \eta}_\alpha,{\tilde M}\right)\right).
\label{ignoreXiosc}
\end{align}
Defining the one-particle Hamiltonian ${\hat H}_k^{\left(q,{\tilde q}\right)}\left({\hat x},{\hat p};\eta_\alpha,M;{\tilde \eta}_\alpha,{\tilde M}\right)$ as
\begin{align}
{\hat\rho}_k^{\left(q,{\tilde q}\right)}\left({\hat x},{\hat p};\eta_\alpha,M;{\tilde \eta}_\alpha,{\tilde M}\right)
=e^{-\frac{1}{2}U\left({\hat x}\right)}
e^{-T\left({\hat p}\right)}
e^{-\frac{1}{2}U\left({\hat x}\right)}
=
e^{-{\hat H}_k^{\left(q,{\tilde q}\right)}\left({\hat x},{\hat p};\eta_\alpha,M;{\tilde \eta}_\alpha,{\tilde M}\right)},
\label{Hhatfromrho}
\end{align}
with
\begin{subequations}
\begin{align}
&U(x)=\sum_{\alpha=1}^q\log 2\cosh\frac{x+\pi\eta_\alpha}{2}-\frac{iqMx}{2},\\
&T(p)=\sum_{\alpha=1}^{\tilde q}\log 2\cosh\frac{p+\pi{\tilde \eta}_\alpha}{2}-\frac{i{\tilde q}{\tilde M}p}{2},
\end{align}
\end{subequations}
we can write the right-hand side of \eqref{ignoreXiosc} in terms of the number of eigenstates of ${\hat H}_k^{\left(q,{\tilde q}\right)}\left({\hat x},{\hat p};\eta_\alpha,M;{\tilde \eta}_\alpha,{\tilde M}\right)$
with $E_n \le E$
\begin{align}
n(E)=\text{Tr}\theta\left(
E-{\hat H}_k^{\left(q,{\tilde q}\right)}\left({\hat x},{\hat p};\eta_\alpha,M;{\tilde \eta}_\alpha,{\tilde M}\right)
\right),
\end{align}
where
\begin{align}
\theta(z)=\begin{cases}
0&\quad (z< 0)\\
1&\quad (z\ge 0)
\end{cases},
\end{align}
as\footnote{
Here we performed the partial integration because after performing it we can ignore the non-perturbative part ${\cal O}\left(e^{-E}\right)$ in \eqref{nEatlargeE}.
}
\begin{align}
J_k^{\left(q,{\tilde q}\right)}\left(\mu;\eta_\alpha,M;{\tilde \eta}_\alpha,{\tilde M}\right)
\approx 
\int_0^\infty dE\frac{dn\left(E\right)}{dE}\log\left(1+e^{\mu-E}\right)
=\int_0^\infty dEn\left(E\right)\frac{e^{\mu-E}}{1+e^{\mu-E}}.
\label{JfromnE}
\end{align}
Here in the second equality in \eqref{JfromnE} we have assumed $n\left(0\right)=0$ and that $\lim_{E\rightarrow\infty} n\left(E\right)e^{-E}\rightarrow 0$, both of which are true in the current setup.
The same calculation of the modified grand potential was performed for the ABJM theory with or without mass deformations in \cite{Marino:2011eh,Nosaka:2015iiw}.
Since our Hamiltonian \eqref{Hhatfromrho} has the same structure as that for the (mass deformed) ABJM theory, the calculation for our setup is parallel to that in \cite{Nosaka:2015iiw} as we demonstrate below, and we obtain the number of states in the large $E$ expansion as
\begin{align}
n(E)=C^{\left(q,{\tilde q}\right)}_k\left(M;{\tilde M}\right)E^2+B^{\left(q,{\tilde q}\right)}_k\left(\eta_\alpha,M;{\tilde\eta}_\alpha,{\tilde M}\right)
-\frac{\pi^2 C^{\left(q,{\tilde q}\right)}_k\left(M;{\tilde M}\right)}{3}
+{\cal O}\left(e^{-\# E}\right),
\label{nEatlargeE}
\end{align}
with $C_k^{\left(q,{\tilde q}\right)}\left(M,;{\tilde M}\right)$ \eqref{Cofpqmodel} and $B_k^{\left(q,{\tilde q}\right)}\left(\eta_\alpha,M;{\tilde \eta}_\alpha,{\tilde M}\right)$ \eqref{Bofpqmodel}.
Plugging this into \eqref{JfromnE} and using the integration formula
\begin{align}
\int_0^\infty dEE^a\frac{e^{\mu-E}}{1+e^{\mu-E}}&=
-\Gamma\left(a+1\right)\text{Li}_{a+1}\left(-e^{\mu}\right)\nonumber \\
&=
\frac{\left(2\pi i\right)^{a+1}}{a+1}B_{a+1}\left(\frac{\mu}{2\pi i}+\frac{1}{2}\right)-\left(-1\right)^a\Gamma\left(a+1\right)\text{Li}_{a+1}\left(-e^{-\mu}\right),
\end{align}
where in the second line we have used the formula \eqref{polylogreflectionid},
we obtain the large $\mu$ expansion of the modified grand potential as
\begin{align}
J_k^{\left(q,{\tilde q}\right)}\left(\mu;\eta_\alpha,M;{\tilde \eta}_\alpha,{\tilde M}\right)&=\frac{C_k^{\left(q,{\tilde q}\right)}\left(M;{\tilde M}\right)}{3}\mu^3+
B_k^{\left(q,{\tilde q}\right)}\left(\eta_\alpha,M;{\tilde \eta}_\alpha,{\tilde M}\right)\mu\nonumber \\
&\quad + (\text{const.})+{\cal O}\left(e^{-\# \mu}\right).
\end{align}
Here (const.) is a $\mu$-independent constant.
We cannot determine this constant in the current analysis since ${\cal O}\left(e^{-E}\right)$ corrections in $n\left(E\right)$ also contribute to this constant.

\begin{figure}
\begin{center}
\includegraphics[width=8cm]{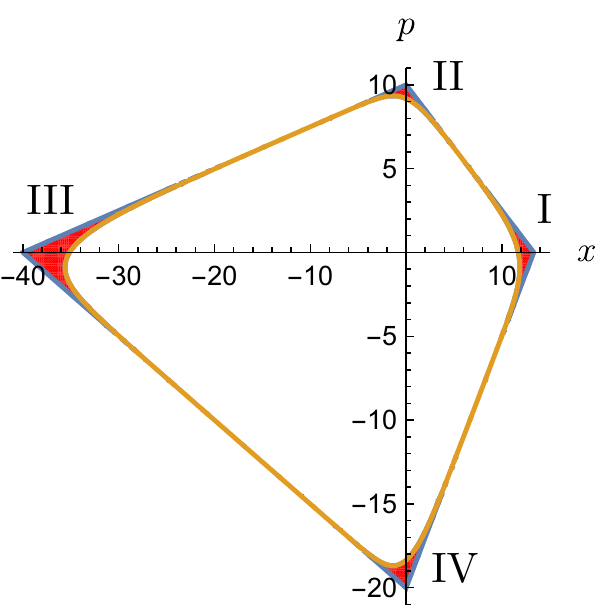}
\caption{
Shape of the Fermi surface $H_W(x,p)=E$ for large $E$ (boundary of yellow region) and the small deviations I, II, III, IV from the squashed diamond \eqref{squasheddiamond} in \eqref{nEbyI_II_III_IV} (red regions).
Here to draw the regions we have chosen the parameters as $q=2$, ${\tilde q}=3$, $M=i/2$, ${\tilde M}=i/3$, $\eta_1=1/3$, ${\tilde \eta}_1=1/4$, ${\tilde \eta}_2=1/5$, $\hbar=1/10$, $E=20$ and approximated $H_W\left(x,p\right)$ with \eqref{HW}.
}
\label{fig_squasheddiamond}
\end{center}
\end{figure}

To calculate the number of states, first we define the Wigner transformation of quantum operators as
\begin{align}
{\cal O}_W=\int_{-\infty}^\infty \frac{dy}{2\pi} e^{\frac{ipy}{\hbar}}\braket{x-\frac{y}{2}|{\hat {\cal O}}|x+\frac{y}{2}},
\end{align}
which satisfies the following properties:
\begin{align}
\left(f\left({\hat x}\right)\right)_W=f\left(x\right),\quad
\left(f\left({\hat p}\right)\right)_W=f\left(p\right),\quad
\left({\hat A}{\hat B}\right)_W=A_W\star B_W,\quad
\text{tr}{\hat {\cal O}}=\int\frac{dxdp}{2\pi\hbar}{\cal O}_W,
\end{align}
with
\begin{align}
\star=e^{\frac{i\hbar}{2}\left(
{\overleftarrow \partial}_x
{\overrightarrow \partial}_p
-
{\overleftarrow \partial}_p
{\overrightarrow \partial}_x
\right)}.
\end{align}
The number of states $n(E)$ is obtained from the Wigner transformation $H_W(x,p)$ of the Hamiltonian ${\hat H}_k^{\left(q,{\tilde q}\right)}\left({\hat x},{\hat p};\eta_\alpha,M;{\tilde\eta}_\alpha,{\tilde M}\right)$ as\footnote{
Precisely speaking, we also have to take into account the deviation of $\theta\left(E-H_W\right)$ from $\theta\left(E-{\hat H}^{\left(q,\tilde{q}\right)}_k\left({\hat x},{\hat p};\eta_\alpha,M;\tilde{\eta}_\alpha,\tilde{M}\right)\right)_W$, which is given by the derivatives of $\theta\left(E-H_W\right)$ with respect to $H_W$ multiplied by ${\cal G}_r$'s with $r\ge 2$ defined by \eqref{calG} (with $v_\alpha=\pi\eta_\alpha/\hbar$ and $\tilde{v}_\alpha=\pi\tilde{\eta}_\alpha/\hbar$).
Since the derivatives of the step function are supported only around the Fermi surface $H_W\left(x,p\right)=E$ and ${\cal G}_r$ in the current setup are exponentially suppressed in $E$ on the Fermi surface, the deviation of $\theta\left(E-H_W\right)$ gives only the non-perturbative corrections in $1/E$ in the large $E$ expansion of the number of states $n\left(E\right)$.
}
\begin{align}
n(E)=\int_{H_W\le E}\frac{dxdp}{2\pi\hbar}.
\end{align}
By using the Baker-Campbell-Hausdorff formula we can calculate the Hamiltonian as
\begin{align}
{\hat H}^{\left(q,{\tilde q}\right)}_k\left({\hat x},{\hat p};\eta_\alpha,M;{\tilde\eta}_\alpha,{\tilde M}\right)&=U\left({\hat x}\right)+T\left({\hat p}\right)-\frac{1}{24}\left[U\left({\hat p}\right),\left[U\left({\hat p}\right),T\left({\hat p}\right)\right]\right]\nonumber \\
&\quad -\frac{1}{12}\left[T\left({\hat p}\right),\left[U\left({\hat p}\right),T\left({\hat p}\right)\right]\right]+\cdots,
\end{align}
from which the Wigner transformation $H_W(x,p)$ is
\begin{align}
H_W(x,p)=U(x)+T(p)+\frac{\hbar^2}{24}\left(\partial_x U(x)\right)^2\partial_p^2T(p)-\frac{\hbar^2}{12}\partial_x^2U(x)\left(\partial_pT(p)\right)^2+\cdots.
\label{HW}
\end{align}
Here we have omitted the terms in $H_W(x,p)$ which contains either $\partial_x^mU(x)\partial_p^nT(p)$ with $m,n\ge 2$, $\partial_x^mU(x)\partial_pT(p)$ with $m\ge 3$ or $\partial_xU(x)\partial_p^mT(p)$ with $m\ge 3$, which do not affect the final result \eqref{nEatlargeE}.
When $\left|x\right|$ or $\left|p\right|$ is large, $U(x)$, $T(p)$ and their derivatives are expanded as
\begin{subequations}
\begin{align}
&U(x)=\frac{q\left|x\right|}{2}-\frac{iqMx}{2}+{\cal O}\left(e^{-\left|x\right|}\right),\\
&\partial_xU(x)=\frac{q\text{sgn}\left(x\right)}{2}-\frac{iqM}{2}+{\cal O}\left(e^{-\left|x\right|}\right),\\
&\partial_x^2U(x)={\cal O}\left(e^{-\left|x\right|}\right),\\
&T(p)=\frac{{\tilde q}\left|p\right|}{2}-\frac{i{\tilde q}{\tilde M}p}{2}+{\cal O}\left(e^{-\left|p\right|}\right),\\
&\partial_pT(p)=\frac{{\tilde q}\text{sgn}\left(p\right)}{2}-\frac{i{\tilde q}{\tilde M}}{2}+{\cal O}\left(e^{-\left|p\right|}\right),\\
&\partial_p^2T(p)={\cal O}\left(e^{-\left|p\right|}\right).
\end{align}
\end{subequations}
When $E$ is large, the points on the Fermi surface $H_W\left(x,p\right)=E$ satisfy either $\left|x\right|\sim E$ or $\left|p\right|\sim E$.
For generic points where both $\left|x\right|$ and $\left|p\right|$ are large, the Fermi surface is approximated by a squashed diamond
\begin{align}
\frac{q\left|x\right|}{2}-\frac{iqMx}{2}+\frac{{\tilde q}\left|p\right|}{2}-\frac{i{\tilde q}{\tilde M}p}{2}=E,
\label{squasheddiamond}
\end{align}
whose volume is $\frac{8E^2}{q{\tilde q}\left(1+M^2\right)\left(1+{\tilde M}^2\right)}$.
We can calculate $n\left(E\right)$ by subtracting the volumes of the deviations at the four corners of the squashed diamond where either $\left|x\right|\ll E$ or $\left|p\right|\ll E$ as
\begin{align}
n\left(E\right)=\frac{1}{2\pi\hbar}
\left[
\frac{8E^2}{q{\tilde q}\left(1+M^2\right)\left(1+{\tilde M}^2\right)}
-\text{vol}_\text{I}
-\text{vol}_\text{II}
-\text{vol}_\text{III}
-\text{vol}_\text{IV}
\right].
\label{nEbyI_II_III_IV}
\end{align}
See figure \ref{fig_squasheddiamond}.
To calculate the volume of region I, where $x\sim E$ while $p$ is not necessarily of ${\cal O}\left(E\right)$, we use only the expansion formulas for $U(x)$ and its derivatives to write $H_W(x,p)$ as
\begin{align}
H_W=
\frac{q(1-iM)x}{2}
+T(p)
+\frac{q^2\left(1-iM\right)^2\hbar^2}{96}\partial_p^2T(p)
+
\frac{q(1-iM)}{2}
f\left(\partial_p^3T(p),\cdots\right),
\end{align}
where we have ignored ${\cal O}\left(e^{-E}\right)$ corrections.
Here $f$ is an unknown function of $p$ containing only the third or higher derivatives of $T(p)$ which we have ignored in \eqref{HW}.
Solving $H_W(x,p)=E$ we obtain the shape $x_\text{in}(p)$ of the inner boundary of the region I (orange line in figure \ref{fig_squasheddiamond}) as
\begin{align}
x_\text{in}(p)=\frac{2}{q\left(1-iM\right)}\left[E-T(p)-\frac{q^2\left(1-iM\right)^2\hbar^2}{96}\partial_p^2T(p)-\frac{q\left(1-iM\right)}{2}f\left(\partial_p^3T(p),\cdots\right)\right].
\end{align}
We can calculate the volume of region I from this $x_\text{in}\left(p\right)$ and the outer boundary $x_\text{out}\left(p\right)=\frac{2}{q\left(1-iM\right)}\left(E-\frac{{\tilde q}\left|p\right|}{2}+\frac{i{\tilde q}{\tilde M}p}{2}\right)$ obtained from \eqref{squasheddiamond} (blue line in figure \ref{fig_squasheddiamond}) as
\begin{align}
\text{vol}_\text{I}&=\int_{-\infty}^\infty dp\left(x_\text{out}\left(p\right)-x_\text{in}\left(p\right)\right)\nonumber \\
&=
\frac{2}{q\left(1-iM\right)}
\int_{-\infty}^\infty dp\left[
T(p)-\frac{{\tilde q}\left|p\right|}{2}+\frac{i{\tilde q}{\tilde M}p}{2}
+\frac{q^2\left(1-iM\right)^2\hbar^2}{96}\partial_p^2T(p)\right.\nonumber \\
&\quad \left.+\frac{q\left(1-iM\right)}{2}f\left(\partial_p^3T(p),\cdots\right)\right].
\end{align}
Here we have set the boundary of the integration, where $p={\cal O}\left(E\right)$, to simply $p=\pm\infty$ since the integrand is exponentially suppressed at large $p$.
Notice that the integration of the last term amounts to a function of the second or higher derivatives of $T(p)$ evaluated at $p=\pm\infty$, which vanishes regardless of the concrete form of the function $f$.
Evaluating the other contributions as
\begin{subequations}
\begin{align}
\int_{-\infty}^\infty dp\left[T(p)-\frac{{\tilde q}\left|p\right|}{2}+\frac{i{\tilde q}{\tilde M}p}{2}\right]
&=\sum_{\alpha=1}^{\tilde q}
\left(
\int_0^\infty dp\log\left(1+e^{-p-\pi{\tilde\eta}_\alpha}\right)
+\int^0_{-\infty} dp\log\left(1+e^{p+\pi{\tilde\eta}_\alpha}\right)
\right)\nonumber \\
&=-\sum_{\alpha=1}^{{\tilde q}}\left(\text{Li}_2\left(-e^{\pi{\tilde\eta}_\alpha}\right)
+\text{Li}_2\left(-e^{-\pi{\tilde\eta}_\alpha}\right)
\right)\nonumber \\
&=\sum_{\alpha=1}^{{\tilde q}}\left(\frac{\pi^2{\tilde\eta}_\alpha^2}{2}+\frac{\pi^2}{6}\right),\\
\int_{-\infty}^\infty dp\partial_p^2T\left(p\right)&=\left[\partial_pT\left(p\right)\right]_{p=-\infty}^\infty={\tilde q},
\end{align}
\end{subequations}
where in the last step in the first integration we have used the formula for the polylogarithm \eqref{polylogreflectionid}, we obtain
\begin{align}
\text{vol}_\text{I}=\frac{\pi^2}{q\left(1-iM\right)}\left(\sum_{\alpha=1}^{{\tilde q}}{\tilde\eta}_\alpha^2+\frac{{\tilde q}}{3}\right)+\frac{\hbar^2q{\tilde q}\left(1-iM\right)}{48}.
\label{volIfinal}
\end{align}
In the same manner we obtain $\text{vol}_\text{II}$, $\text{vol}_\text{III}$ and $\text{vol}_\text{IV}$ as
\begin{subequations}
\label{volII_III_IV}
\begin{align}
&\text{vol}_\text{II}=\frac{\pi^2}{{\tilde q}\left(1-i{\tilde M}\right)}\left(\sum_{\alpha=1}^{q}\eta_\alpha^2+\frac{q}{3}\right)-\frac{\hbar^2q{\tilde q}\left(1-i{\tilde M}\right)}{24},\\
&\text{vol}_\text{III}=\frac{\pi^2}{q\left(1+iM\right)}\left(\sum_{\alpha=1}^{{\tilde q}}{\tilde\eta}_\alpha^2+\frac{{\tilde q}}{3}\right)+\frac{\hbar^2q{\tilde q}\left(1+iM\right)}{48},\\
&\text{vol}_\text{IV}=\frac{\pi^2}{{\tilde q}\left(1+i{\tilde M}\right)}\left(\sum_{\alpha=1}^{q}\eta_\alpha^2+\frac{q}{3}\right)-\frac{\hbar^2q{\tilde q}\left(1+i{\tilde M}\right)}{24}.
\end{align}
\end{subequations}
Plugging \eqref{volIfinal} and \eqref{volII_III_IV} into \eqref{nEbyI_II_III_IV} we finally obtain the large $E$ expansion of the number of states as \eqref{nEatlargeE}.

\subsection{\texorpdfstring{$A_k^{\left(q,{\tilde q}\right)}\left(\eta_\alpha,M;{\tilde \eta}_\alpha,{\tilde M}\right)$}{A} from WKB expansion of spectral traces\label{subsec:LargeN-A}}
In the analysis in the previous section we cannot determine the constant part $A_k^{\left(q,{\tilde q}\right)}\left(\eta_\alpha,M;{\tilde \eta}_\alpha,{\tilde M}\right)$ in the large $\mu$ expansion of the grand potential since it also depends on the non-perturbative corrections in $1/E$ to the number of states $n\left(E\right)$.
Nevertheless, we can still calculate the modified grand potential order by order in $\hbar$ in the following way.
First we write \eqref{ignoreXiosc} as \cite{Hatsuda:2015oaa}
\begin{align}
J^{\left(q,{\tilde q}\right)}_k\left(\mu;\eta_\alpha,M;{\tilde\eta}_\alpha,{\tilde M}\right)&\approx \sum_{n=1}^\infty\frac{(-1)^n}{n}\text{Tr}{\hat\rho}^{\left(q,{\tilde q}\right)}_k\left({\hat x},{\hat p};\eta_\alpha,M;{\tilde\eta}_\alpha,{\tilde M}\right)^ne^{n\mu}\nonumber \\
&=\int^{i\infty+0^+}_{-i\infty+0^+}\frac{ds}{2\pi i}
\Gamma\left(s\right)
\Gamma\left(-s\right)
{\cal Z}^{\left(q,{\tilde q}\right)}\left(s;\eta_\alpha,M;{\tilde\eta}_\alpha,{\tilde M}\right)e^{s\mu},
\label{JbyWKB}
\end{align}
where ${\cal Z}^{\left(q,{\tilde q}\right)}\left(s;\eta_\alpha,M;{\tilde\eta}_\alpha,{\tilde M}\right)$ is the spectral zeta function
\begin{align}
{\cal Z}^{\left(q,{\tilde q}\right)}\left(s;\eta_\alpha,M;{\tilde\eta}_\alpha,{\tilde M}\right)=\text{Tr}{\hat\rho}^{\left(q,{\tilde q}\right)}_k\left({\hat x},{\hat p};\eta_\alpha,M;{\tilde\eta}_\alpha,{\tilde M}\right)^s.
\label{calZ}
\end{align}
As commented below \eqref{ignoreXiosc}, the equality $\approx$ in \eqref{ignoreXiosc} is valid up to the ${\cal O}\left(e^{-4\pi^2C_k^{\left(q,{\tilde q}\right)}\left(M,{\tilde M}\right)\mu}\right)$ corrections to $J^{\left(q,{\tilde q}\right)}_k\left(\mu;\eta_\alpha,M;{\tilde\eta}_\alpha,{\tilde M}\right)$, which are invisible in the small $\hbar$ expansion in any case.
Hence in the following we simply denote the first equality $\approx$ in \eqref{JbyWKB} as $=$.
In \eqref{JbyWKB}, the small $e^{\mu}$ expansion of $J^{\left(q,{\tilde q}\right)}_k\left(\mu;\eta_\alpha,M;{\tilde\eta}_\alpha,{\tilde M}\right)$ is obtained by evaluating the last integration by summing over the residues in $\text{Re}\left[s\right]>0$, which are at $s=1,2,\cdots$, while we can obtain the large $\mu$ expansion of $J^{\left(q,{\tilde q}\right)}_k\left(\mu;\eta_\alpha,M;{\tilde\eta}_\alpha,{\tilde M}\right)$ by closing the contour to the right and summing over the residues in $\text{Re}\left[s\right]\le 0$.
In particular, the perturbative part of the large $\mu$ expansion is given by the residue at $s=0$:
\begin{align}
J^{\left(q,{\tilde q}\right)}_k\left(\mu;\eta_\alpha,M;{\tilde\eta}_\alpha,{\tilde M}\right)
&=-\text{Res}\left[
\Gamma\left(s\right)
\Gamma\left(-s\right)
{\cal Z}^{\left(q,{\tilde q}\right)}\left(s;\eta_\alpha,M;{\tilde\eta}_\alpha,{\tilde M}\right)e^{s\mu},s\rightarrow 0\right]\nonumber \\
&\quad +{\cal O}\left(e^{-\# \mu}\right).
\label{JpertfromcalZ}
\end{align}

The spectral zeta function can be calculated order by order in the $\hbar$ expansion by using the Wigner transformation
\begin{align}
{\cal Z}^{\left(q,{\tilde q}\right)}\left(s;\eta_\alpha,M;{\tilde\eta}_\alpha,{\tilde M}\right)=\int\frac{dxdp}{2\pi\hbar}
\left({\hat\rho}^{\left(q,{\tilde q}\right)}_k\left({\hat x},{\hat p};\eta_\alpha,M;{\tilde\eta}_\alpha,{\tilde M}\right)^s\right)_W.
\end{align}
For our purpose it is convenient to rescale $\eta_\alpha$ and ${\tilde \eta}_\alpha$ as $\eta_\alpha=\hbar v_\alpha/\pi$ and ${\tilde\eta}_\alpha=\hbar {\tilde v}_\alpha/\pi$ and consider the $\hbar$ expansion with $v_\alpha,{\tilde v}_\alpha$ kept as ${\cal O}\left(1\right)$,
\begin{align}
{\cal Z}^{\left(q,{\tilde q}\right)}\left(s;\frac{\hbar v_\alpha}{\pi},M;\frac{\hbar {\tilde v}_\alpha}{\pi},{\tilde M}\right)=\sum_{\ell=0}^\infty \hbar^{2\ell-1}{\cal Z}^{\left(q,{\tilde q}\right)}_{2\ell}\left(s;v_\alpha,M;{\tilde v}_\alpha,{\tilde M}\right).
\end{align}
Under this rescaling it is easy to calculate the leading part of the spectral zeta function as
\begin{align}
{\cal Z}^{\left(q,{\tilde q}\right)}_{0}\left(s;v_\alpha,M;{\tilde v}_\alpha,{\tilde M}\right)=
\frac{
\Gamma\left(\frac{sq\left(1+iM\right)}{2}\right)
\Gamma\left(\frac{sq\left(1-iM\right)}{2}\right)
\Gamma\left(\frac{s{\tilde q}\left(1+i{\tilde M}\right)}{2}\right)
\Gamma\left(\frac{s{\tilde q}\left(1-i{\tilde M}\right)}{2}\right)
}{
2\pi \Gamma\left(sq\right)
\Gamma\left(s{\tilde q}\right)
},
\end{align}
and we can also show that the higher order corrections have the following structure \cite{Moriyama:2014waa,Hatsuda:2015oaa,Nosaka:2015iiw}
\begin{align}
{\cal Z}^{\left(q,{\tilde q}\right)}_{2\ell}\left(s;v_\alpha,M;{\tilde v}_\alpha,{\tilde M}\right)=D^{\left(q,{\tilde q}\right)}_{2\ell}\left(s;v_\alpha,M;{\tilde v}_\alpha,{\tilde M}\right){\cal Z}^{\left(q,{\tilde q}\right)}_{0}\left(s;v_\alpha,M;{\tilde v}_\alpha,{\tilde M}\right),
\label{DrelationofcalZ}
\end{align}
with $D^{\left(q,{\tilde q}\right)}_{2\ell}\left(s;v_\alpha,M;{\tilde v}_\alpha,{\tilde M}\right)$ some rational functions of $s$.
See appendix \ref{app_WignerKirkwood} for the detailed argument.
Once we assume this structure \eqref{DrelationofcalZ}, we can determine $D^{\left(q,{\tilde q}\right)}_{2\ell}\left(s;v_\alpha,M;{\tilde v}_\alpha,{\tilde M}\right)$ by extrapolating the spectral traces for $s=1,2,\cdots$ \cite{Okuyama:2015auc}.
These data can be generated efficiently by calculating $\left({\hat\rho}^{\left(q,{\tilde q}\right)}_k\left({\hat x},{\hat p};\frac{\hbar v_\alpha}{\pi},M;\frac{\hbar {\tilde v}_\alpha}{\pi},{\tilde M}\right)^s\right)_W$ recursively as
\begin{align}
&\left({\hat\rho}^{\left(q,{\tilde q}\right)}_k\left({\hat x},{\hat p};\frac{\hbar v_\alpha}{\pi},M;\frac{\hbar {\tilde v}_\alpha}{\pi},{\tilde M}\right)^{s+1}\right)_W\nonumber \\
&=\left({\hat\rho}^{\left(q,{\tilde q}\right)}_k\left({\hat x},{\hat p};\frac{\hbar v_\alpha}{\pi},M;\frac{\hbar {\tilde v}_\alpha}{\pi},{\tilde M}\right)^s\right)_W
\star \rho^{\left(q,{\tilde q}\right)}_{k,W}\left(x,p;\frac{\hbar v_\alpha}{\pi},M,\frac{\hbar {\tilde v}_\alpha}{\pi},{\tilde M}\right),\\
&\rho^{\left(q,{\tilde q}\right)}_{k,W}\left(x,p;\frac{\hbar v_\alpha}{\pi},M;\frac{\hbar {\tilde v}_\alpha}{\pi},{\tilde M}\right)\nonumber \\
&=
\prod_{\alpha=1}^q
\frac{e^{\frac{iM}{2}x}}{F(x)\cosh\frac{\hbar v_\alpha}{2}+2F'(x)\sinh\frac{\hbar v_\alpha}{2}}
\star \prod_{\alpha=1}^{\tilde q}
\frac{e^{\frac{i{\tilde M}}{2}p}}{F(p)\cosh\frac{\hbar {\tilde v}_\alpha}{2}+2F'(p)\sinh\frac{\hbar {\tilde v}_\alpha}{2}},
\end{align}
with $F\left(x\right)=2\cosh\frac{x}{2}$, whose $\hbar$ expansion can be simplified by using the following functional identities at each step
\begin{align}
\partial_x^{2n}F\left(x\right)=2^{-2n}F\left(x\right),\quad
\partial_x^{2n-1}F\left(x\right)=2^{-2n+2}F'\left(x\right),\quad
F'\left(x\right)^2=\frac{F\left(x\right)^2}{4}-1,
\label{derivativerelationofF}
\end{align}
and then using the following integration formulas
\begin{subequations}
\label{I1andI2}
\begin{align}
I_1\left(\alpha,n\right)&=\int_{-\infty}^\infty dx\frac{e^{\alpha x}}{F(x)^n}=\frac{\Gamma(\frac{n}{2}+\alpha)\Gamma(\frac{n}{2}-\alpha)}{\Gamma(n)},\\
I_2\left(\alpha,n\right)&=\int_{-\infty}^\infty dx\frac{e^{\alpha x}F'(x)}{F(x)^n}=\frac{\alpha \Gamma(\frac{n-1}{2}+\alpha)\Gamma(\frac{n-1}{2}-\alpha)}{\Gamma(n)}.
\end{align}
\end{subequations}
Note that these $I_1\left(\alpha,n\right)$ and $I_2\left(\alpha,n\right)$ satisfy the following relations
\begin{align}
I_1\left(\alpha,n+2\right)=\frac{n^2-4\alpha^2}{4n\left(n+1\right)}I_1\left(\alpha,n\right),\quad
I_2\left(\alpha,n\right)=\frac{\alpha I_1\left(n-1,\alpha\right)}{n-1}.
\label{recursionrelationofI1I2}
\end{align}
We have calculated the explicit expressions of $D^{\left(q,{\tilde q}\right)}_{2\ell}\left(s;v_\alpha,M;{\tilde v}_\alpha,{\tilde M}\right)$ for various setups up to $\ell=5$, part of which are listed in appendix \ref{app_listofD} (see the Mathematica notebook attached to this paper in arXiv.org for more results).
Plugging those results into \eqref{JpertfromcalZ} we observe that the small $\hbar$ expansion of the modified grand potential is given as
\begin{align}
&J^{\left(q,{\tilde q}\right)}_k\left(\mu;\frac{\hbar v_\alpha}{\pi},M;\frac{\hbar {\tilde v}_\alpha}{\pi},{\tilde M}\right)=
\frac{C^{\left(q,{\tilde q}\right)}_k\left(M;{\tilde M}\right)}{3}\mu^3
+B^{\left(q,{\tilde q}\right)}_k\left(\frac{\hbar v_\alpha}{\pi},M;\frac{\hbar {\tilde v}_\alpha}{\pi},{\tilde M}\right)\mu\nonumber \\
&\quad +\frac{1}{4}\sum_\pm\left[
\sum_{\alpha=1}^{\tilde q}
{\cal A}\left(\left(1\pm iM\right)qk,\frac{\hbar\left({\tilde v}_\alpha-{\tilde v}_\beta\right)}{\pi}\right)
+\sum_{\alpha=1}^q
{\cal A}\left(\left(1\pm i{\tilde M}\right){\tilde q}k,\frac{\hbar \left(v_\alpha-v_\beta\right)}{\pi}\right)
\right]\nonumber \\
&\quad +{\cal O}\left(e^{-\# \mu}\right),
\end{align}
with
\begin{equation}
{\cal A}\left(\kappa,\chi\right)
=\sum_{\ell=0}^{\infty}\sum_{n=0}^{\ell}\frac{\left(-1\right)^{\ell+n}}{\left(2n\right)!\left(2\ell-2n\right)!}
B_{2\ell-2}B_{2\ell-2n}\pi^{2\ell-2}\kappa^{2\ell-2n-1}\chi^{2n}.\label{eq:calA-Series}
\end{equation}
Here $B_n$ are the Bernoulli numbers.
In this paper we use the notation $B_{-2}=2\zeta\left(3\right)$.\footnote{This is the reminiscent of \eqref{eq:Zeta-Ber2}.}
Although the expressions of \eqref{eq:calA-Series}  and \eqref{eq:calA-Def} are different,
we will demonstrate in section \ref{subsec:calA-Prop} that by resuming the series expansion \eqref{eq:calA-Series} we actually obtain \eqref{eq:calA-Def}.

\subsection{Properties of \texorpdfstring{$\mathcal{A}\left(\kappa,\chi\right)$}{calA}\label{subsec:calA-Prop}}

In this section we study some aspects of the function ${\cal A}\left(\kappa,\chi\right)$ defined in \eqref{eq:calA-Def}.

We first comment on the range of the arguments. 
The definition clearly shows that the function ${\cal A}\left(\kappa,\chi\right)$ diverges when $\kappa=0$ or the absolute value of the imaginary part of $\chi$ is equal to or larger than $2$. 
To avoid this divergence, we restrict the parameters as
\begin{equation}
\kappa\neq0,\quad\left|\mathrm{Im}\left(\chi\right)\right|<2.\label{eq:calA-NonDiv}
\end{equation}
This is consistent with \eqref{eq:M-Range} and \eqref{eq:eta-Range} under \eqref{Aofpqmodel}.

The parity of the function ${\cal A}\left(\kappa,\chi\right)$ can be easily reed off from the series expansion \eqref{eq:calA-Series} as
\begin{equation}
{\cal A}\left(\kappa,\chi\right)=-{\cal A}\left(-\kappa,\chi\right),\quad{\cal A}\left(\kappa,\chi\right)={\cal A}\left(\kappa,-\chi\right).
\end{equation}

Next, we comment on the relation between the function ${\cal A}\left(\kappa,\chi\right)$ and the function $A^{\mathrm{ABJM}}\left(\kappa\right)$ which appears in the ABJM theory without any deformations.
Since the ABJM theory is the $\left(1,1\right)$ model, the function $A_{k}^{\left(q,\tilde{q}\right)}$ in \eqref{Aofpqmodel} must reduce to the function $A^{\mathrm{ABJM}}\left(\kappa\right)$ as
\begin{equation}
A_{k}^{\left(1,1\right)}\left(M=0;\tilde{M}=0\right)=A^{\mathrm{ABJM}}\left(k\right).
\end{equation}
This relation implies
\begin{equation}
{\cal \mathcal{A}}\left(\kappa,0\right)=A^{\mathrm{ABJM}}\left(\kappa\right).\label{eq:calA-ABJMA}
\end{equation}
We can directly obtain this relation by comparing the small $\kappa$ expansions of both the functions.
The small $\kappa$ expansion of $A^{\mathrm{ABJM}}\left(\kappa\right)$  is known to be \cite{Marino:2011eh,Hanada:2012si}
\begin{equation}
A^{\mathrm{ABJM}}\left(\kappa\right)=\frac{2\zeta\left(3\right)}{\pi^{2}\kappa}+\sum_{\ell=1}^{\infty}\frac{\left(-1\right)^{\ell}}{\left(2\ell\right)!}B_{2\ell-2}B_{2\ell}\pi^{2\ell-2}\kappa^{2\ell-1}.\label{eq:A-ABJM-Series}
\end{equation}
On the other hand, the small $\kappa$ expansion of ${\cal A}\left(\kappa,\chi\right)$ is in \eqref{eq:calA-Series}.
Because $\chi=0$ means focusing on the $n=0$ part, the small $\kappa$ expansion of ${\cal A}\left(\kappa,0\right)$ is the same with \eqref{eq:A-ABJM-Series}, and thus \eqref{eq:calA-ABJMA} holds.

Next, we show that the function ${\cal A}\left(\kappa,\chi\right)$ has several different closed form expressions.
In other words, starting from the small $\kappa$, $\chi$ expansion in \eqref{eq:calA-Series}, there are several ways to resum ${\cal A}\left(\kappa,\chi\right)$.

First, we show that the resummation of this expansion becomes \eqref{eq:calA-Def}.
As a first step, by rearranging the summation as
\begin{equation}
\sum_{\ell=0}^{\infty}\sum_{n=0}^{\ell}f\left(\ell,n\right)
=\sum_{n=0}^{\infty}\sum_{\ell=0}^{\infty}f\left(\ell+n,n\right),
\end{equation}
we obtain
\begin{equation}
{\cal A}\left(\kappa,\chi\right)=\sum_{n=0}^{\infty}\sum_{\ell=0}^{\infty}\frac{\left(-1\right)^{\ell}}{\left(2n\right)!\left(2\ell\right)!}B_{2\ell+2n-2}B_{2\ell}\pi^{2\ell+2n-2}\kappa^{2\ell-1}\chi^{2n}.\label{eq:calA-Series2}
\end{equation}
By using \eqref{eq:Zeta-Ber1} and \eqref{eq:Zeta-Int}, we can express the Bernoulli number in an integral expression
\begin{equation}
B_{2n}=\begin{cases}
2\zeta\left(3\right) & \left(n=-1\right)\\
1 & \left(n=0\right)\\
{\displaystyle \left(-1\right)^{n-1}2\int_{0}^{\infty}dy\left(\frac{1}{e^{2\pi y}-1}\frac{d}{dy}y^{2n}\right)} & \left(n\geq1\right)
\end{cases}.\label{eq:Bernoulli-Int}
\end{equation}
Recall that in this paper we use the notation $B_{-2}=2\zeta\left(3\right)$. 
We apply this formula to the first Bernoulli number $B_{2\ell+2n-2}$ in \eqref{eq:calA-Series2}
\begin{align}
{\cal A}\left(\kappa,\chi\right) & =\frac{2\zeta\left(3\right)}{\pi^{2}\kappa}+\frac{\chi^{2}}{2\kappa}-\frac{\kappa}{12}\nonumber \\
 & \quad+\frac{2}{\pi^{2}\kappa}\int_{0}^{\infty}dy\frac{1}{e^{2\pi y}-1}\frac{d}{dy}\left[\frac{1}{y^{2}}\sum_{n=0}^{\infty}\sum_{\ell=0}^{\infty}\frac{\left(-1\right)^{n}\left(\pi\chi y\right)^{2n}}{\left(2n\right)!}\frac{B_{2\ell}\left(\pi\kappa y\right)^{2\ell}}{\left(2\ell\right)!}-\frac{1}{y^{2}}\right].
\end{align}
In this expression, we can easily perform the summations by using formulas
\begin{equation}
\sum_{n=0}^{\infty}\frac{\left(-1\right)^{n}}{\left(2n\right)!}z^{2n}=\cos z,\quad\sum_{n=0}^{\infty}\frac{B_{2n}}{\left(2n\right)!}z^{2n}=\frac{z}{2\tanh\frac{z}{2}}.\label{eq:Bernoulli-Sum}
\end{equation}
Then, we finally obtain \eqref{eq:calA-Def}.
It is important to note that while the original infinite sum expression for ${\cal A}\left(\kappa,\chi\right)$ \eqref{eq:calA-Series} is an asymptotic series which do not converge for any nonzero $\kappa$ and $\chi$, the integral expression \eqref{eq:calA-Def} is well-defined for finite values of $\kappa$ and $\chi$ within the range of \eqref{eq:calA-NonDiv}.
With the integral expression \eqref{eq:calA-Def} ${\cal A}\left(\kappa,\chi\right)$, we have also confirmed, for $q=1$, $\tilde{q}=2$, $M=\tilde{M}=0$, and $\kappa=1,2,3$, that the Airy form \eqref{eq:LargeN-Gen} with the parameters \eqref{CBAofpqmodel} reproduces the correct behavior of the partition function as $N$ increases.

Second, we obtain another closed form of the function ${\cal A}\left(\kappa,\chi\right)$ which respects the relation \eqref{eq:calA-ABJMA}. We can perform the resummation for the difference of ${\cal A}\left(\kappa,\chi\right)$ and $A^{\mathrm{ABJM}}\left(\kappa\right)$. Namely, we consider
\begin{equation}
{\cal A}\left(\kappa,\chi\right)=A^{\mathrm{ABJM}}\left(\kappa\right)+\sum_{\ell=1}^{\infty}\sum_{n=1}^{\ell}\frac{\left(-1\right)^{\ell+n}}{\left(2n\right)!\left(2\ell-2n\right)!}B_{2\ell-2}B_{2\ell-2n}\kappa^{2\ell-2n-1}\chi^{2n}\pi^{2\ell-2}.\label{eq:calA-A-Series}
\end{equation}
We can perform the resummation in a similar way, and we finally obtain
\begin{equation}
{\cal A}\left(\kappa,\chi\right)
=A^{\mathrm{ABJM}}\left(\kappa\right)+\frac{\chi^{2}}{2\kappa}
-\frac{2}{\pi}\int_{0}^{\infty}dy\frac{1}{e^{2\pi y}-1}\frac{d}{dy}\left(\frac{\sin^{2}\frac{\pi\chi y}{2}}{y\tanh\frac{\pi\kappa y}{2}}\right).
\label{eq:calA-Closed2}
\end{equation}
Note that the closed form of $A^{\mathrm{ABJM}}\left(\kappa\right)$ is known to be \cite{Hanada:2012si,Hatsuda:2014vsa}
\begin{equation}
A^{\mathrm{ABJM}}\left(\kappa\right)=\frac{2\zeta\left(3\right)}{\pi^{2}\kappa}\left(1-\frac{\kappa^{3}}{16}\right)+\frac{\kappa^{2}}{\pi^{2}}\int_{0}^{\infty}dy\frac{y}{e^{\kappa y}-1}\log\left(1-e^{-2y}\right).\label{eq:A-ABJM-Closed}
\end{equation}
The expression of ${\cal A}\left(\kappa,\chi\right)$ in \eqref{eq:calA-Closed2} with \eqref{eq:A-ABJM-Closed} is more suitable for evaluating the function ${\cal A}\left(\kappa,\chi\right)$ with the numerical integration.
Furthermore, in this expression we can calculate $\partial_\chi\mathcal{A}\left(\kappa,\chi\right)$ explicitly for integer values of $\kappa$ as follows.
First we rewrite \eqref{eq:calA-Closed2} by integrating the third term by parts as
\begin{align}
\mathcal{A}\left(\kappa,\chi\right)=A^{\text{ABJM}}\left(\kappa\right)+\frac{\chi^2}{2\kappa}-\int_0^\infty dy\frac{1}{\sinh^2\pi y}\left(\frac{\sin^2\frac{\pi\chi y}{2}}{y\tanh\frac{\pi\kappa y}{2}}-\frac{\pi\chi^2}{2\kappa}\right),
\end{align}
where we have introduced a constant $-\pi\chi^2/\left(2\kappa\right)$ under the derivative in \eqref{eq:calA-Closed2} in order to make the boundary terms vanish.
Hence the derivative $\partial_\chi{\cal A}\left(\kappa,\chi\right)$ is given as
\begin{align}
\partial_\chi\mathcal{A}\left(\kappa,\chi\right)=\frac{\chi}{\kappa}-\int_0^\infty dy\frac{1}{\sinh^2\pi y}\left(\frac{\pi \sin \pi\chi y}{2\tanh\frac{\pi\kappa y}{2}}-\frac{\pi\chi}{\kappa}\right).
\end{align}
Since the integrand is an even function of $y$ and finite at $y\rightarrow 0$ in total, we can evaluate the integration term by term by the principal value integral over $\left(-\infty,\infty\right)$
\begin{align}
\partial_\chi\mathcal{A}\left(\kappa,\chi\right)=\frac{\chi}{\kappa}
-\frac{1}{2}
\fint_{-\infty}^\infty 
dy\frac{1}{\sinh^2\pi y}\frac{\pi \sin \pi\chi y}{2\tanh\frac{\pi\kappa y}{2}}
+\frac{1}{2}\fint_{-\infty}^\infty dy\frac{1}{2\sinh^2\pi y}\frac{\pi\chi}{\kappa}.
\label{dAstep1}
\end{align}
Here we have denoted the principal value integral as $\fint$.
These integrations can be performed systematically with the following formula after an appropriate regularization
\begin{align}
&\fint_{-\infty}^\infty dy\frac{e^{\alpha y}}{\prod_{a=1}^n \sinh\pi \nu_a\left(y-\beta_a\right)}\nonumber \\
&=\frac{1}{1-\left(-1\right)^{\sum_a\nu_a}e^{i\alpha}}\fint_\gamma dy\frac{e^{\alpha y}}{\prod_{a=1}^n \sinh\pi \nu_a
\left(y-\beta_a
\right)}\nonumber \\
&=\frac{2\pi i}{1-\left(-1\right)^{\sum_a\nu_a}e^{i\alpha}}\sum_{a=1}^n\left(
\frac{1}{2}\sum_{j=0,\nu_a}\frac{\left(-1\right)^j e^{\alpha \left(\beta_a+\frac{ij}{\nu_a}\right)}}{\pi \nu_a\prod_{a'\left(\neq a\right)}\sinh\pi\nu_{a'}\left(\beta_a+\frac{ij}{\nu_a}-\beta_{a'}\right)}
\right.\nonumber \\
&\quad \left.
+\sum_{j=1}^{\nu_a-1}
\frac{\left(-1\right)^j e^{\alpha \left(\beta_a+\frac{ij}{\nu_a}\right)}}{\pi \nu_a\prod_{a'\left(\neq a\right)}\sinh\pi\nu_{a'}\left(\beta_a+\frac{ij}{\nu_a}-\beta_{a'}\right)}\right),
\label{Boxintegralformula}
\end{align}
where $\nu_a$ are positive integers with $\text{GCD}\left(\nu_a\right)=1$, $\alpha$ is a generic complex number with $|\text{Re}\left[\alpha\right]|<\sum a\pi\nu_a$, $\beta_a$ are real numbers and $\gamma$ is a contercrockwise contour surrounding the regime $0\le\text{Im}[y]\le 1$.
Here the second expression in \eqref{Boxintegralformula} is obtained due to the quasi-periodicity of the integrand under $y\rightarrow y+i$, while the third expression is obtained by applying the Cauchy's residue theorem to the second expression.
From the second term in the integrand in \eqref{dAstep1}, we obtain
\begin{align}
\frac{1}{2}\fint_{-\infty}^\infty dy\frac{1}{\sinh^2\pi y}\frac{\pi\chi}{\kappa}=\frac{\pi\chi}{2\kappa}\lim_{\epsilon_1,\epsilon_2\rightarrow 0}\fint dy\frac{e^{\epsilon_1y}}{\sinh\pi y\sinh \pi(y-\epsilon_2)}
=-\frac{\chi}{\kappa},
\end{align}
which cancels with the first term $\chi/\kappa$ in \eqref{dAstep1}.
Hence $\partial_\chi\mathcal{A}\left(\kappa,\chi\right)$ is given by the contribution from the first term in the integrand of \eqref{dAstep1}, which we can calculate as
\begin{align}
&\partial_\chi\mathcal{A}\left(\kappa,\chi\right)\\
&=
-\frac{1}{2}\fint_{-\infty}^\infty dy\frac{1}{\sinh^2\pi y}\left(\frac{\pi\sin \pi\chi y}{2\tanh\frac{\pi\kappa y}{2}}\right)\nonumber \\
&=
\sum_\pm \frac{\pi i}{8}
\lim_{\epsilon_1,\epsilon_2\rightarrow 0}
\fint_{-\infty}^\infty dy\frac{e^{\pi\left(i\chi\pm\frac{\kappa}{2}\right)y}}{\sinh \pi y\sinh\pi(y-\epsilon_1)\sinh\frac{\pi\kappa(y-\epsilon_3)}{2}}\nonumber \vspace{0.1cm}\\
&=\begin{cases}
\displaystyle \frac{\pi}{1-e^{-\pi\chi}}\left[\frac{\left(1+e^{-\pi\chi}\right)\left(4-\kappa^2+6\chi^2\right)}{24\kappa}+\sum_{a=1}^{\frac{\kappa}{2}-1}\frac{e^{-\frac{2\pi\chi a}{\kappa}}}{\kappa\sin^2\frac{2\pi a}{\kappa}}\right]&\quad \left(\kappa\text{: even}\right)\vspace{0.1cm}\\
\displaystyle \frac{\pi}{1-e^{-2\pi\chi}}\left[\frac{\left(1+e^{-2\pi\chi}\right)\left(4-\kappa^2+6\chi^2\right)}{24\kappa}-\frac{\kappa}{4}e^{-\pi\chi}
+\sum_{a=1}^{\kappa-1}\frac{e^{-\frac{2\pi\chi a}{\kappa}}}{\kappa\sin^2\frac{2\pi a}{\kappa}}
\right]&\quad \left(\kappa\text{: odd}\right)
\end{cases}.
\end{align}
In particular, from these results we find the small $\chi$ expansion of $\mathcal{A}\left(\kappa,\chi\right)$ with $\kappa=1$ as
\begin{equation}
{\cal A}\left(1,\chi\right)=A^{\mathrm{ABJM}}\left(1\right)+\frac{1}{8}\left(1+\frac{\pi^{2}}{4}\right)\chi^{2}+\frac{\pi^{2}}{48}\left(1-\frac{\pi^{2}}{16}\right)\chi^{4}+\cdots.\label{eq:calA-k1}
\end{equation}
We will use this expression later.

Third, we obtain another closed form expression of the function ${\cal A}\left(\kappa,\chi\right)$, which we will use to prove the property of the function ${\cal A}\left(\kappa,\chi\right)$. We start with \eqref{eq:calA-Series2} and divide the summation as
\begin{equation}
\mathcal{A}\left(\kappa,\chi\right)=\sum_{\ell=0}^{\infty}\frac{\left(-1\right)^{\ell}}{\left(2\ell\right)!}B_{2\ell}\left(\pi\kappa\right)^{2\ell-1}a_{\ell}\left(\chi\right),\label{eq:calA-Series3}
\end{equation}
where
\begin{equation}
a_{\ell}\left(z\right)=\frac{1}{\pi}\sum_{n=0}^{\infty}\frac{1}{\left(2n\right)!}B_{2\ell+2n-2}\left(\pi z\right)^{2n}.\label{eq:al-Def}
\end{equation}
Notice that $a_{\ell}\left(z\right)$ with $\ell\geq1$ is given as derivatives of $a_{0}\left(z\right)$ as
\begin{equation}
a_{\ell}\left(z\right)=\left(\frac{1}{\pi}\frac{d}{dz}\right)^{2\ell}a_{0}\left(z\right).
\label{eq:al-a0}
\end{equation}
After plugging this expression into \eqref{eq:calA-Series3}, we can perform the resummation by using \eqref{eq:Bernoulli-Sum}. Namely, we can formally write the function $\mathcal{A}\left(\kappa,\chi\right)$ as
\begin{equation}
\mathcal{A}\left(\kappa,\chi\right)=\frac{i}{2\pi\tanh\frac{i\kappa}{2}\frac{d}{d\chi}}\frac{d}{d\chi}a_{0}\left(\chi\right).
\end{equation}
Interestingly, the function $\left(\tanh\pi z\right)^{-1}$ is related to the double sine function as \eqref{eq:DDS-b1}. Then, we can use the integral expression of the double sine function in \eqref{eq:DS-Integral}, and we obtain
\begin{equation}
\mathcal{A}\left(\kappa,\chi\right)
=-\frac{i}{2\pi}\frac{d}{d\chi}a_{0}\left(\chi\right)
-\frac{2}{\pi\kappa}\int_{\mathbb{R}+i0^+}\frac{a_{0}\left(\chi+\frac{\kappa}{\pi}t\right)}{\left(2\sinh t\right)^{2}}dt.
\end{equation}
Our remaining task is to obtain a closed form expression of $a_{0}\left(\chi\right)$. 
The function $a_{0}\left(\chi\right)$ can be resummed by using a formula which we can obtain by integrating \eqref{eq:Bernoulli-Sum} twice (see \eqref{eq:PL-Der})
\begin{equation}
\sum_{n=0}^{\infty}\frac{B_{2n-2}}{\left(2n\right)!}z^{2n}=-\frac{\pi^{2}}{6}z-\frac{1}{12}z^{3}-z\mathrm{Li}_{2}\left(e^{z}\right)+2\mathrm{Li}_{3}\left(e^{z}\right).
\end{equation}
Note that in this paper $B_{-2}=2\zeta\left(3\right)$. 
By comparing this with $a_{0}\left(\chi\right)$ in small $\chi$ expansion \eqref{eq:al-Def}, we find
\begin{equation}
a_{0}\left(z\right)=-\frac{\pi^{2}}{6}z-\frac{\pi^{2}}{12}z^{3}-z\mathrm{Li}_{2}\left(e^{\pi z}\right)+\frac{2}{\pi}\mathrm{Li}_{3}\left(e^{\pi z}\right).\label{eq:a0-Closed}
\end{equation}
By using this expression and \eqref{eq:DS-b1} we also find
\begin{equation}
\frac{d}{dz}a_{0}\left(z\right)=-2\pi i\log s_{b=1}\left(\frac{z}{2}\right).
\end{equation}
Therefore, we finally obtain the third closed form expression of $\mathcal{A}\left(\kappa,\chi\right)$
\begin{equation}
\mathcal{A}\left(\kappa,\chi\right)
=-\log s_{b=1}\left(\frac{\chi}{2}\right)
-\frac{2}{\pi\kappa}\int_{\mathbb{R}+i0^+}\frac{a_{0}\left(\chi+\frac{\kappa}{\pi}t\right)}{\left(2\sinh t\right)^{2}}dt,
\label{eq:calA-Closed3}
\end{equation}
with \eqref{eq:a0-Closed} and \eqref{eq:DS-b1}.

We have obtained three closed form expressions, \eqref{eq:calA-Def}, \eqref{eq:calA-Closed2} and \eqref{eq:calA-Closed3}, by resumming \eqref{eq:calA-Series}.
As a test of the equalities of these expressions, we have checked that the values of these functions for $k=1,2,3$ and $\chi=0,0.5,1$ coincide by performing the numerical integrations.

\section{Consistency checks of our result\label{sec:Consistency}}

In this section we perform some consistency checks of our result \eqref{CBAofpqmodel}.

\subsection{Comparison with previous results\label{subsec:Comparison}}

The density matrix we have studied \eqref{eq:DM-qq} includes many density matrices of other theories as special cases.
They were studied for various motivations.
In this section we compare our result \eqref{CBAofpqmodel} with these results.

First, $\left(q,\tilde{q}\right)$ model with $\eta_{\alpha}=\tilde{\eta}_{\alpha}=0$ was studied in \cite{Nosaka:2015iiw}. 
They found that
\begin{align}
C_{k}^{\left(q,\tilde{q}\right)}\left(M;\tilde{M}\right) & =\frac{2}{\pi^{2}kq\tilde{q}\left(1+M^{2}\right)\left(1+\tilde{M}^{2}\right)},\nonumber \\
B_{k}^{\left(q,\tilde{q}\right)}\left(0,M;0,\tilde{M}\right) & =\frac{\pi^{2}}{3}C_{k}^{\left(q,\tilde{q}\right)}\left(M;\tilde{M}\right)
-\frac{q\tilde{q}}{6k}\left(\frac{1}{q^{2}\left(1+M^{2}\right)}
+\frac{1}{\tilde{q}^{2}\left(1+\tilde{M}^{2}\right)}\right)
+\frac{kq\tilde{q}}{24},\nonumber \\
A_{k}^{\left(q,\tilde{q}\right)}\left(0,M;0,\tilde{M}\right) & =\frac{1}{4}\sum_{\pm}\left[\tilde{q}^{2}A^{\mathrm{ABJM}}\left(\left(1\pm iM\right)qk\right)+q^{2}A^{\mathrm{ABJM}}\left(\left(1\pm i\tilde{M}\right)\tilde{q}k\right)\right].
\end{align}
This is consistent with our result \eqref{CBAofpqmodel} with $\eta_{\alpha}=\tilde{\eta}_{\alpha}=0$. 
Note that $\mathcal{A}\left(\kappa,0\right)$ reduces to $A^{\mathrm{ABJM}}\left(\kappa\right)$ as \eqref{eq:calA-ABJMA}.

Second, the large $N$ behavior of the density matrix \eqref{eq:DM-qq} with $\left(q,\tilde{q}\right)=\left(2N_{\mathrm{f}},1\right)$, $M=0$, $\tilde{M}=i/3$ and
\begin{equation}
\eta_{\alpha}=\begin{cases}
{\displaystyle \frac{1}{3}i} & \left(1\leq\alpha\leq N_{\mathrm{f}}\right)\\
{\displaystyle -\frac{1}{3}i} & \left(N_{\mathrm{f}}+1\leq\alpha\leq2N_{\mathrm{f}}\right)
\end{cases},
\end{equation}
was studied in the context of the squashing in \cite{Hatsuda:2016uqa}. 
(This would be an accidental coincidence between the density matrices. We will see a similar coincidence in section \ref{sec:SYM-Squash}.)
They found that
\begin{equation}
C_{k}^{\left(2N_{\mathrm{f}},1\right)}\left(0;\frac{i}{3}\right)=\frac{9}{8\pi^{2}kN_{\mathrm{f}}},\quad B_{k}^{\left(2N_{\mathrm{f}},1\right)}\left(\eta_{\alpha},0;\frac{i}{3}\right)=\frac{7}{24kN_{\mathrm{f}}}+\frac{kN_{\mathrm{f}}}{12}-\frac{N_{\mathrm{f}}}{4k},\label{eq:CB-Hatsuda}
\end{equation}
and\footnote{
We correct a typo there and thank Yasuyuki Hatsuda, who is the author of \cite{Hatsuda:2016uqa}, for confirming this typo.
}
\begin{align}
A_{k}^{\left(2N_{\mathrm{f}},1\right)}\left(\eta_{\alpha},0;\frac{i}{3}\right) & =\frac{1}{2\pi k}\left[\frac{\zeta\left(3\right)}{\pi N_{\mathrm{f}}}+N_{\mathrm{f}}^{2}\left(\frac{5\zeta\left(3\right)}{2\pi}+\frac{\psi^{\left(1\right)}\left(\frac{1}{3}\right)-\psi^{\left(1\right)}\left(\frac{2}{3}\right)}{4\sqrt{3}}\right)\right]\nonumber \\
 & \quad +2\pi k\left(-\frac{N_{\mathrm{f}}}{24\pi}-N_{\mathrm{f}}^{2}\left(\frac{1}{24\pi}+\frac{1}{72\sqrt{3}}\right)\right)+\mathcal{O}\left(k^{3}\right),\label{eq:A-Hatsuda}
\end{align}
where $\psi^{\left(n\right)}\left(z\right)=\partial_{z}^{n+1}\log\Gamma\left(z\right)$ is the polygamma function. 
For $C_{k}^{\left(2N_{\mathrm{f}},1\right)}$ and $B_{k}^{\left(2N_{\mathrm{f}},1\right)}$ , it is easy to see that our results \eqref{Cofpqmodel} and \eqref{Bofpqmodel} are consistent with \eqref{eq:CB-Hatsuda}. 
For $A_{k}^{\left(2N_{\mathrm{f}},1\right)}$, our result \eqref{Aofpqmodel} reads
\begin{align}
 & A_{k}^{\left(2N_{\mathrm{f}},1\right)}\left(\eta_{\alpha},0;\frac{i}{3}\right)\nonumber \\
 & =\frac{1}{2}A^{\mathrm{ABJM}}\left(2N_{\mathrm{f}}k\right)+\frac{N_{\mathrm{f}}^{2}}{2}\sum_{\pm}\left[A^{\mathrm{ABJM}}\left(\left(1\pm\frac{1}{3}\right)k\right)+{\cal A}\left(\left(1\pm\frac{1}{3}\right)k,\frac{2i}{3}\right)\right].
\end{align}
For comparing this expression with \eqref{eq:A-Hatsuda}, we use the small $\kappa$ expansion of $A^{\mathrm{ABJM}}\left(\kappa\right)$ in \eqref{eq:A-ABJM-Series} and the one of ${\cal A}\left(\kappa,\chi\right)$ in \eqref{eq:calA-Series3}.
Here the function $a_{0}\left(z\right)$ is given in \eqref{eq:a0-Closed} and $a_{1}\left(z\right)$ is obtained from \eqref{eq:al-a0} as
\begin{equation}
a_{1}\left(z\right)=\frac{z}{2\tanh\left(\frac{\pi z}{2}\right)}.
\end{equation}
By using these closed forms we obtain
\begin{equation}
a_{0}\left(\pm\frac{2i}{3}\right)=-\frac{8\zeta\left(3\right)}{9\pi}+\frac{\psi^{(1)}\left(\frac{1}{3}\right)-\psi^{(1)}\left(\frac{2}{3}\right)}{9\sqrt{3}},
\quad a_{1}\left(\pm\frac{2i}{3}\right)=\frac{1}{3\sqrt{3}}.
\end{equation}
Therefore, our result \eqref{Aofpqmodel} is also consistent with \eqref{eq:A-Hatsuda}.

Third, the large $N$ behavior of an $\mathcal{N}=4$ $\mathrm{U}\left(N\right)$ SYM theory with one adjoint and $N_{\mathrm{f}}$ fundamental hypermultiplets was studied in \cite{Chester:2023qwo}. 
This theory is the IR dual to the $\left(N_{\mathrm{f}},1\right)$ model with $k=1$. 
As we will discuss in section \ref{subsec:MM-FGF-SYM}, the matrix model of the $\left(N_{\mathrm{f}},1\right)$ model with $k=1$ in \eqref{eq:MM-pq-Def} is equal to \eqref{eq:MM-SYM-Def} with $b=1$ under the parameter identification \eqref{eq:SYM-q1-Para}. 
Then, in terms of the $\left(q,\tilde{q}\right)$ model, the result in \cite{Chester:2023qwo} is given as
\begin{align}
C_{k=1}^{\left(q,1\right)}\left(0;\tilde{M}\right) & =\frac{2}{\pi^{2}q\left(1+\tilde{M}^{2}\right)},\nonumber \\
B_{k=1}^{\left(q,1\right)}\left(\eta_{\alpha},0;\tilde{M}\right) & =\frac{2}{3q\left(1+\tilde{M}^{2}\right)}+\frac{q}{24}-\frac{1}{6q}-\frac{1}{1+\tilde{M}^{2}}\left(\frac{q}{6}+\sum_{I=1}^{q-1}\mu_{I}^{2}\right),\nonumber \\
A_{k=1}^{\left(q,1\right)}\left(\eta_{\alpha},0;\tilde{M}\right) & =\frac{q^{2}}{4}\sum_{\pm}A^{\mathrm{ABJM}}\left(1\pm i\tilde{M}\right)+\frac{1}{2}A^{\mathrm{ABJM}}\left(q\right)+q\sum_{I=1}^{q-1}\mu_{I}^{2}\left[\frac{1}{1+\tilde{M}^{2}}-\frac{\pi^{2}}{72}\right.\nonumber \\
 & \quad\left.-\frac{\pi^{4}\left(3\tilde{M}^{2}-1\right)}{21600}-\frac{\pi^{6}\left(5\tilde{M}^{4}-10\tilde{M}^{2}+1\right)}{1270080}+\cdots\right]+\mathcal{O}\left(\mu_{I}^{4}\right).\label{eq:CBA-Chester}
\end{align}
Here $\mu_{I}$ are mass parameters associated with a set of $q-1$ Cartan generators $T^{I}$ 
given by
\begin{equation}
T^{I}
=\frac{1}{\sqrt{2I\left(I+1\right)}}
\mathrm{diag}
(
\underset{I}{\underbrace{1,\ldots,1}},
-I,
\underset{q-I-1}{\underbrace{0,\ldots,0}}
).
\label{eq:T-Def}
\end{equation}
Note that $T^{I}$ satisfy
\begin{equation}
\mathrm{tr}\left(T^{I}\right)=0,\quad
\mathrm{tr}\left(T^{I}T^{J}\right)=\frac{1}{2}\delta_{IJ}.\label{eq:T-Prop}
\end{equation}
The relation between $\mu_{I}$ and our parameters is
\begin{equation}
\sum_{I=1}^{q-1}\mu_{I}\left(T^{I}\right)_{\alpha,\alpha}=\frac{1}{2}\eta_{\alpha}.\label{eq:mu-eta}
\end{equation}
They also evaluated the order $\mu^{4}$ terms at $\tilde{M}=0$ as\footnote{
Although this is different from the one which is explicitly written in the main text of \cite{Chester:2023qwo}, we confirmed this result (with the series expansion $\mathfrak{b}=\pi^{2}/12+\pi^{4}/720-\cdots$ in footnote 25 of \cite{Chester:2023qwo}) by using materials in appendix D of \cite{Chester:2023qwo}.
}
\begin{equation}
\left.\partial_{\mu_{1}}^{4}A\right|_{\mu_{I},\tilde{M}=0}
=4\left(6+q\right)\mathfrak{b},\quad
\left.\partial_{\mu_{3}}^{4}A\right|_{\mu_{I},\tilde{M}=0}
=4\left(6+\frac{7}{6}q\right)\mathfrak{b},
\label{eq:A-Chester}
\end{equation}
where
\begin{equation}
\mathfrak{b}
=-\frac{1}{2}\pi^{2}\left(\frac{\pi^{2}}{32}-\frac{1}{2}\right).\label{eq:frakb-Def}
\end{equation}
We have confirmed that \eqref{eq:CBA-Chester} is consistent with our result \eqref{CBAofpqmodel}. 
Here we have used \eqref{eq:calA-A-Series} and
\begin{equation}
\sum_{\alpha=1}^{q}\eta_{\alpha}^{2}
=2\sum_{I=1}^{q-1}\mu_{I}^{2},\quad
\sum_{\alpha,\beta=1}^{q}\left(\eta_{\alpha}-\eta_{\beta}\right)^{2}
=4q\sum_{I=1}^{q-1}\mu_{I}^{2}.
\end{equation}
This is obtained from \eqref{eq:mu-eta} and \eqref{eq:T-Prop}.
We have also confirmed that \eqref{eq:A-Chester} is consistent with our result \eqref{Aofpqmodel}. 
Here we have used \eqref{eq:calA-k1}.\footnote{
In \cite{Chester:2023qwo}, the authors also found that
\begin{equation}
\mathfrak{b}=-\frac{\pi^{2}}{2}\left.\left(xA^{\mathrm{ABJM}}\left(x\right)\right)''\right|_{x=1}.
\end{equation}
This comes from the fact that the coefficient of $\chi^{4}$ of ${\cal A}\left(1,\chi\right)$ (namely, $\frac{1}{48}\pi^{2}\left(1-\frac{1}{16}\pi^{2}\right)$ in \eqref{eq:calA-k1}) is given by $-\frac{1}{24}\pi^{2}\left.\left(xA^{\mathrm{ABJM}}\left(x\right)\right)''\right|_{x=1}$. This can be easily checked by comparing the series expansions of them, \eqref{eq:calA-Series} and \eqref{eq:A-ABJM-Series}.
}

\subsection{Identities from density matrix\label{subsec:ExpectInDM}}

The density matrices \eqref{eq:DM-qq} with different length of quiver diagrams (namely, different values of $\left(q,\tilde{q}\right)$) sometimes coincide with each other if the parameters are adjusted. 
This accidental coincidence leads to equalities between $C_{k}^{\left(q,\tilde{q}\right)}$, $B_{k}^{\left(q,\tilde{q}\right)}$ and $A_{k}^{\left(q,\tilde{q}\right)}$ in \eqref{CBAofpqmodel}. 
In this section we study this point.

A simple but important identity here is
\begin{equation}
\prod_{n=1}^{L}2\cosh\left(z+\frac{\pi i}{L}\left(\frac{L+1}{2}-n\right)\right)
=2\cosh\left(Lz\right).\label{eq:N-Angle}
\end{equation}
This identity leads to an accidental coincidence between the density matrices \eqref{eq:DM-qq} of $\left(rq,\tilde{r}\tilde{q}\right)_{k}$ and $\left(q,\tilde{q}\right)_{r\tilde{r}k}$ theories when
\begin{align}
\left(rq,\tilde{r}\tilde{q}\right)_{k}\text{ with }\left\{ \eta_{\alpha}=\xi_{\alpha}',M;\tilde{\eta}_{\alpha}=\tilde{\xi}_{\alpha}',\tilde{M}\right\}  & \leftrightarrow\left(q,\tilde{q}\right)_{r\tilde{r}k}\text{ with }\left\{ \eta_{\alpha}=\xi_{\alpha},M;\tilde{\eta}_{\alpha}=\tilde{\xi}_{\alpha},\tilde{M}\right\} ,\label{eq:qq-qq-rel}
\end{align}
where
\begin{equation}
\xi_{\alpha}'=\begin{cases}
{\displaystyle \frac{\xi_{1}}{r}+\frac{2i}{r}\left[\frac{1+r}{2}-\alpha\right]} & \left(1\leq\alpha\leq r\right)\vspace{0.1cm}\\
{\displaystyle \frac{\xi_{2}}{r}+\frac{2i}{r}\left[\frac{1+r}{2}-\left(\alpha-r\right)\right]} & \left(r+1\leq\alpha\leq2r\right)\\
\quad\quad\quad\vdots & \quad\quad\vdots\\
{\displaystyle \frac{\xi_{q}}{r}+\frac{2i}{r}\left[\frac{1+r}{2}-\left(\alpha-\left(q-1\right)r\right)\right]} & \left(\left(q-1\right)r+1\leq\alpha\leq qr\right)
\end{cases},
\end{equation}
and
\begin{equation}
\tilde{\xi}_{\alpha}'=\begin{cases}
{\displaystyle \frac{\tilde{\xi}_{1}}{\tilde{r}}+\frac{2i}{\tilde{r}}\left[\frac{1+\tilde{r}}{2}-\alpha\right]} & \left(1\leq\alpha\leq\tilde{r}\right)\\
{\displaystyle \frac{\tilde{\xi}_{2}}{\tilde{r}}+\frac{2i}{\tilde{r}}\left[\frac{1+\tilde{r}}{2}-\left(\alpha-\tilde{r}\right)\right]} & \left(\tilde{r}+1\leq\alpha\leq2\tilde{r}\right)\\
\quad\quad\quad\vdots & \quad\quad\vdots\\
{\displaystyle \frac{\tilde{\xi}_{\tilde{q}}}{\tilde{r}}+\frac{2i}{\tilde{r}}\left[\frac{1+\tilde{r}}{2}-\left(\alpha-\left(\tilde{q}-1\right)\tilde{r}\right)\right]} & \left(\left(\tilde{q}-1\right)\tilde{r}+1\leq\alpha\leq\tilde{q}\tilde{r}\right)
\end{cases}.
\end{equation}

One can easily see that the functions $C_{k}^{\left(q,\tilde{q}\right)}$ and $B_{k}^{\left(q,\tilde{q}\right)}$ in \eqref{Cofpqmodel}, \eqref{Bofpqmodel} indeed coincide under \eqref{eq:qq-qq-rel}
\begin{equation}
C_{k}^{\left(rq,\tilde{r}\tilde{q}\right)}\left(M;\tilde{M}\right)=C_{r\tilde{r}k}^{\left(q,\tilde{q}\right)}\left(M;\tilde{M}\right),\quad B_{k}^{\left(rq,\tilde{r}\tilde{q}\right)}\left(\xi_{\alpha}',M;\tilde{\xi}_{\alpha}',\tilde{M}\right)=B_{r\tilde{r}k}^{\left(q,\tilde{q}\right)}\left(\xi_{\alpha},M;\tilde{\xi}_{\alpha},\tilde{M}\right).
\end{equation}
On the other hand, for the function $A_{k}^{\left(q,\tilde{q}\right)}$ the relation \eqref{eq:qq-qq-rel} implies
\begin{equation}
A_{k}^{\left(rq,\tilde{r}\tilde{q}\right)}\left(\xi_{\alpha}',M;\tilde{\xi}_{\alpha}',\tilde{M}\right)=A_{r\tilde{r}k}^{\left(q,\tilde{q}\right)}\left(\xi_{\alpha},M;\tilde{\xi}_{\alpha},\tilde{M}\right).
\label{eq:qq-qq-A}
\end{equation}
This relation is not trivial. 
Remember that the function $A_{k}^{\left(q,\tilde{q}\right)}$ is written in terms of the function $\mathcal{A}\left(\kappa,\chi\right)$ as \eqref{Aofpqmodel}, and thus \eqref{eq:qq-qq-A} is satisfied if the following relation holds for an arbitrary $k\neq0$, $\left|\mathrm{Im}\left(\chi\right)\right|<2$ and an arbitrary positive integer $L$
\begin{equation}
\sum_{\alpha,\beta=1}^{L}{\cal \mathcal{A}}\left(\kappa,\frac{\chi}{L}+\frac{2i}{L}\left(\alpha-\beta\right)\right)
=\mathcal{A}\left(L\kappa,\chi\right).\label{eq:calA-Sum}
\end{equation}
Note that the first two restrictions are the same with \eqref{eq:calA-NonDiv}. 
In subsection \ref{subsec:calA-Sum} we prove this identity.

\subsubsection{Proof of summation formula for \texorpdfstring{$\mathcal{ A}\left(\kappa,\chi\right)$}{A}\label{subsec:calA-Sum}}

In this section we prove the summation formula \eqref{eq:calA-Sum}. 
The expression \eqref{eq:calA-Closed3} is useful for this purpose. 
Indeed, \eqref{eq:calA-Sum} holds if the following two relations hold
\begin{subequations}
\label{DS-a-Sum}
\begin{align}
\sum_{\alpha,\beta=1}^{L}a_{0}\left(z+\frac{2i}{L}\left(\alpha-\beta\right)\right) & =\frac{1}{L}a_{0}\left(Lz\right),\label{eq:a-Sum}\\
\sum_{\alpha,\beta=1}^{L}i\log s_{b=1}\left(z+\frac{i}{L}\left(\alpha-\beta\right)\right) & =i\log s_{b=1}\left(Lz\right).\label{eq:DS-Sum}
\end{align}
\end{subequations}

We first verify \eqref{eq:a-Sum}. 
We use the closed form expression \eqref{eq:a0-Closed} to prove this identity. 
By using this expression, we can evaluate the left-hand side of \eqref{eq:a-Sum}. 
The first two terms can be directly calculated as
\begin{equation}
\sum_{\alpha,\beta=1}^{L}\left[\left(z+\frac{2i}{L}\left(\alpha-\beta\right)\right)+\frac{1}{2}\left(z+\frac{2i}{L}\left(\alpha-\beta\right)\right)^{3}\right]=z+\frac{1}{2}L^{2}z^{3}.\label{eq:a0-Sum1}
\end{equation}
In order to evaluate the third term, we perform the rearrangement of the summation for an arbitrary function $f\left(z\right)$
\begin{equation}
\sum_{\alpha,\beta=1}^{L}\left(\alpha-\beta\right)f\left(\alpha-\beta\right)=\sum_{\alpha=1}^{L-1}\alpha\left(L-\alpha\right)\left(f\left(\alpha\right)-f\left(\alpha-L\right)\right).\label{eq:Sum-Rearrange}
\end{equation}
We also use a summation formula for the polylogarithm \eqref{eq:PL-Sum}, and then we obtain
\begin{equation}
\sum_{\alpha,\beta=1}^{L}\left(z+\frac{2i}{L}\left(\alpha-\beta\right)\right)\mathrm{Li}_{2}\left(e^{\pi\left(z+\frac{2i}{L}\left(\alpha-\beta\right)\right)}\right)=z\mathrm{Li}_{2}\left(e^{\pi Lz}\right).\label{eq:a0-Sum2}
\end{equation}
We can also compute the fourth term by using \eqref{eq:PL-Sum} as
\begin{equation}
\sum_{\alpha,\beta=1}^{L}\mathrm{Li}_{3}\left(e^{\pi\left(z+\frac{2i}{L}\left(\alpha-\beta\right)\right)}\right)=L^{-1}\mathrm{Li}_{3}\left(e^{\pi Lz}\right).\label{eq:a0-Sum3}
\end{equation}
By using \eqref{eq:a0-Sum1}, \eqref{eq:a0-Sum2} and \eqref{eq:a0-Sum3}, we can compute the left hand side of \eqref{eq:a-Sum} as
\begin{equation}
\sum_{\alpha,\beta=1}^{L}a_{0}\left(\frac{1}{L}z+\frac{2i}{L}\left(\alpha-\beta\right)\right)=\frac{1}{L}\left(-\frac{\pi^{3}}{6}z-\frac{\pi^{3}}{12}z^{3}-\pi z\mathrm{Li}_{2}\left(e^{\pi z}\right)+2\mathrm{Li}_{3}\left(e^{\pi z}\right)\right).
\end{equation}
This is the right-hand side of \eqref{eq:a-Sum} in the expression \eqref{eq:a0-Closed}.

Next, we show \eqref{eq:DS-Sum}. 
To prove this, we use the closed form expression \eqref{eq:DS-b1}. 
The flow of the computation is the same as the above computation. 
The first two terms can be directly calculated as
\begin{equation}
\sum_{\alpha,\beta=1}^{L}\left[\frac{1}{6}+\left(z+\frac{i}{L}\left(\alpha-\beta\right)\right)^{2}\right]=\frac{1}{6}+L^{2}z^{2}.\label{eq:Da0-Sum1}
\end{equation}
We can compute the third term by rearranging the summation as \eqref{eq:Sum-Rearrange} and using \eqref{eq:PL-Sum} (see \eqref{eq:PL-Log})
\begin{equation}
\sum_{\alpha,\beta=1}^{L}\left(z+\frac{2i}{L}\left(\alpha-\beta\right)\right)\log\left(1-e^{2\pi\left(z+\frac{i}{L}\left(\alpha-\beta\right)\right)}\right)=Lz\log\left(1-e^{2\pi Lz}\right).\label{eq:Da0-Sum2}
\end{equation}
We can also compute the fourth term by using \eqref{eq:PL-Sum} as
\begin{equation}
\sum_{\alpha,\beta=1}^{L}\mathrm{Li}_{2}\left(e^{2\pi\left(z+\frac{i}{L}\left(\alpha-\beta\right)\right)}\right)=\mathrm{Li}_{2}\left(e^{2\pi Lz}\right).\label{eq:Da0-Sum3}
\end{equation}
By using \eqref{eq:Da0-Sum1}, \eqref{eq:Da0-Sum2} and \eqref{eq:Da0-Sum3}, we can compute the left hand side of \eqref{eq:DS-Sum} as
\begin{equation}
\sum_{\alpha,\beta=1}^{L}i\log s_{b=1}\left(z+\frac{i}{L}\left(\alpha-\beta\right)\right)
=\frac{\pi}{12}+\frac{\pi}{2}L^{2}z^{2}-Lz\log\left(1-e^{2\pi Lz}\right)-\frac{1}{2\pi}\mathrm{Li}_{2}\left(e^{2\pi Lz}\right).
\end{equation}
This is the right-hand side of \eqref{eq:DS-Sum} in the expression \eqref{eq:DS-b1}.

\section{Application to \texorpdfstring{$S_{b}^{3}$}{S3b} partition function\label{sec:SYM-Squash}}

In this section we study the partition function of an $\mathcal{N}=4$ $\mathrm{U}\left(N\right)$ super Yang-Mills (SYM) theory with one adjoint and $N_{\mathrm{f}}$ fundamental hypermultiplets on squashed three sphere $S_{b}^{3}$.

\subsection{Matrix model and Fermi gas formalism\label{subsec:MM-FGF-SYM}}

The SYM theory admits a mass deformation for the adjoint hypermultiplet and mass deformations for the fundamental matters.
We can also add an FI term for the $\mathrm{U}\left(1\right)$ factor of the $\mathrm{U}\left(N\right)$ gauge group.

Thanks to the supersymmetric localization for the squashed three sphere \cite{Jafferis:2010un,Hama:2010av,Hama:2011ea,Imamura:2011wg}, the partition function reduces to a matrix model. 
The matrix model describing the partition function of the SYM theory on $S_{b}^{3}$ is
\begin{align}
Z_{b,N_{\mathrm{f}}}^{\mathrm{SYM}}\left(N;\zeta,m,y_{\alpha}\right) & 
=\frac{1}{N!}\int_{-\infty}^{\infty}\prod_{i=1}^{N}\frac{d\lambda_{i}}{2\pi}e^{i\zeta\sum_{i=1}^{N}\lambda_{i}}
\prod_{i<j}^{N}2\sinh\frac{b\lambda_{ij}}{2}\prod_{i<j}^{N}2\sinh\frac{\lambda_{ij}}{2b}\nonumber \\
 & \quad\quad\times\prod_{i,j}^{N}{\cal D}_{b}\left(\frac{\lambda_{ij}}{2\pi}+m\right)\prod_{\alpha=1}^{N_{\mathrm{f}}}\prod_{i=1}^{N}{\cal D}_{b}\left(\frac{\lambda_{i}}{2\pi}+y_{\alpha}\right),
 \label{eq:MM-SYM-Def}
\end{align}
where $\lambda_{ij}=\lambda_{i}-\lambda_{j}$.
$\zeta$ and $m$ denote the FI parameter and the mass parameter for the adjoint matter, respectively. 
$y_{\alpha}$ denote the mass parameters for the fundamental matters, which, by shifting the integration variables $\lambda_{i}$, we have set as
\begin{equation}
\sum_{\alpha=1}^{N_{\mathrm{f}}}y_{\alpha}=0.
\end{equation}
(We have ignored additional factors appearing in this step since they are independent of the integration variables and thus they are just overall factors.)
The function ${\cal D}_{b}$ is defined as \eqref{eq:calD-Def}.
Note that the squashing parameter is symmetric under $b \leftrightarrow b^{-1}$.
Correspondingly, the matrix model is invariant under this symmetry (thanks to the property of the double sine function \eqref{eq:DS-Integral}).
Thus, in the following we assume $b\geq 1$ without loss of generality.

In \cite{Kubo:2024qhq}, the authors showed that the SYM matrix model with $N_{\mathrm{f}}=1$ is drastically simplified when $b^{2}$ is a positive odd integer and $m=m_{b}$ with
\begin{equation}
m_{b}=\frac{b^{2}-3}{4b}i.\label{eq:mb-Def}
\end{equation}
Utilizing the simplification, they further applied the Fermi gas formalism.
The same simplification occurs for the general $N_{\mathrm{f}}$ case. 
Furthermore, we find that the assumption that $b^{2}$ is a positive odd integer is not necessary to apply the Fermi gas formalism.\footnote{
The Fermi gas formalism with a general $N_{\mathrm{f}}$ and a positive odd integer $b^{2}$ was discussed in \cite{Bobev:2025ltz}.
} 
Now we demonstrate these arguments.
We start with the matrix model with $m=m_{b}$
\begin{align}
Z_{b,N_{\mathrm{f}}}^{\mathrm{SYM}}\left(N;\zeta,m_{b},y_{\alpha}\right) & 
=\frac{1}{N!}\int_{-\infty}^{\infty}\left(\prod_{i=1}^{N}\frac{d\lambda_{i}}{2\pi}\right)e^{i\zeta\sum_{i=1}^{N}\lambda_{i}}\prod_{i<j}^{N}2\sinh\frac{b\lambda_{ij}}{2}\prod_{i<j}^{N}2\sinh\frac{\lambda_{ij}}{2b}\nonumber \\
 & \quad\quad\times\prod_{i,j}^{N}\frac{s_{b}\left(\frac{\lambda_{ij}}{2\pi}+\frac{i}{2}\left(b-b^{-1}\right)\right)}{s_{b}\left(\frac{\lambda_{ij}}{2\pi}-ib^{-1}\right)}
 \prod_{\alpha=1}^{N_{\mathrm{f}}}\prod_{i=1}^{N}{\cal D}_{b}\left(\frac{\lambda_{i}}{2\pi}+y_{\alpha}\right).
\end{align}
Here we have used the definition of $\mathcal{D}_b$ in \eqref{eq:calD-Def}. 
Note that when we choose $m=m_{b}$ we assume $b>1$ since $s_{b}\left(z\right)^{-1}$ has poles at $z\in\left\{ nb+mb^{-1}-iQ/2|n,m\in\mathbb{Z}_{\geq0}\right\} $. 
By using \eqref{eq:DS-Prop} and \eqref{eq:DS-Cosh}, we obtain
\begin{align}
\prod_{i,j}^{N}s_{b}\left(\frac{\lambda_{ij}}{2\pi}+\frac{i}{2}\left(b-b^{-1}\right)\right) & 
=\left(\sqrt{\frac{s_{b}\left(\frac{i}{2}\left(b-b^{-1}\right)\right)}{s_{b}\left(-\frac{i}{2}\left(b-b^{-1}\right)\right)}}\right)^{N}
\prod_{i<j}^{N}\frac{s_{b}\left(\frac{\lambda_{ij}}{2\pi}+\frac{i}{2}\left(b-b^{-1}\right)\right)}{s_{b}\left(\frac{\lambda_{ij}}{2\pi}-\frac{i}{2}\left(b-b^{-1}\right)\right)}\nonumber \\
 & =b^{-N}\prod_{i<j}^{N}\frac{2\sinh\left(\frac{\lambda_{ij}}{2b}\right)}{2\sinh\left(\frac{b\lambda_{ij}}{2}\right)},
\end{align}
and
\begin{equation}
\prod_{i,j}^{N}\frac{1}{s_{b}\left(\frac{\lambda_{ij}}{2\pi}-ib^{-1}\right)}=\left(\sqrt{\frac{s_{b}\left(ib^{-1}\right)}{s_{b}\left(-ib^{-1}\right)}}\right)^{N}\prod_{i<j}^{N}\frac{s_{b}\left(\frac{\lambda_{ij}}{2\pi}+ib^{-1}\right)}{s_{b}\left(\frac{\lambda_{ij}}{2\pi}-ib^{-1}\right)}=\prod_{i,j}^{N}\frac{1}{2\cosh\left(\frac{\lambda_{ij}}{2b}+\frac{i}{2b^{2}}\right)}.
\end{equation}
These formulas allow us to simplify the SYM matrix model as
\begin{align}
 & Z_{b,N_{\mathrm{f}}}^{\mathrm{SYM}}\left(N;\zeta,m_{b},y_{\alpha}\right)\nonumber \\
 & =\frac{1}{N!}\int_{-\infty}^{\infty}\left(\prod_{i=1}^{N}\frac{d\lambda_{i}}{2\pi}\right)e^{ib\zeta\sum_{i=1}^{N}\lambda_{i}}\frac{\prod_{i<j}^{N}\left(2\sinh\frac{\lambda_{ij}}{2}\right)^{2}}{\prod_{i,j}^{N}2\cosh\left(\frac{\lambda_{ij}}{2}+\frac{\pi i}{2b^{2}}\right)}\prod_{\alpha=1}^{N_{\mathrm{f}}}\prod_{i=1}^{N}{\cal D}_{b}\left(\frac{b}{2\pi}\lambda_{i}+y_{\alpha}\right).\label{eq:MM-SYM}
\end{align}
Here we have rescaled the integration variables as $\lambda_{i}\rightarrow b\lambda_{i}$. 
In this expression, we can apply the Fermi gas formalism by using \eqref{eq:CauchyDet} and \eqref{eq:Cosh-op}. 
After absorbing the remaining factors into the determinant, we finally obtain
\begin{equation}
Z_{b,N_{\mathrm{f}}}^{\mathrm{SYM}}\left(N;\zeta,m_{b},y_{\alpha}\right)=\frac{1}{N!}\int_{-\infty}^{\infty}\left(\prod_{i=1}^{N}\frac{d\lambda_{i}}{2\pi}\right)\det\left(\left[\braket{\lambda_{i}|\hat{\rho}_{b,N_{\mathrm{f}}}^{\mathrm{SYM}}\left(\hat{x},\hat{p};\zeta,m_{b},y_{\alpha}\right)|\lambda_{j}}\right]_{i,j}^{N\times N}\right),
\label{eq:FGF-SYM}
\end{equation}
where the density matrix for the SYM theory is
\begin{equation}
\hat{\rho}_{b,N_{\mathrm{f}}}^{\mathrm{SYM}}\left(\hat{x},\hat{p};\zeta,m_{b},y_{\alpha}\right)=e^{ib\zeta\hat{x}}\left(\prod_{\alpha=1}^{N_{\mathrm{f}}}\mathcal{D}_{b}\left(\frac{b}{2\pi}\hat{x}+y_{\alpha}\right)\right)\frac{e^{-\frac{1}{2b^{2}}\hat{p}}}{2\cosh\frac{\hat{p}}{2}}.\label{eq:DM-SYM}
\end{equation}
Notice that in this case the commutation relation is $\left[\hat{x},\hat{p}\right]=2\pi i$.

Before closing this section, we comment on a duality, which we have used in section \ref{subsec:Comparison}. 
The SYM theory studied here is expected to be IR dual to the $\left(q,\tilde{q}\right)$ model with $\left(q,\tilde{q}\right)=\left(N_{\mathrm{f}},1\right)$ and $k=1$. 
Indeed, both of them describe the low-energy effective theories of M2-branes probing the space $\mathbb{C}^{2}\times\left(\mathbb{C}^{2}/\mathbb{Z}_{q}\right)$ \cite{Benini:2009qs,Imamura:2008ji}. 
When $b=1$, we can show that the Fermi gas density matrices of these theories indeed match. 
In this case, we can apply the Fermi gas formalism to \eqref{eq:MM-SYM-Def} (for general $m$) by using \eqref{eq:DS-Cosh}, \eqref{eq:CauchyDet} and \eqref{eq:Cosh-op}. 
The density matrix is then given as
\begin{equation}
\hat{\rho}_{b=1,N_{\mathrm{f}}}^{\mathrm{SYM}}\left(\hat{x},\hat{p};\zeta,m,y_{\alpha}\right)=\frac{e^{i\zeta\hat{x}}}{\prod_{\alpha=1}^{N_{\mathrm{f}}}2\cosh\frac{\hat{x}+2\pi y_{\alpha}}{2}}\frac{e^{im\hat{p}}}{2\cosh\frac{\hat{p}}{2}}.
\label{eq:DM-SYM-b1}
\end{equation}
On the other hand, the dual density matrix is given by \eqref{eq:DM-qq} with $\left(q,\tilde{q}\right)=\left(N_{\mathrm{f}},1\right)$ and $k=1$. 
By comparing this density matrix with \eqref{eq:DM-SYM-b1}, we find that these density matrices indeed match under the parameter identification
\begin{equation}
\frac{\zeta}{N_{\mathrm{f}}}=\frac{1}{2}M,\quad 
m=\frac{1}{2}\tilde{M},\quad
y_{\alpha}=\frac{1}{2}\eta_{\alpha}.\label{eq:SYM-q1-Para}
\end{equation}

\subsection{Large \texorpdfstring{$N$}{N}\label{subsec:LargeN-SYM}}

When $b=\sqrt{2n-1}$ with positive integer $n$ (we further assume $n\geq 2$ since we assumed $b>1$), the function $\mathcal{D}_{b}$ is simplified as \eqref{eq:D-Odd}. 
Then, one can easily find that the density matrix for the SYM theory \eqref{eq:DM-SYM} is the special case of the one for the $\left(q,\tilde{q}\right)$ model \eqref{eq:DM-qq}. 
More explicitly, the $\left(nN_{\mathrm{f}},1\right)$ model with the Chern-Simons level $k=1$ corresponds to the SYM theory with $b=\sqrt{2n-1}$ and general $N_{\mathrm{f}}$ as
\begin{equation}
\left(nN_{\mathrm{f}},1\right)_{k=1}\text{ with }\left\{ M=\frac{2b}{nN_{\mathrm{f}}}\zeta,\eta_{\alpha}=y_{\alpha}';\tilde{M}=ib^{-2}\right\} \leftrightarrow\mathrm{SYM}_{b=\sqrt{2n-1},N_{\mathrm{f}}}\text{ with }\left\{ \zeta,m_{b},y_{\alpha}\right\} ,\label{eq:q1-SYMF-para}
\end{equation}
where
\begin{equation}
y_{\alpha}'=\begin{cases}
{\displaystyle \frac{2}{b}y_{1}+\frac{2i}{b^{2}}\left[\frac{1+n}{2}-\alpha\right]} & \left(1\leq\alpha\leq n\right)\vspace{0.1cm}\\
{\displaystyle \frac{2}{b}y_{2}+\frac{2i}{b^{2}}\left[\frac{1+n}{2}-\left(\alpha-n\right)\right]} & \left(n+1\leq\alpha\leq2n\right)\\
\quad\quad\quad\vdots & \quad\quad\vdots\\
{\displaystyle \frac{2}{b}y_{N_{\mathrm{f}}}+\frac{2i}{b^{2}}\left[\frac{1+n}{2}-\left(\alpha-\left(N_{\mathrm{f}}-1\right)n\right)\right]} & \left(\left(N_{\mathrm{f}}-1\right)n+1\leq\alpha\leq N_{\mathrm{f}}n\right)
\end{cases}.
\end{equation}
By substituting this into \eqref{CBAofpqmodel}, we obtain the large $N$ behavior of the SYM theory with $N_{\mathrm{f}}$ flavors.
Remark that we do not claim that this is a duality like the duality discussed around \eqref{eq:DM-SYM-b1}.
We emphasise that this is the correspondence between the $\left(q,\tilde{q}\right)$ model on the round three sphere and the SYM theory on the squashed three sphere.

\begin{figure}[t]
\begin{centering}
\includegraphics[scale=0.7]{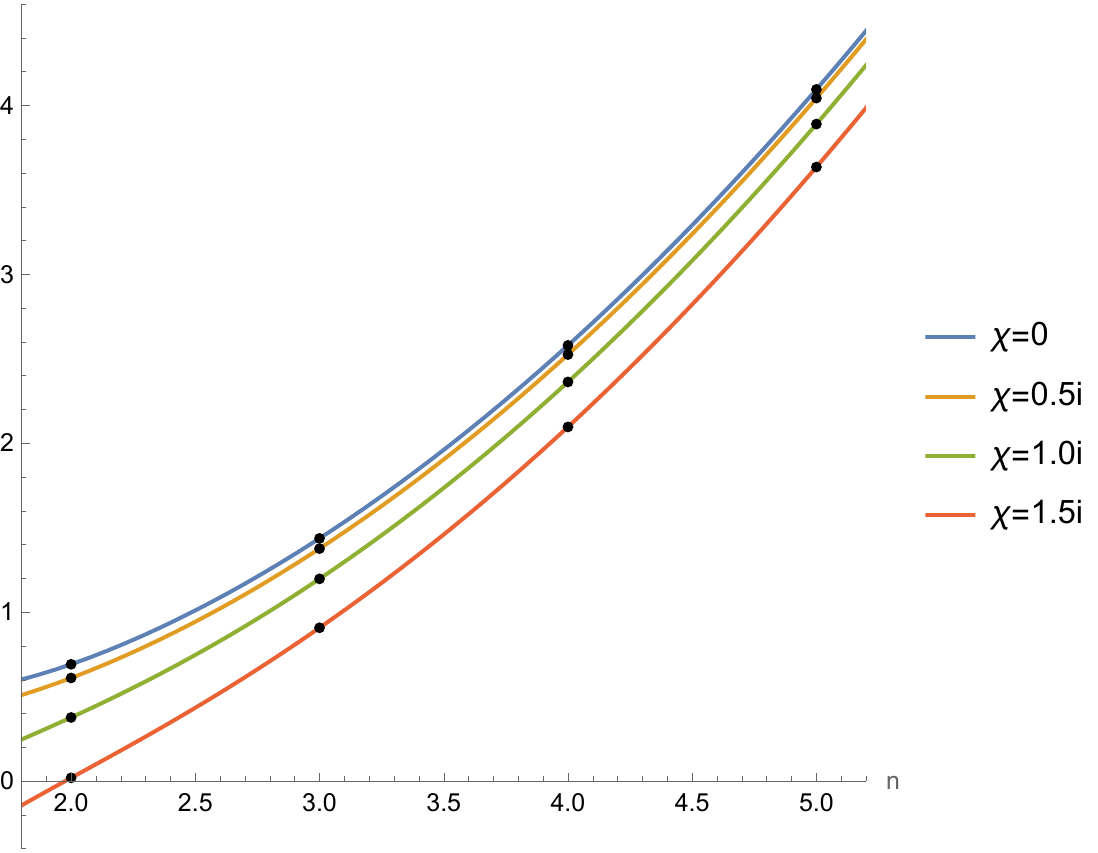}
\par\end{centering}
\caption{Numerical checks of the summation formula \eqref{eq:calA-Sum2} as a function of $n$ with various $\chi$. Dots represent the right-hand side while lines represent the left-hand side. 
Note that in the left-hand side $n$ must be an integer while in the right-hand side $n$ can take values in $\mathbb{R}_{>1}$.}
\label{fig:calA-Sum2}
\end{figure}

We first study the function $A_{k}^{\left(q,\tilde{q}\right)}$ in \eqref{Aofpqmodel}. By using \eqref{eq:q1-SYMF-para}, we find
\begin{align}
A_{b=\sqrt{2n-1},N_{\mathrm{f}}}^{\mathrm{SYM}}\left(\zeta,m_{b},y_{\alpha}\right) & =\frac{1}{4}\sum_{\pm}\left[A^{\mathrm{ABJM}}\left(\frac{\left(b^{2}+1\right)N_{\mathrm{f}}\pm4ib\zeta}{2}\right)\right.\nonumber \\
 & \quad\quad\left.+\sum_{\alpha,\beta=1}^{N_{\mathrm{f}}}\sum_{\alpha',\beta'=1}^{n}{\cal \mathcal{A}}\left(1\pm b^{-2},\frac{2}{b}\left(y_{\alpha}-y_{\beta}\right)+\frac{2i}{b^{2}}\left(\alpha'-\beta'\right)\right)\right].\label{AofSYM0}
\end{align}
Here we have used \eqref{eq:calA-ABJMA}. On the other hand, when $N_{\mathrm{f}}=1$, the exact large $N$ expansion of the SYM theory was studied with the help of a quantum curve \cite{Kubo:2024qhq}, where it was found that
\begin{align}
 & A_{b=\sqrt{2n-1},N_{\mathrm{f}}=1}^{\mathrm{SYM}}\left(\zeta,m_{b}\right)\nonumber \\
 & =\frac{1}{4}\left[\sum_{\pm}A^{\mathrm{ABJM}}\left(\frac{b^{2}+1\pm4ib\zeta}{2}\right)+A^{\mathrm{ABJM}}\left(2n-2\right)-A^{\mathrm{ABJM}}\left(4n-2\right)\right].\label{AofSYMNf1}
\end{align}
Then, the equality between \eqref{AofSYM0} with $N_{\mathrm{f}}=1$ and \eqref{AofSYMNf1} implies the following relation for positive integers $n\geq2$
\begin{align}
 & \sum_{\alpha,\beta=1}^{n}\left[{\cal \mathcal{A}}\left(\frac{2n}{2n-1},\frac{2i}{2n-1}\left(\alpha-\beta\right)\right)+{\cal \mathcal{A}}\left(\frac{2n-2}{2n-1},\frac{2i}{2n-1}\left(\alpha-\beta\right)\right)\right]\nonumber \\
 & =A^{\mathrm{ABJM}}\left(2n-2\right)-A^{\mathrm{ABJM}}\left(4n-2\right).
 \label{eq:SYMNf1-calA}
\end{align}
This can be regarded as another summation formula. Note that 
the previous summation formula \eqref{eq:calA-Sum} is valid for a general parameter $\chi$ (with $\left|\mathrm{Im}\left(\chi\right)\right|<2$ ). 
Motivated by this fact, we expect that the following generalization holds with $\left|\mathrm{Im}\left(\chi\right)\right|<2$
\begin{align}
 & \sum_{\alpha,\beta=1}^{n}\left[{\cal \mathcal{A}}\left(\frac{2n}{2n-1},\frac{\chi}{2n-1}+\frac{2i}{2n-1}\left(\alpha-\beta\right)\right)+{\cal \mathcal{A}}\left(\frac{2n-2}{2n-1},\frac{\chi}{2n-1}+\frac{2i}{2n-1}\left(\alpha-\beta\right)\right)\right]\nonumber \\
 & ={\cal \mathcal{A}}\left(2n-2,\chi\right)-{\cal \mathcal{A}}\left(4n-2,\chi\right).\label{eq:calA-Sum2}
\end{align}
We have checked this relation for various $\chi$ up to $n=5$ by performing the integration \eqref{eq:calA-Closed2} with \eqref{eq:A-ABJM-Closed} numerically, see figure \ref{fig:calA-Sum2}.

The summation formula \eqref{eq:calA-Sum2} simplifies $A_{b,N_{\mathrm{f}}}^{\mathrm{SYM}}$ in \eqref{AofSYM0}. 
We can also obtain $C_{b,N_{\mathrm{f}}}^{\mathrm{SYM}}$ and $B_{b,N_{\mathrm{f}}}^{\mathrm{SYM}}$ by plugging \eqref{eq:q1-SYMF-para} into \eqref{Cofpqmodel} and \eqref{Bofpqmodel}. 
The results are
\begin{subequations}
\label{CBAofSYM}
\begin{align}
C_{b,N_{\mathrm{f}}}^{\mathrm{SYM}}\left(\zeta,m_{b}\right) & =\frac{4b^{4}N_{\mathrm{f}}}{\pi^{2}\left(b^{2}-1\right)\left(\left(b^{2}+1\right)^{2}N_{\mathrm{f}}^{2}+16b^{2}\zeta^{2}\right)},\label{CofSYM}\\
B_{b,N_{\mathrm{f}}}^{\mathrm{SYM}}\left(\zeta,m_{b},y_{\alpha}\right) & =\frac{1}{b^{2}-1}\left[\frac{\left(3b^{4}+1\right)N_{\mathrm{f}}}{3\left(\left(b^{2}+1\right)^{2}N_{\mathrm{f}}^{2}+16b^{2}\zeta^{2}\right)}-b^{2}\sum_{\alpha=1}^{N_{\mathrm{f}}}y_{\alpha}^{2}-\frac{N_{\mathrm{f}}}{24}\left(b^{4}-b^{2}+2\right)\right],\label{BofSYM}\\
A_{b,N_{\mathrm{f}}}^{\mathrm{SYM}}\left(\zeta,m_{b},y_{\alpha}\right) & =\frac{1}{4}\left[\sum_{\pm}A^{\mathrm{ABJM}}\left(\frac{\left(b^{2}+1\right)N_{\mathrm{f}}\pm4ib\zeta}{2}\right)\right.\nonumber \\
 & \quad\quad\left.+\sum_{\alpha,\beta=1}^{N_{\mathrm{f}}}\left({\cal \mathcal{A}}\left(b^{2}-1,2b\left(y_{\alpha}-y_{\beta}\right)\right)-{\cal \mathcal{A}}\left(2b^{2},2b\left(y_{\alpha}-y_{\beta}\right)\right)\right)\right].\label{AofSYM}
\end{align}
\end{subequations}
Namely, we find that the exact large $N$ expansion of the matrix model \eqref{eq:MM-SYM-Def} up to non-perturbative corrections is given as \eqref{eq:LargeN-Gen} with $C^{\mathcal{T}}\left(\boldsymbol{\xi}\right)=C_{b,N_{\mathrm{f}}}^{\mathrm{SYM}}\left(\zeta,m_{b}\right)$, $B^{\mathcal{T}}\left(\boldsymbol{\xi}\right)=B_{b,N_{\mathrm{f}}}^{\mathrm{SYM}}\left(\zeta,m_{b},y_{\alpha}\right)$ and $A^{\mathcal{T}}\left(\boldsymbol{\xi}\right)=A_{b,N_{\mathrm{f}}}^{\mathrm{SYM}}\left(\zeta,m_{b},y_{\alpha}\right)$.
Interestingly, in this representation $n$, which is a positive integer, no longer appears explicitly, and thus we can regard these functions as an analytic function with respect to $b$. 
Therefore, we expect that \eqref{CBAofSYM} holds for general $b>1$, and thus we have omitted $b=\sqrt{2n-1}$.

\subsection{Consistency checks\label{subsec:SYM-Consistency}}

In this section we perform consistency checks of our result \eqref{CBAofSYM} in various ways. We emphasize that these results are expected to hold for arbitrary $b>1$.

First, the functions $C_{b,N_{\mathrm{f}}}^{\mathrm{SYM}}$, $B_{b,N_{\mathrm{f}}}^{\mathrm{SYM}}$ and $A_{b,N_{\mathrm{f}}}^{\mathrm{SYM}}$ with $y_{\alpha}=0$ were recently proposed in \cite{Bobev:2025ltz} as
\begin{align}
C_{b,N_{\mathrm{f}}}^{\mathrm{SYM}}\left(\zeta,m\right) & =\left(\frac{2}{\pi\left(b+b^{-1}\right)^{2}\sqrt{2N_{\mathrm{f}}\Delta_{1}\Delta_{2}\Delta_{3}\Delta_{4}}}\right)^{2},\nonumber \\
B_{b,N_{\mathrm{f}}}^{\mathrm{SYM}}\left(\zeta,m\right) & =\frac{N_{\mathrm{f}}}{24}-\frac{N_{\mathrm{f}}}{12}\left(\frac{1}{\Delta_{1}}+\frac{1}{\Delta_{2}}\right)-\frac{1}{12N_{\mathrm{f}}}\left(\frac{1}{\Delta_{3}}+\frac{1}{\Delta_{4}}\right)\nonumber \\
 & \quad-\frac{4}{3\left(b+b^{-1}\right)^{2}}\left(-\frac{N_{\mathrm{f}}}{8\Delta_{1}\Delta_{2}}+\frac{\Delta_{1}^{2}+\Delta_{2}^{2}-2\left(\Delta_{1}+\Delta_{2}\right)+\Delta_{1}\Delta_{2}}{8N_{\mathrm{f}}\Delta_{1}\Delta_{2}\Delta_{3}\Delta_{4}}\right),\nonumber \\
A_{b,N_{\mathrm{f}}}^{\mathrm{SYM}}\left(\zeta,m_b\right) & =\frac{1}{4}\sum_{\pm}A^{\mathrm{ABJM}}\left(\frac{\left(b^{2}+1\right)N_{\mathrm{f}}\pm4ib\zeta}{2}\right) \nonumber \\
& \quad+\frac{N_{\mathrm{f}}^{2}}{4}\left(A^{\mathrm{ABJM}}\left(b^{2}-1\right)-A^{\mathrm{ABJM}}\left(2b^{2}\right)\right),
\label{eq:CBA-SYM-Bobev}
\end{align}
where
\begin{equation}
\Delta_{1}=\frac{1}{2}+\frac{2mi}{b+b^{-1}},\quad\Delta_{2}=\frac{1}{2}-\frac{2mi}{b+b^{-1}},\quad\Delta_{3}=\frac{1}{2}-\frac{2\zeta i}{\left(b+b^{-1}\right)N_{\mathrm{f}}},\quad\Delta_{4}=\frac{1}{2}+\frac{2\zeta i}{\left(b+b^{-1}\right)N_{\mathrm{f}}}.
\end{equation}
One can easily see that our result \eqref{CBAofSYM} with $y_{\alpha}=0$ exactly matches \eqref{eq:CBA-SYM-Bobev} with $m=m_{b}$.

When $N_{\mathrm{f}}=1$, the exact $1/N$ expansion was also studied in \cite{Kubo:2024qhq}. 
The equality between our $A_{b,N_{\mathrm{f}}}^{\mathrm{SYM}}$ in \eqref{CofSYM} with $N_{\mathrm{f}}=1$ and their result has already been discussed around \eqref{eq:SYMNf1-calA}. 
For $C_{b,N_{\mathrm{f}}=1}^{\mathrm{SYM}}$ and $B_{b,N_{\mathrm{f}}=1}^{\mathrm{SYM}}$, they found
\begin{align}
C_{b,N_{\mathrm{f}}=1}^{\mathrm{SYM}}\left(\zeta,m_{b}\right) & =\frac{4b^{4}}{\pi^{2}\left(b^{2}-1\right)\left(\left(b^{2}+1\right)^{2}+16b^{2}\zeta^{2}\right)},\nonumber \\
B_{b,N_{\mathrm{f}}=1}^{\mathrm{SYM}}\left(\zeta,m_{b}\right) & =-\frac{b^{8}+b^{6}\left(16\zeta^{2}+1\right)-b^{4}\left(16\zeta^{2}+23\right)+b^{2}\left(32\zeta^{2}+3\right)-6}{24\left(b^{2}-1\right)\left(\left(b^{2}+1\right)^{2}+16b^{2}\zeta^{2}\right)}.
\end{align}
It is clear that our results \eqref{CofSYM} and \eqref{BofSYM} with $N_{\mathrm{f}}=1$ reduce to these results.

\begin{figure}[t]
\begin{centering}
\includegraphics[scale=0.6]{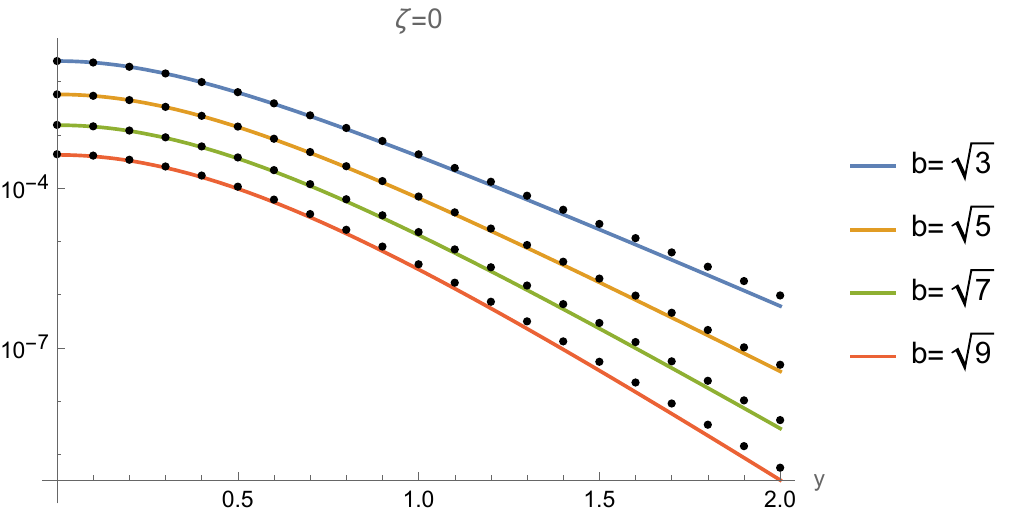}
\par\end{centering}
\smallskip{}

\begin{centering}
\includegraphics[scale=0.6]{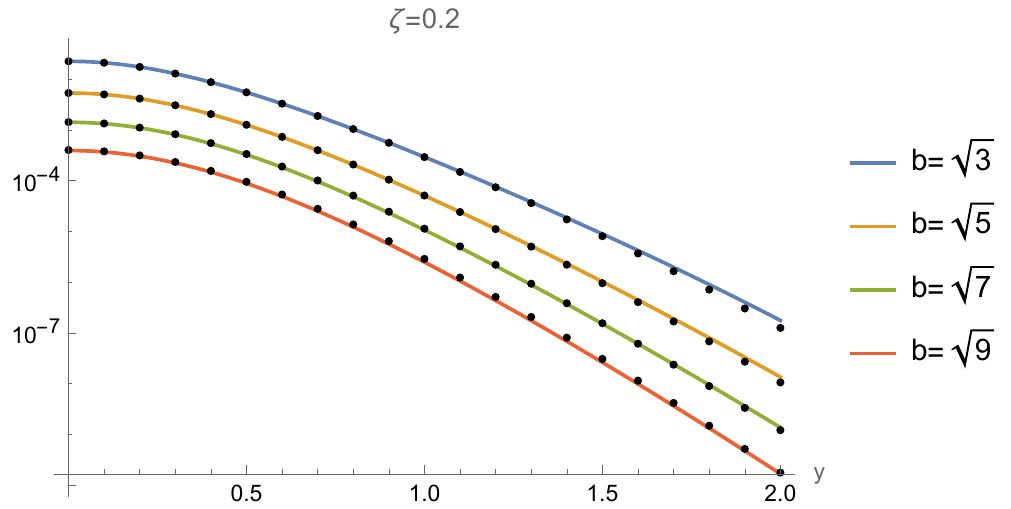}
\par\end{centering}
\caption{$Z_{b,N_{\mathrm{f}}=2}^{\mathrm{SYM},\mathrm{pert}}$ in \eqref{eq:Zpert} vs $Z_{b,N_{\mathrm{f}}=2}^{\mathrm{SYM}}$ in \eqref{eq:Zexact} at $N=1$, $\zeta=0,0.2$ as functions of $y$, where $\left(y_{1},y_{2}\right)=\left(y,-y\right)$. They are compatible with each other. The difference should come from the non-perturbative part ${\cal O}\left(e^{-\#\sqrt{N}}\right)$ in \eqref{eq:LargeN-Gen}.}
\label{fig:PvsE}
\end{figure}

Second, we compare our result \eqref{CBAofSYM} with the exact values at $N=1$, $N_{\mathrm{f}}=2$ with various $b$. Namely, we compare the perturbative function
\begin{align}
 & Z_{b,N_{\mathrm{f}}=2}^{\mathrm{SYM},\mathrm{pert}}\left(1;\zeta,m_{b},y_{\alpha}\right)\nonumber \\
 & =e^{A_{b,2}^{\mathrm{SYM}}\left(\zeta,m_{b},y_{\alpha}\right)}C_{b,2}^{\mathrm{SYM}}\left(\zeta,m_{b}\right)^{-\frac{1}{3}}\text{Ai}\left[C_{b,2}^{\mathrm{SYM}}\left(\zeta,m_{b}\right)^{-\frac{1}{3}}\left(1-B_{b,2}^{\mathrm{SYM}}\left(\zeta,m_{b},y_{\alpha}\right)\right)\right],\label{eq:Zpert}
\end{align}
and the exact function
\begin{equation}
Z_{b,N_{\mathrm{f}}=2}^{\mathrm{SYM}}\left(1;\zeta,m_{b},y_{\alpha}\right)={\cal D}_{b}\left(m_{b}\right)\int_{-\infty}^{\infty}\frac{d\lambda}{2\pi}e^{i\zeta\sum_{i=1}^{N}\lambda_{i}}\prod_{\alpha=1}^{2}{\cal D}_{b}\left(\frac{\lambda}{2\pi}+y_{\alpha}\right).\label{eq:Zexact}
\end{equation}
Figure \ref{fig:PvsE} shows the numerical result of these functions for $\zeta=0$ and $\zeta=0.2$.
We have chosen $b^{2}$ to be odd to use \eqref{eq:D-Odd} for numerically evaluating \eqref{eq:Zexact}. 
For evaluating \eqref{eq:Zpert}, we have used \eqref{eq:calA-Closed2} with \eqref{eq:A-ABJM-Closed}. 
We also evaluated these functions with $b=\sqrt{2}$ using \eqref{eq:calD-b2}. Table \ref{tab:fig:PvsEb2} shows the numerical result.
\begin{table}[t]
\begin{centering}
\begin{tabular}{|c|c|c|}
\hline 
$y$ & $Z_{\sqrt{2},2}^{\mathrm{SYM},\mathrm{pert}}\left(\zeta=0\right)$ & $Z_{\sqrt{2},2}^{\mathrm{SYM}}\left(\zeta=0\right)$\tabularnewline
\hline 
\hline 
0 & 0.0596079 & 0.059275\tabularnewline
\hline 
0.1 & 0.0559911 & 0.0555738\tabularnewline
\hline 
0.2 & 0.046786 & 0.0461615\tabularnewline
\hline 
0.3 & 0.0354902 & 0.0346407\tabularnewline
\hline 
0.4 & 0.0250305 & 0.0240313\tabularnewline
\hline 
\end{tabular}\quad{}%
\begin{tabular}{|c|c|c|}
\hline 
$y$ & $Z_{\sqrt{2},2}^{\mathrm{SYM},\mathrm{pert}}\left(\zeta=0.2\right)$ & $Z_{\sqrt{2},2}^{\mathrm{SYM}}\left(\zeta=0.2\right)$\tabularnewline
\hline 
\hline 
0 & 0.0557911 & 0.0555738\tabularnewline
\hline 
0.1 & 0.0522504 & 0.0519804\tabularnewline
\hline 
0.2 & 0.0432631 & 0.0428701\tabularnewline
\hline 
0.3 & 0.0322991 & 0.031788\tabularnewline
\hline 
0.4 & 0.0222461 & 0.0216819\tabularnewline
\hline 
\end{tabular}
\par\end{centering}
\caption{$Z_{b=\sqrt{2},N_{\mathrm{f}}=2}^{\mathrm{SYM},\mathrm{pert}}$ in \eqref{eq:Zpert} vs $Z_{b=\sqrt{2},N_{\mathrm{f}}=2}^{\mathrm{SYM}}$ in \eqref{eq:Zexact} at $N=1$, $\zeta=0,0.2$ as functions of $y$, where $\left(y_{1},y_{2}\right)=\left(y,-y\right)$. They are again compatible with each other, and the difference should come from the non-perturbative part ${\cal O}\left(e^{-\#\sqrt{N}}\right)$ in \eqref{eq:LargeN-Gen}.}
\label{tab:fig:PvsEb2}
\end{table}

Third, according to the summation formula \eqref{eq:calA-Sum}, by choosing $y_{\alpha}$ appropriately
we can simplify the function $A_{b,N_{\mathrm{f}}}^{\mathrm{SYM}}$ in \eqref{AofSYM} as
\begin{align}
 & A_{b,N_{\mathrm{f}}}^{\mathrm{SYM}}\left(\zeta,m_{b},y_{\alpha}=y_{\alpha}\left(b\right)\right)\nonumber \\
 & =\frac{1}{4}\left[\sum_{\pm}A^{\mathrm{ABJM}}\left(\frac{\left(b^{2}+1\right)N_{\mathrm{f}}\pm4ib\zeta}{2}\right)+A^{\mathrm{ABJM}}\left(\left(b^{2}-1\right)N_{\mathrm{f}}\right)-A^{\mathrm{ABJM}}\left(2b^{2}N_{\mathrm{f}}\right)\right],
 \label{eq:A-SYM-Nfto1}
\end{align}
where
\begin{equation}
y_{\alpha}\left(b\right)=\frac{i}{bN_{\mathrm{f}}}\left(\frac{1+N_{\mathrm{f}}}{2}-\alpha\right).
\label{eq:y-by}
\end{equation}
In section \ref{subsec:ExpectInDM}, we have seen that the summation formula \eqref{eq:calA-Sum} comes from the identity between the density matrices for the $\left(q,\tilde{q}\right)$ model with the help of \eqref{eq:N-Angle}.
Here we demonstrate that the density matrix for the SYM theory \eqref{eq:DM-SYM} is simplified under \eqref{eq:N-Angle}, and the simplification is consistent with \eqref{eq:A-SYM-Nfto1}.
To see this, it is convenient to introduce a quantum curve
\begin{align}
&\hat{\mathcal{O}}_{b,N_{\mathrm{f}}}^{\mathrm{SYM}}\left(\hat{x},\hat{p};\zeta,m_{b},y_{\alpha}\right) \nonumber \\
& =e^{-\frac{ib\zeta}{2}\hat{x}}\prod_{\alpha=1}^{N_{\mathrm{f}}}s_{b}\left(\frac{b}{2\pi}\hat{x}+y_{\alpha}-\frac{i}{4}Q\right) 
 \left(e^{\left(\frac{1}{2}+\frac{1}{2b^{2}}\right)\hat{p}}+e^{-\left(\frac{1}{2}-\frac{1}{2b^{2}}\right)\hat{p}}\right)\prod_{\alpha=1}^{N_{\mathrm{f}}} \frac{e^{-\frac{ib\zeta}{2}\hat{x}}}{s_{b}\left(\frac{b}{2\pi}\hat{x}+y_{\alpha}+\frac{i}{4}Q\right)}.
 \label{eq:QC-SYM-Gen}
\end{align}
This is the inverse of the density matrix with an appropriate similarity transformation
\begin{equation}
\hat{\mathcal{O}}_{b,N_{\mathrm{f}}}^{\mathrm{SYM}}\left(\hat{x},\hat{p};\zeta,m_{b},y_{\alpha}\right)=\hat{A}\hat{\rho}_{b,N_{\mathrm{f}}}^{\mathrm{SYM}}\left(\hat{x},\hat{p};\zeta,m_{b},y_{\alpha}\right)^{-1}\hat{A}^{-1},
\end{equation}
where
\begin{equation}
\hat{A}
=e^{-\frac{ib\zeta}{2}\hat{x}}\prod_{\alpha=1}^{N_{\mathrm{f}}}s_{b}\left(\frac{b}{2\pi}\hat{x}+y_{\alpha}-\frac{i}{4}Q\right) .
\end{equation}
Note that the partition function of the free Fermi gas system \eqref{eq:FGF-SYM} is invariant under similarity transformations, and thus the large $N$ behavior is also invariant. 
Thanks to the property of the double sine function \eqref{eq:calD-Def} and the Baker-Campbell-Hausdorff formula for $\left(\hat{x},\hat{p}\right)$
\begin{equation}
e^{c_{1}\hat{x}}e^{c_{2}\hat{p}}=e^{c_{1}\pi i}e^{c_{1}\hat{x}+c_{2}\hat{p}}=e^{2c_{1}c_{2}\pi i}e^{c_{2}\hat{p}}e^{c_{1}\hat{x}},
\end{equation}
we obtain
\begin{align}
 & \hat{\mathcal{O}}_{b,N_{\mathrm{f}}}^{\mathrm{SYM}}\left(\hat{x},\hat{p};\zeta,m_{b},y_{\alpha}\right)\nonumber \\
 & =e^{-ib\zeta\hat{x}+\left(\frac{1}{2}+\frac{1}{2b^{2}}\right)\hat{p}}+e^{-\frac{1}{2}\left(\frac{1}{2}-\frac{1}{2b^{2}}\right)\hat{p}}e^{-ib\zeta\hat{x}}\left( \prod_{\alpha=1}^{N_{\mathrm{f}}}2\cosh\left(\frac{b^{2}}{2}\hat{x}+\pi by_{\alpha}\right)\right) e^{-\frac{1}{2}\left(\frac{1}{2}-\frac{1}{2b^{2}}\right)\hat{p}}.
\end{align}
Now, when $y_{\alpha}$ are chosen as \eqref{eq:y-by}, by using \eqref{eq:N-Angle} we can simplify the product of the cosh functions. 
The quantum curve becomes a three-term one, and for changing it to the canonical form we introduce new variables
\begin{equation}
\hat{X}=\left(\frac{b^{2}N_{\mathrm{f}}}{2}-ib\zeta\right)\hat{x}-\left(\frac{1}{2}-\frac{1}{2b^{2}}\right)\hat{p},\quad\hat{P}=-ib\zeta\hat{x}+\left(\frac{1}{2}+\frac{1}{2b^{2}}\right)\hat{p}.
\end{equation}
Then we obtain
\begin{equation}
\hat{\mathcal{O}}_{b,N_{\mathrm{f}}}^{\mathrm{SYM}}\left(\hat{x},\hat{p};\zeta,m_{b},y_{\alpha}\left(b\right)\right)=\hat{\mathcal{O}}^{\mathbb{P}\left(1,\mathfrak{m},\mathfrak{n}\right)}\left(\hat{X},\hat{P}\right),
\end{equation}
where\footnote{
This curve describes the anti-canonical bundle of the weighted projective space $\mathbb{P}\left(1,\mathfrak{m},\mathfrak{n}\right)$ as a mirror curve \cite{Kashaev:2015kha,Marino:2015ixa}.
}
\begin{equation}
\hat{\mathcal{O}}^{\mathbb{P}\left(1,\mathfrak{m},\mathfrak{n}\right)}\left(\hat{X},\hat{P}\right)=e^{\hat{X}}+e^{\hat{P}}+e^{-\mathfrak{m}\hat{X}-\mathfrak{n}\hat{P}}.
\end{equation}
This curve is parameterized by $\left(\mathfrak{m},\mathfrak{n}\right)$ and the value of the commutation relation $\left[\hat{X},\hat{P}\right]=i\hbar'$. 
Given the data $\left(\mathfrak{m},\mathfrak{n},\hbar'\right)$, the large $N$ behavior is known to be \cite{Hatsuda:2015oaa}
\begin{align}
C^{\mathbb{P}\left(1,\mathfrak{m},\mathfrak{n}\right)}\left(\hbar'\right) & =\frac{\left(\mathfrak{m}+\mathfrak{n}+1\right)^{2}}{\mathfrak{m}\mathfrak{n}}\frac{1}{4\pi\hbar'},\nonumber \\
B^{\mathbb{P}\left(1,\mathfrak{m},\mathfrak{n}\right)}\left(\hbar'\right) & =\frac{\mathfrak{m}^{2}+\mathfrak{m}\mathfrak{n}+\mathfrak{n}^{2}+\mathfrak{m}+\mathfrak{n}+1}{12\mathfrak{m}\mathfrak{n}}\frac{\pi}{\hbar'}-\frac{\mathfrak{m}+\mathfrak{n}+1}{48\pi}\hbar',\nonumber \\
A^{\mathbb{P}\left(1,\mathfrak{m},\mathfrak{n}\right)}\left(\hbar'\right) & =\frac{1}{4}\left[A^{\mathrm{ABJM}}\left(\frac{\hbar'}{\pi}\right)+A^{\mathrm{ABJM}}\left(\frac{\mathfrak{m}\hbar'}{\pi}\right)\right.\nonumber \\
 & \quad\quad\left.+A^{\mathrm{ABJM}}\left(\frac{\mathfrak{n}\hbar'}{\pi}\right)-A^{\mathrm{ABJM}}\left(\frac{\left(\mathfrak{m}+\mathfrak{n}+1\right)\hbar'}{\pi}\right)\right].
\end{align}
In our case,
\begin{equation}
\mathfrak{m}=\frac{\left(b+b^{-1}\right)N_{\mathrm{f}}+4i\zeta}{\left(b+b^{-1}\right)N_{\mathrm{f}}-4i\zeta},\quad\mathfrak{n}=\frac{2\left(b-b^{-1}\right)N_{\mathrm{f}}}{\left(b+b^{-1}\right)N_{\mathrm{f}}-4i\zeta},\quad\hbar'=\frac{\left(b^{2}+1\right)N_{\mathrm{f}}-4ib\zeta}{2}\pi.\label{eq:mnh-Ours}
\end{equation}
Now one can easily see that $A^{\mathbb{P}\left(1,\mathfrak{m},\mathfrak{n}\right)}$ with \eqref{eq:mnh-Ours} exactly matches \eqref{eq:A-SYM-Nfto1}. 
$C^{\mathbb{P}\left(1,\mathfrak{m},\mathfrak{n}\right)}$ and $B^{\mathbb{P}\left(1,\mathfrak{m},\mathfrak{n}\right)}$ with \eqref{eq:mnh-Ours} also match  $C_{b,N_{\mathrm{f}}}^{\mathrm{SYM}}$ and $B_{b,N_{\mathrm{f}}}^{\mathrm{SYM}}$ in \eqref{CofSYM}, \eqref{BofSYM} with \eqref{eq:y-by}.

\section{Conclusion\label{sec:Conclusion}}

In this paper we have studied the large $N$ behavior of the $S^3$ partition function of the $\left(q,\tilde{q}\right)$ model with mass and FI deformations.
We have found its exact large $N$ expansion \eqref{CBAofpqmodel}, which supports the universality of the Airy form \eqref{eq:LargeN-Gen}.
We have especially found the closed form expression of the coefficient $A_{k}^{\left(q,\tilde{q}\right)}$ in \eqref{Aofpqmodel}.
Unlike the previous cases, this function cannot be written as a linear combination of $A^{\text{ABJM}}(k)$, and a new function $\mathcal{A}\left(\kappa,\chi\right)$ has been introduced.
Although we have computed the function $\mathcal{A}\left(\kappa,\chi\right)$ in the small $\kappa$ espansion as \eqref{eq:calA-Series}, we have also found the closed form expressions of $\mathcal{A}\left(\kappa,\chi\right)$  as \eqref{eq:calA-Def}, \eqref{eq:calA-Closed2} and \eqref{eq:calA-Closed3}.
We have then studied the properties of $\mathcal{A}\left(\kappa,\chi\right)$, and especially we have found the summation formulas \eqref{eq:calA-Sum} and \eqref{eq:calA-Sum2} and proved the former one.
We have also studied the exact large $N$ expansion of the $S^3_b$ partition function of the SYM theory with multiple fundamental matters by utilizing an accidental coincidence between the Fermi gas density matrices of the $\left(q,\tilde{q}\right)$ model and the SYM theory.
Our result is \eqref{CBAofSYM}, which passes various non-trivial consistency checks.
This result again supports the universality of the Airy form \eqref{eq:LargeN-Gen} even for the squashing case.
Since the SYM theory with $N_{\mathrm{f}}$ flavors describes the M2-branes proving $\mathbb{C}^2\times \mathbb{C}^2/\mathbb{Z}_{N_{\mathrm{f}}}$, the Airy universality is again consistent with the universality expected in the holographic side as discussed in the introduction. 

There are various interesting directions for further study.

The $\left(q,\tilde{q}\right)$ model and the SYM theory we have studied are the worldvolume theories of the M2-branes. 
As discussed in the introduction, there are various M2-brane theories which (are expected to) obey the Airy form universality. 
It is important to find the Airy coefficients of these theories, especially $A^{\mathcal{T}}\left(\boldsymbol{\xi}\right)$. 
We hope that for a wide class of theories with various mass and FI deformations the coefficient $A^{\mathcal{T}}\left(\boldsymbol{\xi}\right)$ can be written as a linear combination of the function ${\cal A}\left(\kappa,\chi\right)$.

It would also be nice if the function $\mathcal{A}\left(\kappa,\chi\right)$ appears not only in the mass and FI deformations but also for other deformations. 
For example, it was recently pointed out that for the flavored ABJM theory the mass deformation corresponds to the rank deformation in terms of the Fermi gas density matrices \cite{Kubo:2025jxi}. 
In the same paper the Fermi gas formalism was applied to flavored circular quiver theories. 
Therefore, it would be worth studying whether the function $\mathcal{A}\left(\kappa,\chi\right)$ can also describe the rank deformations for the circular quiver theories, and possibly for more general quiver theories.

In this work we have identified the density matrix for the $\left(q,\tilde{q}\right)$ model in \eqref{eq:DM-qq}, whose inverse is clearly the quantum curve. 
(Namely, it is a linear combination of terms of the form $\exp\left(m\hat{x}+n\hat{p}\right)$.) It was conjectured in \cite{Grassi:2014zfa,Codesido:2015dia} (see also a review \cite{Marino:2015nla} and references therein) that the spectral determinant of a quantum curve (which is \eqref{Fredholmdet} in our case) is non-perturbatively computed by the free energy of a topological string whose mirror curve corresponds to the quantum curve. 
This is called the topological string/spectral theory (TS/ST) correspondence, and in this correspondence the function $A^{\mathcal{T}}\left(\boldsymbol{\xi}\right)$ would correspond to the resummation of the all-genus constant maps. 
This was tested for the ABJM theory, where $A^{\mathcal{T}}\left(\boldsymbol{\xi}\right)=A^{\text{ABJM}}(k)$ \cite{Hanada:2012si}. Therefore, it would be natural to expect that the function $\mathcal{A}\left(\kappa,\chi\right)$ also has the topological string interpretation, and it would be interesting to figure out this point.

As we have utilized in section \ref{subsec:ExpectInDM}, a quantum curve associated with some Newton polygon can be equivalent, after tuning the moduli parameters appropriately, to a quantum curve associated with a simpler Newton polygon and with different Planck constant.
Under the TS/ST correspondence this implies that the topological string free energies on the two different Calabi-Yau threefolds are related under the rescaling of the string coupling constant.
It would be interesting to find an interpretation of these simplifications of the quantum curves from the viewpoint of the topological string theory.

It would be interesting to ask whether there is a physical interpretation for the relation between the density matrices of the $\left(q,\tilde{q}\right)$ model on the round three sphere and the one of the SYM theory on the squashed three sphere which we used in section \ref{sec:SYM-Squash}.
Notice that we have also obtained the quantum curve for the SYM theory as \eqref{eq:QC-SYM-Gen}, and thus not only the partition function of the $\left(q,\tilde{q}\right)$ but also that of the SYM theory would be related to the topological string free energies under the TS/ST correspondence.
Thus, we anticipate that the TS/ST correspondence provides an answer to the question.

\subsection*{Acknowledgments}

We are grateful to Yasuyuki Hatsuda for valuable discussions.
NK and YP are supported by National key research and development program under grant No.~2022YFE0134300 and the National Natural Science Foundation of China (NSFC) under grants No.~12175164 and No.~12247103.
The work of T.N.~was supported by the Startup Funding no.~2302-SRFP-2024-0012 of Shanghai Institute for Mathematics and Interdisciplinary Sciences.

\appendix

\section{Special functions\label{sec:SpecialFunc}}

In this section we enumerate properties of special functions.

\subsection{Riemann zeta function and polylogarithm\label{subsec:ZetaFunc}}

The Riemann zeta function $\zeta\left(s\right)$ is defined as
\begin{equation}
\zeta\left(s\right)=\sum_{n=1}^{\infty}\frac{1}{n^{s}}.\label{eq:Zeta-Def}
\end{equation}
The integral expression for $\mathrm{Re}\left(s\right)>1$ is
\begin{equation}
\zeta\left(s\right)=\frac{1}{\Gamma\left(s\right)}\int_{0}^{\infty}\frac{x^{s-1}}{e^{x}-1}dx.\label{eq:Zeta-Int}
\end{equation}
The Riemann zeta function with an integer argument is related to the Bernoulli number $B_{n}$. 
For any positive integer $n\geq1$,
\begin{equation}
\zeta\left(2n\right)=\frac{\left(-1\right)^{n+1}B_{2n}\left(2\pi\right)^{2n}}{2\left(2n\right)!}.\label{eq:Zeta-Ber1}
\end{equation}
For a non-positive integer $n\geq0$,
\begin{equation}
\zeta\left(-n\right)=-\frac{B_{n+1}}{n+1}.\label{eq:Zeta-Ber2}
\end{equation}

The polylogarithm is defined as
\begin{equation}
{\rm Li}_{s}\left(z\right)=\sum_{n=1}^{\infty}\frac{z^{n}}{n^{s}}.
\label{eq:PL-Def}
\end{equation}
The Riemann zeta function is a special case of the polylogarithm function when $\mathrm{Re}\left(s\right)>1$
\begin{equation}
{\rm Li}_{s}\left(1\right)=\zeta\left(s\right).\label{eq:PL-Zeta}
\end{equation}
When $s=1$, the polylogarithm reduces to the logarithm
\begin{equation}
{\rm Li}_{1}\left(z\right)=-\log\left(1-z\right).\label{eq:PL-Log}
\end{equation}
The derivative of the polylogarithm is given by
\begin{equation}
\frac{\partial}{\partial z}{\rm Li}_{s}\left(z\right)=\frac{1}{z}{\rm Li}_{s-1}\left(z\right).\label{eq:PL-Der}
\end{equation}
We use the following summation formula in section \ref{subsec:calA-Sum}
\begin{equation}
\sum_{n=1}^{p}{\rm Li}_{s}\left(ze^{2\pi i\frac{n}{p}}\right)=p^{1-s}{\rm Li}_{s}\left(z^{p}\right).\label{eq:PL-Sum}
\end{equation}
The sum of two polylogarithms is simplified as 
\begin{align}
\text{Li}_n\left(-e^x\right)+\left(-1\right)^n\text{Li}_n\left(-e^{-x}\right)=-\frac{\left(2\pi i\right)^n}{n!}B_n\left(\frac{x}{2\pi i}+\frac{1}{2}\right),
\label{polylogreflectionid}
\end{align}
where $B_n(x)$ is the Bernoulli polynomial.

\subsection{Double sine function\label{subsec:DoubleSine}}

The double sine function is defined as
\begin{equation}
s_{b}\left(z\right)=\prod_{\ell,m=0}^{\infty}\frac{\ell b+mb^{-1}+\frac{Q}{2}-iz}{\ell b+mb^{-1}+\frac{Q}{2}+iz},\label{eq:DS-Def}
\end{equation}
where
\begin{equation}
Q=b+b^{-1}.\label{eq:Q-Def}
\end{equation}
This function has an integral representation in the strip $\left|\mathrm{Im}\left(z\right)\right|<Q/2$
\begin{equation}
i\log s_{b}\left(z\right)=\frac{\pi}{2}z^{2}+\frac{\pi}{24}\left(b^{2}+b^{-2}\right)+i\int_{\mathbb{R}+i0^+}\frac{dt}{t}\frac{e^{-2izt}}{4\sinh\left(bt\right)\sinh\left(b^{-1}t\right)}.\label{eq:DS-Integral}
\end{equation}
The double sine function satisfies the following relations
\begin{equation}
s_{b}\left(0\right)=1,\quad s_{b}\left(z\right)=s_{b^{-1}}\left(z\right),\quad s_{b}\left(z\right)s_{b}\left(-z\right)=1,\quad\overline{s_{b}\left(z\right)}=s_{b}\left(-\bar{z}\right).\label{eq:DS-Prop}
\end{equation}
Especially, the following ratio of the double sine functions is simplified as
\begin{equation}
\frac{s_{b}\left(z+\frac{i}{2}b^{\pm1}\right)}{s_{b}\left(z-\frac{i}{2}b^{\pm1}\right)}=\frac{1}{2\cosh\left(\pi b^{\pm1}z\right)}.\label{eq:DS-Cosh}
\end{equation}
The double sine function with $b=1$ is written as
\begin{equation}
i\log s_{b=1}\left(z\right)=\frac{\pi}{12}+\frac{\pi}{2}z^{2}-z\log\left(1-e^{2\pi z}\right)-\frac{1}{2\pi}\mathrm{Li}_{2}\left(e^{2\pi z}\right).\label{eq:DS-b1}
\end{equation}
By using \eqref{eq:PL-Der}, one obtains the derivative of this function
\begin{equation}
\frac{d}{dz}i\log s_{b=1}\left(z\right)=-\frac{\pi z}{\tanh\pi z}.\label{eq:DDS-b1}
\end{equation}

In this paper, $\mathcal{D}_{b}\left(z\right)$ denotes the ratio of the double sine functions ($Q$ is defined in \eqref{eq:Q-Def})
\begin{equation}
\mathcal{D}_{b}\left(z\right)=\frac{s_{b}\left(z+\frac{i}{4}Q\right)}{s_{b}\left(z-\frac{i}{4}Q\right)}.\label{eq:calD-Def}
\end{equation}
When $b^{2}$ is a positive odd integer, say $b^{2}=2n-1$, thanks to the relation \eqref{eq:DS-Cosh} this function is simplified as
\begin{equation}
\mathcal{D}_{b=\sqrt{2n-1}}\left(\mu\right)=\prod_{j=1}^{n}\frac{1}{2\cosh\left(\frac{\pi}{b}\mu+\frac{\pi i}{b^{2}}\left(\frac{n+1}{2}-j\right)\right)}.\label{eq:D-Odd}
\end{equation}
When $b=\sqrt{2}$, $\mathcal{D}_{b}$ is also simplified as \cite{Hatsuda:2016uqa}
\begin{align}
\mathcal{D}_{\sqrt{2}}\left(\mu\right) & =\frac{1}{2^{\frac{1}{4}}\left(2\cosh\left(2\sqrt{2}\pi\mu\right)\right)^{\frac{1}{8}}\left(\sqrt{2}\cosh\left(\sqrt{2}\pi\mu\right)+1\right)^{\frac{1}{2}}}\nonumber \\
 & \quad\times\exp\left[-\sqrt{2}\mu\arctan\left(e^{-2\sqrt{2}\pi\mu}\right)+\frac{i}{4\pi}\left(\mathrm{Li}_{2}\left(ie^{-2\sqrt{2}\pi\mu}\right)-\mathrm{Li}_{2}\left(-ie^{-2\sqrt{2}\pi\mu}\right)\right)\right].\label{eq:calD-b2}
\end{align}

\section{Proof of \texorpdfstring{\eqref{DrelationofcalZ}}{(calZ)} through Wigner-Kirkwood expansion}
\label{app_WignerKirkwood}

In section \ref{subsec:LargeN-A} we have explained the strategy to determine the coefficients ${\cal Z}_{2\ell}^{\left(q,{\tilde q}\right)}\left(s;v_\alpha,M;{\tilde v}_\alpha,{\tilde M}\right)$ of the $\hbar$-expansion of the spectral zeta function from the extrapolation of $s=1,2,\cdots$, by assuming the structure ${\cal Z}_{2\ell}^{\left(q,{\tilde q}\right)}\left(s;v_\alpha,M;{\tilde v}_\alpha,{\tilde M}\right)
=D_{2\ell}^{\left(q,{\tilde q}\right)}\left(s;v_\alpha,M;{\tilde v}_\alpha,{\tilde M}\right){\cal Z}_0^{\left(q,{\tilde q}\right)}\left(s;v_\alpha,M;{\tilde v}_\alpha,{\tilde M}\right)$ with $D_{2\ell}^{\left(q,{\tilde q}\right)}\left(s;v_\alpha,M;{\tilde v}_\alpha,{\tilde M}\right)$ some rational function of $s$ \eqref{DrelationofcalZ}.
In this appendix we explain an altanative algorithm to calculate ${\cal Z}_{2\ell}^{\left(q,{\tilde q}\right)}\left(s;v_\alpha,M;{\tilde v}_\alpha,{\tilde M}\right)$ with $s$ kept as a free parameter, and verify the assumption \eqref{DrelationofcalZ} is indeed correct.

For this purpose we expand the powers of the densit matrix ${\hat\rho}^{\left(q,{\tilde q}\right)}_k\left({\hat x},{\hat p};\frac{\hbar v_\alpha}{\pi},M;\frac{\hbar{\tilde v}_\alpha}{\pi},{\tilde M}\right)^s$ $=$ $e^{-s{\hat H}^{\left(q,{\tilde q}\right)}_k\left({\hat x},{\hat p};\frac{\hbar v_\alpha}{\pi},M;\frac{\hbar{\tilde v}_\alpha}{\pi},{\tilde M}\right)}$ around ${\hat H}^{\left(q,{\tilde q}\right)}_k\left({\hat x},{\hat p};\frac{\hbar v_\alpha}{\pi},M;\frac{\hbar{\tilde v}_\alpha}{\pi},{\tilde M}\right)=H_W(x,p)$ as
\begin{align}
&{\hat\rho}^{\left(q,{\tilde q}\right)}_k\left({\hat x},{\hat p};\frac{\hbar v_\alpha}{\pi},M;\frac{\hbar{\tilde v}_\alpha}{\pi},{\tilde M}\right)^s\nonumber \\
&=\sum_{r=0}^\infty e^{-sH_W\left(x,p\right)}\frac{\left(-s\right)^r}{r!}\left(
{\hat H}^{\left(q,{\tilde q}\right)}_k\left({\hat x},{\hat p};\frac{\hbar v_\alpha}{\pi},M;\frac{\hbar{\tilde v}_\alpha}{\pi},{\tilde M}\right)
-H_W\left(x,p\right)\right)^r.
\end{align}
Hence we can expand the Wigner transformation of the powers of the density matrix as
\begin{align}
\left(
{\hat\rho}^{\left(q,{\tilde q}\right)}_k\left({\hat x},{\hat p};\frac{\hbar v_\alpha}{\pi},M;\frac{\hbar{\tilde v}_\alpha}{\pi},{\tilde M}\right)^s
\right)_W
=\sum_{r=0}^\infty e^{-sH_W\left(x,p\right)}\frac{\left(-s\right)^r}{r!}{\cal G}_r\left(x,p\right),
\end{align}
with
\begin{align}
{\cal G}_r\left(x,p\right)=
\left(\left({\hat H}^{\left(q,{\tilde q}\right)}_k\left({\hat x},{\hat p};\frac{\hbar v_\alpha}{\pi},M;\frac{\hbar{\tilde v}_\alpha}{\pi},{\tilde M}\right)
-H_W\left(x,p\right)\right)^r\right)_W.
\label{calG}
\end{align}
We can calculate ${\cal G}_r\left(x,p\right)$ from $H_W\left(x,p\right)$ and the star product as
\begin{subequations}
\begin{align}
{\cal G}_0\left(x,p\right)&=1,\\
{\cal G}_1\left(x,p\right)&=0,\\
{\cal G}_r\left(x,p\right)&=\left(-1\right)^{r-1}\left(r-1\right)H_W\left(x,p\right)^r\nonumber \\
&\quad +\sum_{r'=2}^r
\binom{r}{r'}
\left(-1\right)^{r-r'}H_W\left(x,p\right)^{r-r'}(\underbrace{H_W\left(x,p\right)\star\cdots \star H_W\left(x,p\right)}_{r'}),\quad(r\ge 2)
\end{align}
\end{subequations}
It is known that the leading power of $\hbar$ in each ${\cal G}_r\left(x,p\right)$ is at least $\hbar^{2\lfloor \frac{r+2}{3}\rfloor}$ \cite{Marino:2011eh}:
\begin{align}
{\cal G}_r\left(x,p\right)=\sum_{a=2\lfloor \frac{r+2}{3}\rfloor}\hbar^a{\cal G}_r^{\left(a\right)}.
\end{align}
Expanding also $H_W\left(x,p\right)$ in $\hbar$ as
\begin{align}
H_W\left(x,p\right)=\sum_{a=0}^\infty \hbar^aH_W^{\left(a\right)},
\end{align}
we can write the spectral trace $\text{Tr}{\hat\rho}^{\left(q,{\tilde q}\right)}_k\left({\hat x},{\hat p};\frac{\hbar v_\alpha}{\pi},M;\frac{\hbar{\tilde v}_\alpha}{\pi},{\tilde M}\right)^s$ \eqref{calZ} as
\begin{align}
\text{Tr}{\hat\rho}^{\left(q,{\tilde q}\right)}_k\left({\hat x},{\hat p};\frac{\hbar v_\alpha}{\pi},M;\frac{\hbar{\tilde v}_\alpha}{\pi},{\tilde M}\right)^s
=
\int \frac{dxdp}{2\pi\hbar}e^{-sH_W^{\left(0\right)}}
\Delta_\text{WK},
\label{calZinWK}
\end{align}
with
\begin{align}
\Delta_\text{WK}=\left(1+\sum_{r=1}^\infty \frac{\left(-s\right)^r}{r!}\left(\sum_{a=1}^\infty\hbar^a H_W^{\left(a\right)}\right)^r\right)
\left(
1+\sum_{r'=2}^\infty\frac{\left(-s\right)^{r'}}{r!}\sum_{a'=2\lfloor\frac{r'+2}{3}\rfloor}^\infty \hbar^{a'}{\cal G}^{\left(a'\right)}_{r'}\right).
\label{DeltaWK}
\end{align}
We can calculate the coefficients ${\cal Z}_{2\ell}^{\left(q,{\tilde q}\right)}\left(s;v_\alpha,M,{\tilde v}_\alpha,{\tilde M}\right)$ of the $\hbar$ expansion of the spectral zeta function analytically in $s$, where for each $\ell$ we can truncate the summation over $(r,a,r',a')$ at some finite orders.

For example, let us consider the sub-leading correction to the spectal trace ${\cal Z}_{2}^{\left(q,{\tilde q}\right)}\left(s;v_\alpha,M;{\tilde v}_\alpha,{\tilde M}\right)$
\begin{align}
{\cal Z}_{2}^{\left(q,{\tilde q}\right)}\left(s;v_\alpha,M;{\tilde v}_\alpha,{\tilde M}\right)
&=\int\frac{dxdp}{2\pi}e^{-sH_W^{\left(0\right)}}\left(-sH_W^{\left(2\right)}+\frac{s^2}{2}\left(H_W^{\left(1\right)}\right)^2+\frac{s^2}{2}{\cal G}_2^{\left(2\right)}-\frac{s^3}{6}{\cal G}_3^{\left(2\right)}\right),
\label{calZ2inWK}
\end{align}
where the expansion coefficients $H_W^{\left(a\right)}$ and ${\cal G}_r^{\left(a\right)}$ are obtained as
\begin{subequations}
\begin{align}
H_W^{\left(0\right)}&=q\log 2\cosh\frac{x}{2}-\frac{iqMx}{2}+{\tilde q}\log 2\cosh\frac{p}{2}-\frac{i{\tilde q}{\tilde M}x}{2},\\
H_W^{\left(1\right)}&=0,\\
H_W^{\left(2\right)}&=
\frac{1}{F\left(x\right)^2}\left(\frac{1}{2}\sum_{\alpha=1}^qv_\alpha^2-\frac{q{\tilde q}^2\left(1-{\tilde M}^2\right)}{48}\right)
+\frac{1}{F\left(p\right)^2}\left(\frac{1}{2}\sum_{\alpha=1}^{\tilde q}{\tilde v}_\alpha^2+\frac{q^2{\tilde q}\left(1-M^2\right)}{96}\right)\nonumber \\
&\quad +\frac{iq{\tilde q}^2{\tilde M}F'\left(p\right)}{12F\left(x\right)^2F\left(p\right)}
-\frac{iq^2{\tilde q}MF'\left(x\right)}{24F\left(x\right)F\left(p\right)^2}
+\frac{1}{F\left(x\right)^2F\left(p\right)^2}\left(-\frac{q^2{\tilde q}}{24}+\frac{q{\tilde q}^2}{12}\right),\\
{\cal G}^{\left(2\right)}_2&=
\frac{1}{4}\left[
\left(\partial_x\partial_pH_W^{\left(0\right)}\right)^2
-\partial_x^2H_W^{\left(0\right)}\partial_p^2H_W^{\left(0\right)}
\right]
=-\frac{q{\tilde q}}{4F\left(x\right)^2F\left(p\right)^2},\\
{\cal G}^{\left(2\right)}_3&=
\frac{1}{4}\left[
2\partial_xH_W^{\left(0\right)}
\partial_pH_W^{\left(0\right)}
\partial_x\partial_pH_W^{\left(0\right)}
-\partial_x^2H_W^{\left(0\right)}\left(\partial_pH_W^{\left(0\right)}\right)^2
-\left(\partial_xH_W^{\left(0\right)}\right)^2\partial_p^2H_W^{\left(0\right)}
\right]\nonumber \\
&=
-\frac{q{\tilde q}^2\left(1-{\tilde M}^2\right)}{16F\left(x\right)^2}
-\frac{q^2{\tilde q}\left(1-M^2\right)}{16F\left(p\right)^2}
+\frac{iq{\tilde q}^2{\tilde M}F'\left(p\right)}{4F\left(x\right)^2F\left(p\right)}
+\frac{iq^2{\tilde q}MF'\left(x\right)}{4F\left(x\right)F\left(p\right)^2}
+\frac{q{\tilde q}\left(q+{\tilde q}\right)}{4F\left(x\right)^2F\left(p\right)^2},
\end{align}
\end{subequations}
where $F\left(x\right)=2\cosh\frac{x}{2}$.
Plugging these into \eqref{calZ2inWK}, we find that the integration of the phase space can be performed term by term by using the formula \eqref{I1andI2}
\begin{align}
&{\cal Z}_2^{\left(q,{\tilde q}\right)}\left(s;v_\alpha,M;{\tilde v}_\alpha,{\tilde M}\right)\nonumber \\
&=\frac{1}{2\pi}\left[
-s\left(
\left(\frac{1}{2}\sum_{\alpha=1}^qv_\alpha^2-\frac{q{\tilde q}^2\left(1-{\tilde M}^2\right)}{48}\right)
I_1\left(\frac{iqMs}{2},qs+2\right)
I_1\left(\frac{i{\tilde q}{\tilde M}s}{2},{\tilde q}s\right)
\right.\right.\nonumber \\
&\quad +
\left(\frac{1}{2}\sum_{\alpha=1}^{\tilde q}{\tilde v}_\alpha^2+\frac{q^2{\tilde q}\left(1-M^2\right)}{96}\right)
I_1\left(\frac{iqMs}{2},qs\right)
I_1\left(\frac{i{\tilde q}{\tilde M}s}{2},{\tilde q}s+2\right)\nonumber \\
&\quad +
\frac{iq{\tilde q}^2{\tilde M}}{12}
I_1\left(\frac{iqMs}{2},qs+2\right)
I_2\left(\frac{i{\tilde q}{\tilde M}s}{2},{\tilde q}s+1\right)
\nonumber \\
&\quad
-
\frac{iq^2{\tilde q}M}{24}
I_2\left(\frac{iqMs}{2},qs+1\right)
I_1\left(\frac{i{\tilde q}{\tilde M}s}{2},{\tilde q}s+2\right)\nonumber \\
&\quad +
\left.
\left(-\frac{q^2{\tilde q}}{24}+\frac{q{\tilde q}^2}{12}\right)
I_1\left(\frac{iqMs}{2},qs+2\right)
I_1\left(\frac{i{\tilde q}{\tilde M}s}{2},{\tilde q}s+2\right)
\right)\nonumber \\
&\quad +\frac{s^2}{2}\left(
-
\frac{q{\tilde q}}{4}
I_1\left(\frac{iqMs}{2},qs+2\right)
I_1\left(\frac{i{\tilde q}{\tilde M}s}{2},{\tilde q}s+2\right)
\right)\nonumber \\
&\quad -\frac{s^3}{6}\left(
-
\frac{q{\tilde q}^2\left(1-{\tilde M}^2\right)}{16}
I_1\left(\frac{iqMs}{2},qs+2\right)
I_1\left(\frac{i{\tilde q}{\tilde M}s}{2},{\tilde q}s\right)
\right.\nonumber \\
&\quad -
\frac{q^2{\tilde q}\left(1-M^2\right)}{16}
I_1\left(\frac{iqMs}{2},qs\right)
I_1\left(\frac{i{\tilde q}{\tilde M}s}{2},{\tilde q}s+2\right)\nonumber \\
&\quad +
\frac{iq{\tilde q}^2{\tilde M}}{4}
I_1\left(\frac{iqMs}{2},qs+2\right)
I_2\left(\frac{i{\tilde q}{\tilde M}s}{2},{\tilde q}s+1\right)\nonumber \\
&\quad +
\frac{iq^2{\tilde q}M}{4}
I_2\left(\frac{iqMs}{2},qs+1\right)
I_1\left(\frac{i{\tilde q}{\tilde M}s}{2},{\tilde q}s+2\right)\nonumber \\
&\quad \left.\left.+
\frac{q{\tilde q}\left(q+{\tilde q}\right)}{4}
I_1\left(\frac{iqMs}{2},qs+2\right)
I_1\left(\frac{i{\tilde q}{\tilde M}s}{2},{\tilde q}s+2\right)
\right)
\right].
\end{align}
Using the relations satisfied by $I_1\left(\alpha,n\right)$ and $I_2\left(\alpha,n\right)$ \eqref{recursionrelationofI1I2}, we can factorize ${\cal Z}_2^{\left(q,{\tilde q}\right)}\left(s;v_\alpha,M;{\tilde v}_\alpha,{\tilde M}\right)$ as
\begin{align}
{\cal Z}_2^{\left(q,{\tilde q}\right)}\left(s;v_\alpha,M;{\tilde v}_\alpha,{\tilde M}\right)
=
\frac{
I_1\left(\frac{iqMs}{2},qs\right)I_1\left(\frac{i{\tilde q}{\tilde M}s}{2},{\tilde q}s\right)
}{2\pi}D_2^{\left(q,{\tilde q}\right)}\left(s;v_\alpha,M;{\tilde v}_\alpha,{\tilde M}\right),
\end{align}
where the first factor is ${\cal Z}_0^{\left(q,{\tilde q}\right)}\left(s;v_\alpha,M;{\tilde v}_\alpha,{\tilde M}\right)$, and the second factor is a rational function of $s$ which is given explicitly as
\begin{align}
D_2^{\left(q,{\tilde q}\right)}\left(s;v_\alpha,M;{\tilde v}_\alpha,{\tilde M}\right)
&=\frac{
s^2
}{384 (1 + q s) (1 + {\tilde q} s)}
\left(
q^2 {\tilde q}^2 \left(1 + M^2\right) \left(1 + {\tilde M}^2\right) \left(1 - s^2\right)\right.\nonumber \\
&\quad \left.- 48 q \left(1 + M^2\right) \left(1 + {\tilde q} s\right)\sum_{\alpha=1}^qv_\alpha^2
- 48 {\tilde q} \left(1 + {\tilde M}^2\right) \left(1 + q s\right)\sum_{\alpha=1}^{\tilde q}{\tilde v}_\alpha^2
\right).
\end{align}

Now let us argue that the structure \eqref{DrelationofcalZ} is universal for all order $\ell$.
The crucial point in the above calculation for $\ell=1$ is the fact that the coefficient of $e^{-sH_W^{\left(0\right)}}$ in the integrand \eqref{calZinWK} is a linear combination of the following terms
\begin{align}
\frac{1}{F\left(x\right)^{2a_1}F\left(p\right)^{2a_2}},\quad
\frac{F'\left(x\right)}{F\left(x\right)^{2a_1+1}F\left(p\right)^{2a_2}},\quad
\frac{F'\left(p\right)}{F\left(x\right)^{2a_1}F\left(p\right)^{2a_2+1}},\quad
\frac{F'\left(x\right)F'\left(p\right)}{F\left(x\right)^{2a_1+1}F\left(p\right)^{2a_2+1}},
\label{goodterms}
\end{align}
with $F\left(x\right)=2\cosh\frac{x}{2}$ and $a_1,a_2\in\mathbb{Z}_{\ge 0}$.
This has allowed us to replace all $I_1\left(\alpha,n\right)$ functions and $I_2\left(\alpha,n\right)$ functions obtained by perform the phase space integration into $I_1\left(\frac{iqMs}{2},qs\right)$ and $I_1\left(\frac{i{\tilde q}{\tilde M}s}{2},{\tilde q}s\right)$ times some rational functions of $s$, and hence factorize ${\cal Z}_2^{\left(q,{\tilde q}\right)}\left(s;v_\alpha,M,{\tilde v}_\alpha,{\tilde M}\right)$ into the form \eqref{DrelationofcalZ}.
Therefore, in order to prove \eqref{DrelationofcalZ} for general $\ell$, it is sufficient to prove the same property for the entire coefficient $\Delta_\text{WK}$ of $e^{-sH_W^{\left(0\right)}}$ in the integrand of the spectral trace \eqref{calZinWK}.

For this purpose, first we notice that by construction $H_W^{\left(a\right)}$ ($a\ge 1$) and ${\cal G}_r^{\left(a\right)}$ are written in some polynomial of $\partial_x^{a_1}U\left(x\right)$ and $\partial_p^{a_2}T\left(x\right)$ with $a_1,a_2\ge 1$, and that the vector space spanned by $\frac{1}{F\left(x\right)^{2a_1}}$ and $\frac{F'\left(x\right)}{F\left(x\right)^{2a_1+1}}$ is the same as the set of polynomials of $\frac{F'\left(x\right)}{F\left(x\right)}$
\begin{align}
\text{Vec}\left(\frac{1}{F\left(x\right)^{2a_1}},\frac{F'\left(x\right)}{F\left(x\right)^{2a_2+1}};a_1,a_2\ge 0\right)
=
\text{Pol}\left[\frac{F'\left(x\right)}{F\left(x\right)}\right].
\end{align}
Therefore, in order to prove that the $\hbar$-expansion coefficients of $\Delta_\text{WK}$ \eqref{DeltaWK} are finite linear combination of the terms \eqref{goodterms}, it is sufficient to prove that $\partial_x^aU\left(x\right)\in\text{Pol}\left[\frac{F'\left(x\right)}{F\left(x\right)}\right]$ for $a\ge 1$.
This is indeed the case since $\partial_xU\left(x\right)$ can be written as
\begin{align}
\partial_xU\left(x\right)=\frac{qF'\left(x\right)}{F\left(x\right)}-\frac{iqM}{2}+\sum_{\alpha=1}^q\partial_x\log\left[\cosh\frac{\hbar v_\alpha}{2}+2\sinh\frac{\hbar v_\alpha}{2}\frac{F'\left(x\right)}{F\left(x\right)}\right]
\end{align}
and $\partial_x\text{Pol}\left[\frac{F'\left(x\right)}{F\left(x\right)}\right]\subset \text{Pol}\left[\frac{F'\left(x\right)}{F\left(x\right)}\right]$, which holds due to the derivative relations of $F\left(x\right)$ \eqref{derivativerelationofF}.

\section{List of \texorpdfstring{$D_{2\ell}^{\left(q,{\tilde q}\right)}\left(s;v_\alpha,M;{\tilde v}_\alpha,{\tilde M}\right)$}{D}}
\label{app_listofD}

In this appendix we list the ratio $D^{\left(q,{\tilde q}\right)}_{2\ell}\left(s;v_\alpha,M;{\tilde v}_\alpha,{\tilde M}\right)$ between the higher order coefficients of the spectral trace in the $\hbar$ expansion and the spectral trace in the classical limit, which we have used in section \ref{subsec:LargeN-A} to guess the expansion formula for the coefficient 
$A_k^{\left(q,{\tilde q}\right)}\left(\eta_\alpha,M;{\tilde\eta}_\alpha,{\tilde M}\right)$ \eqref{Aofpqmodel} of the Airy form of the partition function.
See the Mathematica notebook attached to this paper in arXiv.org for more results.

Below we use the following abbreviations
\begin{align}
V_\nu=\sum_{\alpha=1}^qv_\alpha^\nu,\quad
{\tilde V}_\nu=\sum_{\alpha=1}^{\tilde q}{\tilde v}_\alpha^\nu.
\end{align}

\subsection{\texorpdfstring{$D_{2\ell}^{\left(q,{\tilde q}\right)}\left(s;v_\alpha,0;{\tilde v}_\alpha,0\right)$}{Dqq}}
For $\ell=2$, we find
{\fontsize{9pt}{1pt}\selectfont
\begin{align}
D_4^{\left(q,{\tilde q}\right)}\left(s;v_\alpha,0;{\tilde v}_\alpha,0\right)
&=f_{4,\emptyset,\emptyset}^{\left(q,{\tilde q}\right)}\left(s;0;0\right)
+f_{4,\{2\},\emptyset}^{\left(q,{\tilde q}\right)}\left(s;0;0\right)V_2
+f_{4,\{2\},\emptyset}^{\left({\tilde q},q\right)}\left(s;0;0\right){\tilde V}_2
+f_{4,\{4\},\emptyset}^{\left(q,{\tilde q}\right)}\left(s;0;0\right)V_4\nonumber \\
&\quad +f_{4,\{2,2\},\emptyset}^{\left(q,{\tilde q}\right)}\left(s;0;0\right)V_2^2
+f_{4,\{2\},\{2\}}^{\left(q,{\tilde q}\right)}\left(s;0;0\right)V_2{\tilde V}_2
+f_{4,\{4\},\emptyset}^{\left({\tilde q},q\right)}\left(s;0;0\right){\tilde V}_4\nonumber \\
&\quad+f_{4,\{2,2\},\emptyset}^{\left({\tilde q},q\right)}\left(s;0;0\right){\tilde V}_2^2,
\end{align}
}
with
{\fontsize{9pt}{1pt}\selectfont
\begin{subequations}
\label{f4IJM0Mtilde0}
\begin{align}
f^{\left(q,{\tilde q}\right)}_{4,\emptyset,\emptyset}\left(s;0;0\right)&=
-
\frac{q^3 {\tilde q}^3 s^3\left(1 - s^2\right) \left(24 q+24{\tilde q}+n\left(80+17q{\tilde q}\right)+n^2\left(24q+24{\tilde q}\right)+7q{\tilde q}n^3\right)
}{1474560 \left(1 + q s\right)\left(3 + q s\right) \left(1 + {\tilde q} s\right)\left(3 + {\tilde q} s\right)},\\
f^{\left(q,{\tilde q}\right)}_{4,\{2\},\emptyset}\left(s;0;0\right)&=-\frac{q^2 {\tilde q}^2 s^3\left(1 - s^2\right) \left(4 + q s\right)}{3072 \left(1 + q s\right) \left(1 + {\tilde q} s\right) \left(3 + q s\right)},\\
f^{\left(q,{\tilde q}\right)}_{4,\{4\},\emptyset}\left(s;0;0\right)&=\frac{q^2s^3}{192 \left(1 + q s\right) \left(3 + q s\right)},\\
f^{\left(q,{\tilde q}\right)}_{4,\{2,2\},\emptyset}\left(s;0;0\right)&= \frac{qs^3(2+qs)}{128 \left(1 + q s\right) \left(3 + q s\right)},\\
f^{\left(q,{\tilde q}\right)}_{4,\{2\},\{2\}}\left(s;0;0\right)&=\frac{q^2s^4}{64 \left(1 + {\tilde q} s\right) \left(1 + q s\right)}.
\end{align}
\end{subequations}
}

For $\ell=3$, we find
{\fontsize{9pt}{1pt}\selectfont
\begin{align}
D_6^{\left(q,{\tilde q}\right)}\left(s;v_\alpha,0;{\tilde v}_\alpha,0\right)&=
f_{6,\emptyset,\emptyset}^{\left(q,{\tilde q}\right)}\left(s;0;0\right)
+f_{6,\{2\},\emptyset}^{\left(q,{\tilde q}\right)}\left(s;0;0\right)V_2
+f_{6,\{2\},\emptyset}^{\left({\tilde q},q\right)}\left(s;0;0\right){\tilde V}_2
+f_{6,\{4\},\emptyset}^{\left(q,{\tilde q}\right)}\left(s;0;0\right)V_4\nonumber \\
&\quad +f_{6,\{2,2\},\emptyset}^{\left(q,{\tilde q}\right)}\left(s;0;0\right)V_2^2
+f_{6,\{2\},\{2\}}^{\left(q,{\tilde q}\right)}\left(s;0;0\right)V_2{\tilde V}_2
+f_{6,\{4\},\emptyset}^{\left({\tilde q},q\right)}\left(s;0;0\right){\tilde V}_4\nonumber \\
&\quad +f_{6,\{2,2\},\emptyset}^{\left({\tilde q},q\right)}\left(s;0;0\right){\tilde V}_2^2
+f_{6,\{6\},\emptyset}^{\left(q,{\tilde q}\right)}\left(s;0;0\right)V_6
+f_{6,\{4,2\},\emptyset}^{\left(q,{\tilde q}\right)}\left(s;0;0\right)V_4V_2\nonumber \\
&\quad +f_{6,\{3,3\},\emptyset}^{\left(q,{\tilde q}\right)}\left(s;0;0\right)V_3^2
+f_{6,\{2,2,2\},\emptyset}^{\left(q,{\tilde q}\right)}\left(s;0;0\right)V_2^3
+f_{6,\{4\},\{2\}}^{\left(q,{\tilde q}\right)}\left(s;0;0\right)V_4{\tilde V}_2\nonumber \\
&\quad +f_{6,\{2,2\},\{2\}}^{\left(q,{\tilde q}\right)}\left(s;0;0\right)V_2^2{\tilde V}_2
+f_{6,\{4\},\{2\}}^{\left({\tilde q},q\right)}\left(s;0;0\right)V_2{\tilde V}_4
+f_{6,\{2,2\},\{2\}}^{\left({\tilde q},q\right)}\left(s;0;0\right)V_2{\tilde V}_2^2\nonumber \\
&\quad +f_{6,\{6\},\emptyset}^{\left({\tilde q},q\right)}\left(s;0;0\right){\tilde V}_6
+f_{6,\{4,2\},\emptyset}^{\left({\tilde q},q\right)}{\tilde V}_4{\tilde V}_2
+f_{6,\{3,3\},\emptyset}^{\left({\tilde q},q\right)}{\tilde V}_3^2
+f_{6,\{2,2,2\},\emptyset}^{\left({\tilde q},q\right)}{\tilde V}_2^3,
\end{align}
}
with
{\fontsize{9pt}{1pt}\selectfont
\begin{subequations}
\label{f6IJM0Mtilde0}
\begin{align}
f_{6,\emptyset,\emptyset}^{\left(q,{\tilde q}\right)}\left(s;0;0\right)&=
\frac{
q^3 {\tilde q}^3 s^3\left(1 - s^2\right)}{3963617280 \left(1 + q s\right) \left(3 + q s\right) \left(5 + q s\right) \left(1 + {\tilde q} s\right) \left(3 + {\tilde q} s\right) \left(5 + {\tilde q} s\right)}
\left(1920 q^3 + 1920 {\tilde q}^3\right.\nonumber \\
&\quad  + \left(7168 + 5376 q^2 + 14336 q {\tilde q} + 2944 q^3 {\tilde q} + 5376 {\tilde q}^2 + 2304 q^2 {\tilde q}^2 + 2944 q {\tilde q}^3\right) s\nonumber \\
&\quad + \left(14336 q + 576 q^3 + 14336 {\tilde q} + 17920 q^2 {\tilde q} + 17920 q {\tilde q}^2 + 1656 q^3 {\tilde q}^2 + 576 {\tilde q}^3 + 1656 q^2 {\tilde q}^3\right) s^2\nonumber \\
&\quad + \left(5376 q^2 + 25088 q {\tilde q} + 2272 q^3 {\tilde q} + 5376 {\tilde q}^2 + 12272 q^2 {\tilde q}^2 + 2272 q {\tilde q}^3 + 367 q^3 {\tilde q}^3\right) s^3\nonumber \\
&\quad + \left(576 q^3 + 8960 q^2 {\tilde q} + 8960 q {\tilde q}^2 + 1488 q^3 {\tilde q}^2 + 576 {\tilde q}^3 + 1488 q^2 {\tilde q}^3\right) s^4\nonumber \\
&\quad\left. + \left(928 q^3 {\tilde q} + 3088 q^2 {\tilde q}^2 + 928 q {\tilde q}^3 + 178 q^3 {\tilde q}^3\right) s^5 + \left(312 q^3 {\tilde q}^2 + 312 q^2 {\tilde q}^3\right) s^6 + 31 q^3 {\tilde q}^3 s^7\right),\\
f_{6,\{2\},\emptyset}^{\left(q,{\tilde q}\right)}\left(s;0;0\right)&=
\frac{q^2 {\tilde q}^3 s^3 \left(1 - s^2\right)}{11796480 \left(1 + q s\right) \left(3 + q s\right) \left(5 + q s\right) \left(1 + {\tilde q} s\right) \left(3 + {\tilde q} s\right)}\left(192 {\tilde q} + \left(640 + 192 q^2 + 288 q {\tilde q}\right) s\right.\nonumber \\
&\quad + \left(960 q + 24 q^3 + 192 {\tilde q} + 160 q^2 {\tilde q}\right) s^2 + \left(272 q^2 + 288 q {\tilde q} + 17 q^3 {\tilde q}\right) s^3 + \left(24 q^3 + 80 q^2 {\tilde q}\right) s^4\nonumber \\
&\quad\left. + 7 q^3 {\tilde q} s^5\right),\\
f_{6,\{4\},\emptyset}^{\left(q,{\tilde q}\right)}\left(s;0;0\right)&=
\frac{q^2 {\tilde q}^2 s^3 \left(1 - s^2\right) \left(8 + 12 q s + q^2 s^2\right)}{73728 \left(1 + q s\right) \left(3 + q s\right) \left(5 + q s\right) \left(1 + {\tilde q} s\right)},\\
f_{6,\{2,2\},\emptyset}^{\left(q,{\tilde q}\right)}\left(s;0;0\right)&=
\frac{q {\tilde q}^2 s^3 \left(1 - s^2\right) \left(2 + q s\right) \left(8 + 8 q s + q^2 s^2\right)}{49152 \left(1 + q s\right) \left(3 + q s\right) \left(5 + q s\right) \left(1 + {\tilde q} s\right)},\\
f_{6,\{2\},\{2\}}^{\left(q,{\tilde q}\right)}\left(s;0;0\right)&=
\frac{q^2 {\tilde q}^2 s^4 \left(1 - s^2\right) \left(4 + q s\right) \left(4 + {\tilde q} s\right)}{24576 \left(1 + q s\right) \left(3 + q s\right) \left(1 + {\tilde q} s\right) \left(3 + {\tilde q} s\right)},\\
f_{6,\{6\},\emptyset}^{\left(q,{\tilde q}\right)}\left(s;0;0\right)&=
-\frac{q^2 s^3 \left(1 + 2 q s\right)}{5760 \left(1 + q s\right) \left(3 + q s\right) \left(5 + q s\right)},\\
f_{6,\{4,2\},\emptyset}^{\left(q,{\tilde q}\right)}\left(s;0;0\right)&=
-\frac{q s^3 \left(2 + q s\right)^2}{1536 \left(1 + q s\right) \left(3 + q s\right) \left(5 + q s\right)},\\
f_{6,\{3,3\},\emptyset}^{\left(q,{\tilde q}\right)}\left(s;0;0\right)&=
 \frac{q s^3 \left(2 + q s\right)}{1152 \left(1 + q s\right) \left(3 + q s\right) \left(5 + q s\right)},\\
f_{6,\{2,2,2\},\emptyset}^{\left(q,{\tilde q}\right)}\left(s;0;0\right)&=
-\frac{q s^4 \left(2 + q s\right) \left(4 + q s\right)}{3072 \left(1 + q s\right) \left(3 + q s\right) \left(5 + q s\right)},\\
f_{6,\{4\},\{2\}}^{\left(q,{\tilde q}\right)}\left(s;0;0\right)&=
-\frac{q^2 {\tilde q} s^5}{1536 \left(1 + q s\right) \left(3 + q s\right) \left(1 + {\tilde q} s\right)},\\
f_{6,\{2,2\},\{2\}}^{\left(q,{\tilde q}\right)}\left(s;0;0\right)&=
-\frac{q {\tilde q} s^5 \left(2 + q s\right)}{1024 \left(1 + q s\right) \left(3 + q s\right) \left(1 + {\tilde q} s\right)}.
\end{align}
\end{subequations}
}
Note that both $D_4^{\left(q,{\tilde q}\right)}\left(s;v_\alpha,0;{\tilde v}_\alpha,0\right)$ and $D_6^{\left(q,{\tilde q}\right)}\left(s,v_\alpha,0;{\tilde v}_\alpha,0\right)$ contain the terms which cannot be written as a monomial of the form $\sum_{\alpha<\beta}\left(v_\alpha-v_\beta\right)^\nu$ with some power $\nu$.
Since the perturbative part of the grand potential depends linearly on $D_{2\ell}^{\left(q,{\tilde q}\right)}\left(s;v_\alpha,M;{\tilde v}_\alpha,{\tilde M}\right)$ through the residue formula \eqref{JpertfromcalZ}, this fact apparently contradicts to our conjecture for $A_k^{\left(q,{\tilde q}\right)}\left(\eta_\alpha,M;{\tilde\eta}_\alpha,{\tilde M}\right)$ \eqref{Aofpqmodel} which is expanded in the monomials of $\sum_{\alpha<\beta}\left(v_\alpha-v_\beta\right)^\nu$.
However, by looking at the expressions of $D_{2\ell}^{\left(q,{\tilde q}\right)}\left(s;v_\alpha,0;{\tilde v}_\alpha,0\right)$ carefully, we find that the terms which contributes to the residue at $s=0$ and hence to the perturbative part of the grand potential are indeed written in the terms of the form $\sum_{\alpha<\beta}\left(v_\alpha-v_\beta\right)^\nu$ due to the following identities
\begin{subequations}
\begin{align}
&q\sum_\alpha v_\alpha^4+3\left(\sum_\alpha v_\alpha^2\right)^2=\sum_{\alpha<\beta}\left(v_\alpha-v_\beta\right)^4,\\
&q\sum_\alpha v_\alpha^6+15\sum_\alpha v_\alpha^4\sum_\beta v_\beta^2-10\left(\sum_\alpha v_\alpha^3\right)^2
=\sum_{\alpha<\beta}\left(v_\alpha-v_\beta\right)^6.
\end{align}
\end{subequations}

\subsection{\texorpdfstring{$D_{2\ell}^{\left(q,1\right)}\left(s;v_\alpha,0;{\tilde M}\right)$}{Dq1}}

For $\ell=2$ we find
{\fontsize{9pt}{1pt}\selectfont
\begin{align}
D_4^{\left(q,1\right)}\left(s;v_\alpha,0;{\tilde M}\right)
&=
f_{4,\emptyset,\emptyset}^{\left(q,1\right)}\left(s;0;{\tilde M}\right)
+f_{4,\{2\},\emptyset}^{\left(q,1\right)}\left(s;0;{\tilde M}\right)V_2
+f_{4,\{4\},\emptyset}^{\left(q,1\right)}\left(s;0;{\tilde M}\right)V_4
+f_{4,\{2,2\},\emptyset}^{\left(q,1\right)}\left(s;0;{\tilde M}\right)V_2^2,
\end{align}
}
with
{\fontsize{9pt}{1pt}\selectfont
\begin{subequations}
\begin{align}
f_{4,\emptyset,\emptyset}^{\left(q,1\right)}\left(s;0;{\tilde M}\right) &=     \frac{\left(1 + {\tilde M}^2\right) q^3 s^3 \left(1 - s\right) \left(
-8 - 8 q + \left(-24 - 3 q\right) s - 7 q s^2 + {\tilde M}^2 \left(24 + \left(-24 + 5 q\right) s - 7 q s^2\right)
\right)}{1474560 \left(1 + q s\right) \left(3 + q s\right)},\\
f_{4,\{2\},\emptyset}^{\left(q,1\right)}\left(s;0;{\tilde M}\right) &=\left(1 + {\tilde M}^2\right)  f_{4,\{2\},\emptyset}^{\left(q,1\right)}\left(s;0;0\right),\\
f_{4,\{4\},\emptyset}^{\left(q,1\right)}\left(s;0;{\tilde M}\right) &=                               f_{4,\{4\},\emptyset}^{\left(q,1\right)}\left(s;0;0\right),\\
f_{4,\{2,2\},\emptyset}^{\left(q,1\right)}\left(s;0;{\tilde M}\right) &=                             f_{4,\{2,2\},\emptyset}^{\left(q,1\right)}\left(s;0;0\right) .
\end{align}
\end{subequations}
}
Here $f^{\left(q,1\right)}_{4,I,J}\left(s;0;0\right)$ are given as \eqref{f4IJM0Mtilde0}.

For $\ell=3$, we find
{\fontsize{9pt}{1pt}\selectfont
\begin{align}
D_6^{\left(q,1\right)}\left(s;v_\alpha,0;{\tilde v}_\alpha,{\tilde M}\right)&=
 f_{6,\emptyset,\emptyset}^{\left(q,1\right)}\left(s;0;{\tilde M}\right)
    +f_{6,\{2\},\emptyset}^{\left(q,1\right)}\left(s;0;{\tilde M}\right)V_2
    +f_{6,\{4\},\emptyset}^{\left(q,1\right)}\left(s;0;{\tilde M}\right)V_4\nonumber \\
&\quad  +f_{6,\{2,2\},\emptyset}^{\left(q,1\right)}\left(s;0;{\tilde M}\right)V_2^2
    +f_{6,\{6\},\emptyset}^{\left(q,1\right)}\left(s;0;{\tilde M}\right)V_6
  +f_{6,\{4,2\},\emptyset}^{\left(q,1\right)}\left(s;0;{\tilde M}\right)V_4V_2\nonumber \\
&\quad  +f_{6,\{3,3\},\emptyset}^{\left(q,1\right)}\left(s;0;{\tilde M}\right)V_3^2
+f_{6,\{2,2,2\},\emptyset}^{\left(q,1\right)}\left(s;0;{\tilde M}\right)V_2^3,
\end{align}
}
with
{\fontsize{9pt}{1pt}\selectfont
\begin{subequations}
\begin{align}
f_{6,\emptyset,\emptyset}^{\left(q,1\right)}\left(s;0;{\tilde M}\right) &= \frac{\left(1 + {\tilde M}^2\right) q^3 s^3 \left(1 - s\right)}{11890851840 \left(1 + q s\right) \left(3 + q s\right) \left(5 + q s\right)}\left(384 + 384 q^3\right.\nonumber \\
&\quad + \left(2304 + 3456 q + 1536 q^2 + 384 q^3\right) s + \left(1728 + 4608 q + 3096 q^2 + 216 q^3\right) s^2\nonumber \\
&\quad + \left(2784 q + 1776 q^2 + 387 q^3\right) s^3 + \left(936 q^2 + 192 q^3\right) s^4 + 93 q^3 s^5\nonumber \\
&\quad + {\tilde M}^2 \left(-3840 + \left(-1536 - 5888 q - 2304 q^2\right) s + \left(3456 - 1280 q - 240 q^2 - 168 q^3\right) s^2\right.\nonumber \\
&\quad\left. + \left(5568 q + 192 q^2 + 38 q^3\right) s^3 + \left(1872 q^2 + 56 q^3\right) s^4 + 186 q^3 s^5\right)\nonumber \\
&\quad + {\tilde M}^4 \left(1920 + \left(-3840 + 2944 q\right) s + \left(1728 - 5888 q + 504 q^2\right) s^2\right.\nonumber \\
&\quad\left.\left. + \left(2784 q - 1584 q^2 + 35 q^3\right) s^3 + \left(936 q^2 - 136 q^3\right) s^4 + 93 q^3 s^5\right)
\right),\\
f_{6,\{2\},\emptyset}^{\left(q,1\right)}\left(s;0;{\tilde M}\right)   &= \frac{\left(1 + {\tilde M}^2\right) q^2 s^3 \left(1 - s\right)}{11796480 \left(1 + q s\right) \left(3 + q s\right) \left(5 + q s\right)}\left(64 + \left(192 + 96 q + 64 q^2\right) s\right.\nonumber \\
&\quad + \left(288 q + 32 q^2 + 8 q^3\right) s^2 + \left(80 q^2 + 3 q^3\right) s^3 + 7 q^3 s^4\nonumber \\
&\quad\left. + {\tilde M}^2 \left(-192 + \left(192 - 288 q\right) s + \left(288 q - 64 q^2\right) s^2 + \left(80 q^2 - 5 q^3\right) s^3 + 7 q^3 s^4\right)\right),\\
f_{6,\{4\},\emptyset}^{\left(q,1\right)}\left(s;0;{\tilde M}\right)   &= \left(1 + {\tilde M}^2\right)  f_{6,\{4\},\emptyset}^{\left(q,1\right)}\left(s;0;    0\right),\\
f_{6,\{2,2\},\emptyset}^{\left(q,1\right)}\left(s;0;{\tilde M}\right)  &= \left(1 + {\tilde M}^2\right) f_{6,\{2,2\},\emptyset}^{\left(q,1\right)}\left(s;0;  0\right),\\
f_{6,\{6\},\emptyset}^{\left(q,1\right)}\left(s;0;{\tilde M}\right)   &=                                f_{6,\{6\},\emptyset}^{\left(q,1\right)}\left(s;0;    0\right),\\
f_{6,\{4,2\},\emptyset}^{\left(q,1\right)}\left(s;0;{\tilde M}\right)  &=                               f_{6,\{4,2\},\emptyset}^{\left(q,1\right)}\left(s;0;  0\right),\\
f_{6,\{3,3\},\emptyset}^{\left(q,1\right)}\left(s;0;{\tilde M}\right)  &=                               f_{6,\{3,3\},\emptyset}^{\left(q,1\right)}\left(s;0;  0\right),\\
f_{6,\{2,2,2\},\emptyset}^{\left(q,1\right)}\left(s;0;{\tilde M}\right) &=                              f_{6,\{2,2,2\},\emptyset}^{\left(q,1\right)}\left(s;0;0\right). 
\end{align}
\end{subequations}
}
Here $f^{\left(q,1\right)}_{6,I,J}\left(s;0;0\right)$ are given as \eqref{f6IJM0Mtilde0}.

\bibliographystyle{utphys}
\bibliography{References.bib}

\providecommand{\href}[2]{#2}\begingroup\raggedright\begin{thebibliography}{10}

\bibitem{Kapustin:2009kz}
A.~Kapustin, B.~Willett, and I.~Yaakov, ``{Exact Results for Wilson Loops in
  Superconformal Chern-Simons Theories with Matter},''
  \href{http://dx.doi.org/10.1007/JHEP03(2010)089}{{\em JHEP} {\bfseries 03}
  (2010) 089},
\href{http://arxiv.org/abs/0909.4559}{{\ttfamily arXiv:0909.4559 [hep-th]}}.

\bibitem{Herzog:2010hf}
C.~P. Herzog, I.~R. Klebanov, S.~S. Pufu, and T.~Tesileanu, ``{Multi-Matrix
  Models and Tri-Sasaki Einstein Spaces},''
  \href{http://dx.doi.org/10.1103/PhysRevD.83.046001}{{\em Phys. Rev. D}
  {\bfseries 83} (2011) 046001},
  \href{http://arxiv.org/abs/1011.5487}{{\ttfamily arXiv:1011.5487 [hep-th]}}.

\bibitem{Marino:2011eh}
M.~Marino and P.~Putrov, ``{ABJM theory as a Fermi gas},''
  \href{http://dx.doi.org/10.1088/1742-5468/2012/03/P03001}{{\em J. Stat.
  Mech.} {\bfseries 1203} (2012) P03001},
\href{http://arxiv.org/abs/1110.4066}{{\ttfamily arXiv:1110.4066 [hep-th]}}.

\bibitem{Hosomichi:2008jd}
K.~Hosomichi, K.-M. Lee, S.~Lee, S.~Lee, and J.~Park, ``{N=4 Superconformal
  Chern-Simons Theories with Hyper and Twisted Hyper Multiplets},''
  \href{http://dx.doi.org/10.1088/1126-6708/2008/07/091}{{\em JHEP} {\bfseries
  07} (2008) 091},
\href{http://arxiv.org/abs/0805.3662}{{\ttfamily arXiv:0805.3662 [hep-th]}}.

\bibitem{Aharony:2008ug}
O.~Aharony, O.~Bergman, D.~L. Jafferis, and J.~Maldacena, ``{N=6 superconformal
  Chern-Simons-matter theories, M2-branes and their gravity duals},''
  \href{http://dx.doi.org/10.1088/1126-6708/2008/10/091}{{\em JHEP} {\bfseries
  10} (2008) 091},
\href{http://arxiv.org/abs/0806.1218}{{\ttfamily arXiv:0806.1218 [hep-th]}}.

\bibitem{Fuji:2011km}
H.~Fuji, S.~Hirano, and S.~Moriyama, ``{Summing Up All Genus Free Energy of
  ABJM Matrix Model},'' \href{http://dx.doi.org/10.1007/JHEP08(2011)001}{{\em
  JHEP} {\bfseries 08} (2011) 001},
  \href{http://arxiv.org/abs/1106.4631}{{\ttfamily arXiv:1106.4631 [hep-th]}}.

\bibitem{Closset:2012ru}
C.~Closset, T.~T. Dumitrescu, G.~Festuccia, and Z.~Komargodski,
  ``{Supersymmetric Field Theories on Three-Manifolds},''
  \href{http://dx.doi.org/10.1007/JHEP05(2013)017}{{\em JHEP} {\bfseries 05}
  (2013) 017}, \href{http://arxiv.org/abs/1212.3388}{{\ttfamily arXiv:1212.3388
  [hep-th]}}.

\bibitem{Nishioka:2013gza}
T.~Nishioka and K.~Yonekura, ``{On RG Flow of $\tau_{RR}$ for Supersymmetric
  Field Theories in Three-Dimensions},''
  \href{http://dx.doi.org/10.1007/JHEP05(2013)165}{{\em JHEP} {\bfseries 05}
  (2013) 165}, \href{http://arxiv.org/abs/1303.1522}{{\ttfamily arXiv:1303.1522
  [hep-th]}}.

\bibitem{Binder:2018yvd}
D.~J. Binder, S.~M. Chester, and S.~S. Pufu, ``{Absence of $D^4 R^4$ in
  M-Theory From ABJM},'' \href{http://dx.doi.org/10.1007/JHEP04(2020)052}{{\em
  JHEP} {\bfseries 04} (2020) 052},
  \href{http://arxiv.org/abs/1808.10554}{{\ttfamily arXiv:1808.10554
  [hep-th]}}.

\bibitem{Binder:2019mpb}
D.~J. Binder, S.~M. Chester, and S.~S. Pufu, ``{AdS$_{4}$/CFT$_{3}$ from weak
  to strong string coupling},''
  \href{http://dx.doi.org/10.1007/JHEP01(2020)034}{{\em JHEP} {\bfseries 01}
  (2020) 034}, \href{http://arxiv.org/abs/1906.07195}{{\ttfamily
  arXiv:1906.07195 [hep-th]}}.

\bibitem{Chester:2021gdw}
S.~M. Chester, R.~R. Kalloor, and A.~Sharon, ``{Squashing, Mass, and Holography
  for 3d Sphere Free Energy},''
  \href{http://dx.doi.org/10.1007/JHEP04(2021)244}{{\em JHEP} {\bfseries 04}
  (2021) 244}, \href{http://arxiv.org/abs/2102.05643}{{\ttfamily
  arXiv:2102.05643 [hep-th]}}.

\bibitem{Agmon:2017xes}
N.~B. Agmon, S.~M. Chester, and S.~S. Pufu, ``{Solving M-theory with the
  Conformal Bootstrap},'' \href{http://dx.doi.org/10.1007/JHEP06(2018)159}{{\em
  JHEP} {\bfseries 06} (2018) 159},
  \href{http://arxiv.org/abs/1711.07343}{{\ttfamily arXiv:1711.07343
  [hep-th]}}.

\bibitem{Hirano:2019szi}
S.~Hirano, ``{Quantum Holographic Entanglement Entropy to All Orders in $1/N$
  Expansion},'' \href{http://dx.doi.org/10.1093/ptep/ptaa019}{{\em PTEP}
  {\bfseries 2020} no.~4, (2020) 043B02},
  \href{http://arxiv.org/abs/1911.01640}{{\ttfamily arXiv:1911.01640
  [hep-th]}}.

\bibitem{Chester:2020jay}
S.~M. Chester, R.~R. Kalloor, and A.~Sharon, ``{3d $ \mathcal{N} $ = 4 OPE
  coefficients from Fermi gas},''
  \href{http://dx.doi.org/10.1007/JHEP07(2020)041}{{\em JHEP} {\bfseries 07}
  (2020) 041}, \href{http://arxiv.org/abs/2004.13603}{{\ttfamily
  arXiv:2004.13603 [hep-th]}}.

\bibitem{Nosaka:2015iiw}
T.~Nosaka, ``{Instanton effects in ABJM theory with general R-charge
  assignments},'' \href{http://dx.doi.org/10.1007/JHEP03(2016)059}{{\em JHEP}
  {\bfseries 03} (2016) 059},
\href{http://arxiv.org/abs/1512.02862}{{\ttfamily arXiv:1512.02862 [hep-th]}}.

\bibitem{Awata:2012jb}
H.~Awata, S.~Hirano, and M.~Shigemori, ``{The Partition Function of ABJ
  Theory},'' \href{http://dx.doi.org/10.1093/ptep/ptt014}{{\em PTEP} {\bfseries
  2013} (2013) 053B04},
\href{http://arxiv.org/abs/1212.2966}{{\ttfamily arXiv:1212.2966 [hep-th]}}.

\bibitem{Honda:2013pea}
M.~Honda, ``{Direct derivation of "mirror" ABJ partition function},''
  \href{http://dx.doi.org/10.1007/JHEP12(2013)046}{{\em JHEP} {\bfseries 12}
  (2013) 046},
\href{http://arxiv.org/abs/1310.3126}{{\ttfamily arXiv:1310.3126 [hep-th]}}.

\bibitem{Matsumoto:2013nya}
S.~Matsumoto and S.~Moriyama, ``{ABJ Fractional Brane from ABJM Wilson Loop},''
  \href{http://dx.doi.org/10.1007/JHEP03(2014)079}{{\em JHEP} {\bfseries 03}
  (2014) 079},
\href{http://arxiv.org/abs/1310.8051}{{\ttfamily arXiv:1310.8051 [hep-th]}}.

\bibitem{Kubo:2020qed}
N.~Kubo, ``{Fermi gas approach to general rank theories and quantum curves},''
  \href{http://dx.doi.org/10.1007/JHEP10(2020)158}{{\em JHEP} {\bfseries 10}
  (2020) 158}, \href{http://arxiv.org/abs/2007.08602}{{\ttfamily
  arXiv:2007.08602 [hep-th]}}.

\bibitem{Assel:2015hsa}
B.~Assel, N.~Drukker, and J.~Felix, ``{Partition functions of 3d $\hat
  D$-quivers and their mirror duals from 1d free fermions},''
  \href{http://dx.doi.org/10.1007/JHEP08(2015)071}{{\em JHEP} {\bfseries 08}
  (2015) 071}, \href{http://arxiv.org/abs/1504.07636}{{\ttfamily
  arXiv:1504.07636 [hep-th]}}.

\bibitem{Moriyama:2015jsa}
S.~Moriyama and T.~Nosaka, ``{Superconformal Chern-Simons Partition Functions
  of Affine D-type Quiver from Fermi Gas},''
  \href{http://dx.doi.org/10.1007/JHEP09(2015)054}{{\em JHEP} {\bfseries 09}
  (2015) 054},
\href{http://arxiv.org/abs/1504.07710}{{\ttfamily arXiv:1504.07710 [hep-th]}}.

\bibitem{Kubo:2024raz}
N.~Kubo and T.~Nosaka, ``{Fermi gas formalism for D-type quiver Chern-Simons
  theory with non-uniform ranks},''
  \href{http://dx.doi.org/10.1007/JHEP07(2024)079}{{\em JHEP} {\bfseries 07}
  (2024) 079}, \href{http://arxiv.org/abs/2403.12808}{{\ttfamily
  arXiv:2403.12808 [hep-th]}}.

\bibitem{Mezei:2013gqa}
M.~Mezei and S.~S. Pufu, ``{Three-sphere free energy for classical gauge
  groups},'' \href{http://dx.doi.org/10.1007/JHEP02(2014)037}{{\em JHEP}
  {\bfseries 02} (2014) 037}, \href{http://arxiv.org/abs/1312.0920}{{\ttfamily
  arXiv:1312.0920 [hep-th]}}.

\bibitem{Honda:2015rbb}
M.~Honda, ``{Exact relations between M2-brane theories with and without
  Orientifolds},'' \href{http://dx.doi.org/10.1007/JHEP06(2016)123}{{\em JHEP}
  {\bfseries 06} (2016) 123}, \href{http://arxiv.org/abs/1512.04335}{{\ttfamily
  arXiv:1512.04335 [hep-th]}}.

\bibitem{Okuyama:2016xke}
K.~Okuyama, ``{Orientifolding of the ABJ Fermi gas},''
  \href{http://dx.doi.org/10.1007/JHEP03(2016)008}{{\em JHEP} {\bfseries 03}
  (2016) 008}, \href{http://arxiv.org/abs/1601.03215}{{\ttfamily
  arXiv:1601.03215 [hep-th]}}.

\bibitem{Moriyama:2016xin}
S.~Moriyama and T.~Suyama, ``{Orthosymplectic Chern-Simons Matrix Model and
  Chirality Projection},''
  \href{http://dx.doi.org/10.1007/JHEP04(2016)132}{{\em JHEP} {\bfseries 04}
  (2016) 132},
\href{http://arxiv.org/abs/1601.03846}{{\ttfamily arXiv:1601.03846 [hep-th]}}.

\bibitem{Moriyama:2016kqi}
S.~Moriyama and T.~Nosaka, ``{Orientifold ABJM Matrix Model: Chiral Projections
  and Worldsheet Instantons},''
  \href{http://dx.doi.org/10.1007/JHEP06(2016)068}{{\em JHEP} {\bfseries 06}
  (2016) 068},
\href{http://arxiv.org/abs/1603.00615}{{\ttfamily arXiv:1603.00615 [hep-th]}}.

\bibitem{Geukens:2024zmt}
S.~Geukens and J.~Hong, ``{Subleading analysis for S$^{3}$ partition functions
  of $ \mathcal{N} $ = 2 holographic SCFTs},''
  \href{http://dx.doi.org/10.1007/JHEP06(2024)190}{{\em JHEP} {\bfseries 06}
  (2024) 190}, \href{http://arxiv.org/abs/2405.00845}{{\ttfamily
  arXiv:2405.00845 [hep-th]}}.

\bibitem{Gaiotto:2019mmf}
D.~Gaiotto and T.~Okazaki, ``{Sphere correlation functions and Verma
  modules},'' \href{http://dx.doi.org/10.1007/JHEP02(2020)133}{{\em JHEP}
  {\bfseries 02} (2020) 133}, \href{http://arxiv.org/abs/1911.11126}{{\ttfamily
  arXiv:1911.11126 [hep-th]}}.

\bibitem{Kubo:2024qhq}
N.~Kubo, T.~Nosaka, and Y.~Pang, ``{Exact large N expansion of mass deformed
  ABJM theory on squashed sphere},''
  \href{http://dx.doi.org/10.1007/JHEP02(2025)106}{{\em JHEP} {\bfseries 02}
  (2025) 106}, \href{http://arxiv.org/abs/2411.07334}{{\ttfamily
  arXiv:2411.07334 [hep-th]}}.

\bibitem{Bobev:2025ltz}
N.~Bobev, P.-J. De~Smet, J.~Hong, V.~Reys, and X.~Zhang, ``{An Airy tale at
  large N},'' \href{http://dx.doi.org/10.1007/JHEP07(2025)123}{{\em JHEP}
  {\bfseries 07} (2025) 123}, \href{http://arxiv.org/abs/2502.04606}{{\ttfamily
  arXiv:2502.04606 [hep-th]}}.

\bibitem{Bobev:2022jte}
N.~Bobev, J.~Hong, and V.~Reys, ``{Large N Partition Functions, Holography, and
  Black Holes},'' \href{http://dx.doi.org/10.1103/PhysRevLett.129.041602}{{\em
  Phys. Rev. Lett.} {\bfseries 129} no.~4, (2022) 041602},
  \href{http://arxiv.org/abs/2203.14981}{{\ttfamily arXiv:2203.14981
  [hep-th]}}.

\bibitem{Hristov:2022lcw}
K.~Hristov, ``{ABJM at finite N via 4d supergravity},''
  \href{http://dx.doi.org/10.1007/JHEP10(2022)190}{{\em JHEP} {\bfseries 10}
  (2022) 190}, \href{http://arxiv.org/abs/2204.02992}{{\ttfamily
  arXiv:2204.02992 [hep-th]}}.

\bibitem{Bobev:2022eus}
N.~Bobev, J.~Hong, and V.~Reys, ``{Large N partition functions of the ABJM
  theory},'' \href{http://dx.doi.org/10.1007/JHEP02(2023)020}{{\em JHEP}
  {\bfseries 02} (2023) 020}, \href{http://arxiv.org/abs/2210.09318}{{\ttfamily
  arXiv:2210.09318 [hep-th]}}.

\bibitem{Bobev:2023lkx}
N.~Bobev, J.~Hong, and V.~Reys, ``{Large N partition functions of 3d
  holographic SCFTs},'' \href{http://dx.doi.org/10.1007/JHEP08(2023)119}{{\em
  JHEP} {\bfseries 08} (2023) 119},
  \href{http://arxiv.org/abs/2304.01734}{{\ttfamily arXiv:2304.01734
  [hep-th]}}.

\bibitem{Bobev:2020egg}
N.~Bobev, A.~M. Charles, K.~Hristov, and V.~Reys, ``{The Unreasonable
  Effectiveness of Higher-Derivative Supergravity in AdS$_4$ Holography},''
  \href{http://dx.doi.org/10.1103/PhysRevLett.125.131601}{{\em Phys. Rev.
  Lett.} {\bfseries 125} no.~13, (2020) 131601},
  \href{http://arxiv.org/abs/2006.09390}{{\ttfamily arXiv:2006.09390
  [hep-th]}}.

\bibitem{Bobev:2021oku}
N.~Bobev, A.~M. Charles, K.~Hristov, and V.~Reys, ``{Higher-derivative
  supergravity, AdS$_{4}$ holography, and black holes},''
  \href{http://dx.doi.org/10.1007/JHEP08(2021)173}{{\em JHEP} {\bfseries 08}
  (2021) 173}, \href{http://arxiv.org/abs/2106.04581}{{\ttfamily
  arXiv:2106.04581 [hep-th]}}.

\bibitem{Dabholkar:2014wpa}
A.~Dabholkar, N.~Drukker, and J.~Gomes, ``{Localization in supergravity and
  quantum $AdS_4/CFT_3$ holography},''
  \href{http://dx.doi.org/10.1007/JHEP10(2014)090}{{\em JHEP} {\bfseries 10}
  (2014) 090}, \href{http://arxiv.org/abs/1406.0505}{{\ttfamily arXiv:1406.0505
  [hep-th]}}.

\bibitem{Caputa:2018asc}
P.~Caputa and S.~Hirano, ``{Airy Function and 4d Quantum Gravity},''
  \href{http://dx.doi.org/10.1007/JHEP06(2018)106}{{\em JHEP} {\bfseries 06}
  (2018) 106}, \href{http://arxiv.org/abs/1804.00942}{{\ttfamily
  arXiv:1804.00942 [hep-th]}}.

\bibitem{Hanada:2012si}
M.~Hanada, M.~Honda, Y.~Honma, J.~Nishimura, S.~Shiba, and Y.~Yoshida,
  ``{Numerical studies of the ABJM theory for arbitrary N at arbitrary coupling
  constant},'' \href{http://dx.doi.org/10.1007/JHEP05(2012)121}{{\em JHEP}
  {\bfseries 05} (2012) 121}, \href{http://arxiv.org/abs/1202.5300}{{\ttfamily
  arXiv:1202.5300 [hep-th]}}.

\bibitem{Chester:2023qwo}
S.~M. Chester, S.~S. Pufu, Y.~Wang, and X.~Yin, ``{Bootstrapping M-theory
  orbifolds},'' \href{http://dx.doi.org/10.1007/JHEP06(2024)001}{{\em JHEP}
  {\bfseries 06} (2024) 001}, \href{http://arxiv.org/abs/2312.13112}{{\ttfamily
  arXiv:2312.13112 [hep-th]}}.

\bibitem{Hatsuda:2016uqa}
Y.~Hatsuda, ``{ABJM on ellipsoid and topological strings},''
  \href{http://dx.doi.org/10.1007/JHEP07(2016)026}{{\em JHEP} {\bfseries 07}
  (2016) 026},
\href{http://arxiv.org/abs/1601.02728}{{\ttfamily arXiv:1601.02728 [hep-th]}}.

\bibitem{Hatsuda:2015oaa}
Y.~Hatsuda, ``{Spectral zeta function and non-perturbative effects in ABJM
  Fermi-gas},'' \href{http://dx.doi.org/10.1007/JHEP11(2015)086}{{\em JHEP}
  {\bfseries 11} (2015) 086}, \href{http://arxiv.org/abs/1503.07883}{{\ttfamily
  arXiv:1503.07883 [hep-th]}}.

\bibitem{Moriyama:2014gxa}
S.~Moriyama and T.~Nosaka, ``{Partition Functions of Superconformal
  Chern-Simons Theories from Fermi Gas Approach},''
  \href{http://dx.doi.org/10.1007/JHEP11(2014)164}{{\em JHEP} {\bfseries 11}
  (2014) 164},
\href{http://arxiv.org/abs/1407.4268}{{\ttfamily arXiv:1407.4268 [hep-th]}}.

\bibitem{Pestun:2007rz}
V.~Pestun, ``{Localization of gauge theory on a four-sphere and supersymmetric
  Wilson loops},'' \href{http://dx.doi.org/10.1007/s00220-012-1485-0}{{\em
  Commun. Math. Phys.} {\bfseries 313} (2012) 71--129},
  \href{http://arxiv.org/abs/0712.2824}{{\ttfamily arXiv:0712.2824 [hep-th]}}.

\bibitem{Moriyama:2014waa}
S.~Moriyama and T.~Nosaka, ``{ABJM membrane instanton from a pole cancellation
  mechanism},'' \href{http://dx.doi.org/10.1103/PhysRevD.92.026003}{{\em Phys.
  Rev. D} {\bfseries 92} no.~2, (2015) 026003},
  \href{http://arxiv.org/abs/1410.4918}{{\ttfamily arXiv:1410.4918 [hep-th]}}.

\bibitem{Okuyama:2015auc}
K.~Okuyama, ``{Probing non-perturbative effects in M-theory on orientifolds},''
  \href{http://dx.doi.org/10.1007/JHEP01(2016)054}{{\em JHEP} {\bfseries 01}
  (2016) 054}, \href{http://arxiv.org/abs/1511.02635}{{\ttfamily
  arXiv:1511.02635 [hep-th]}}.

\bibitem{Hatsuda:2014vsa}
Y.~Hatsuda and K.~Okuyama, ``{Probing non-perturbative effects in M-theory},''
  \href{http://dx.doi.org/10.1007/JHEP10(2014)158}{{\em JHEP} {\bfseries 10}
  (2014) 158},
\href{http://arxiv.org/abs/1407.3786}{{\ttfamily arXiv:1407.3786 [hep-th]}}.

\bibitem{Jafferis:2010un}
D.~L. Jafferis, ``{The Exact Superconformal R-Symmetry Extremizes Z},''
  \href{http://dx.doi.org/10.1007/JHEP05(2012)159}{{\em JHEP} {\bfseries 05}
  (2012) 159}, \href{http://arxiv.org/abs/1012.3210}{{\ttfamily arXiv:1012.3210
  [hep-th]}}.

\bibitem{Hama:2010av}
N.~Hama, K.~Hosomichi, and S.~Lee, ``{Notes on SUSY Gauge Theories on
  Three-Sphere},'' \href{http://dx.doi.org/10.1007/JHEP03(2011)127}{{\em JHEP}
  {\bfseries 03} (2011) 127}, \href{http://arxiv.org/abs/1012.3512}{{\ttfamily
  arXiv:1012.3512 [hep-th]}}.

\bibitem{Hama:2011ea}
N.~Hama, K.~Hosomichi, and S.~Lee, ``{SUSY Gauge Theories on Squashed
  Three-Spheres},'' \href{http://dx.doi.org/10.1007/JHEP05(2011)014}{{\em JHEP}
  {\bfseries 05} (2011) 014}, \href{http://arxiv.org/abs/1102.4716}{{\ttfamily
  arXiv:1102.4716 [hep-th]}}.

\bibitem{Imamura:2011wg}
Y.~Imamura and D.~Yokoyama, ``{N=2 supersymmetric theories on squashed
  three-sphere},'' \href{http://dx.doi.org/10.1103/PhysRevD.85.025015}{{\em
  Phys. Rev. D} {\bfseries 85} (2012) 025015},
  \href{http://arxiv.org/abs/1109.4734}{{\ttfamily arXiv:1109.4734 [hep-th]}}.

\bibitem{Benini:2009qs}
F.~Benini, C.~Closset, and S.~Cremonesi, ``{Chiral flavors and M2-branes at
  toric CY4 singularities},''
  \href{http://dx.doi.org/10.1007/JHEP02(2010)036}{{\em JHEP} {\bfseries 02}
  (2010) 036}, \href{http://arxiv.org/abs/0911.4127}{{\ttfamily arXiv:0911.4127
  [hep-th]}}.

\bibitem{Imamura:2008ji}
Y.~Imamura and S.~Yokoyama, ``{N=4 Chern-Simons theories and wrapped M-branes
  in their gravity duals},'' \href{http://dx.doi.org/10.1143/PTP.121.915}{{\em
  Prog. Theor. Phys.} {\bfseries 121} (2009) 915--940},
\href{http://arxiv.org/abs/0812.1331}{{\ttfamily arXiv:0812.1331 [hep-th]}}.

\bibitem{Kashaev:2015kha}
R.~Kashaev and M.~Marino, ``{Operators from mirror curves and the quantum
  dilogarithm},'' \href{http://dx.doi.org/10.1007/s00220-015-2499-1}{{\em
  Commun. Math. Phys.} {\bfseries 346} no.~3, (2016) 967--994},
  \href{http://arxiv.org/abs/1501.01014}{{\ttfamily arXiv:1501.01014
  [hep-th]}}.

\bibitem{Marino:2015ixa}
M.~Marino and S.~Zakany, ``{Matrix models from operators and topological
  strings},'' \href{http://dx.doi.org/10.1007/s00023-015-0422-0}{{\em Annales
  Henri Poincare} {\bfseries 17} no.~5, (2016) 1075--1108},
\href{http://arxiv.org/abs/1502.02958}{{\ttfamily arXiv:1502.02958 [hep-th]}}.

\bibitem{Kubo:2025jxi}
N.~Kubo, ``{Five-brane webs, 3d $ \mathcal{N} $ = 2 theories and quantum
  curves},'' \href{http://dx.doi.org/10.1007/JHEP05(2025)103}{{\em JHEP}
  {\bfseries 05} (2025) 103}, \href{http://arxiv.org/abs/2501.04146}{{\ttfamily
  arXiv:2501.04146 [hep-th]}}.

\bibitem{Grassi:2014zfa}
A.~Grassi, Y.~Hatsuda, and M.~Marino, ``{Topological Strings from Quantum
  Mechanics},'' \href{http://dx.doi.org/10.1007/s00023-016-0479-4}{{\em Annales
  Henri Poincare} {\bfseries 17} no.~11, (2016) 3177--3235},
\href{http://arxiv.org/abs/1410.3382}{{\ttfamily arXiv:1410.3382 [hep-th]}}.

\bibitem{Codesido:2015dia}
S.~Codesido, A.~Grassi, and M.~Marino, ``{Spectral Theory and Mirror Curves of
  Higher Genus},'' \href{http://dx.doi.org/10.1007/s00023-016-0525-2}{{\em
  Annales Henri Poincare} {\bfseries 18} no.~2, (2017) 559--622},
  \href{http://arxiv.org/abs/1507.02096}{{\ttfamily arXiv:1507.02096
  [hep-th]}}.

\bibitem{Marino:2015nla}
M.~Marino, ``{Spectral theory and mirror symmetry.},''
  \href{http://dx.doi.org/10.1090/pspum/098/01722}{{\em Proc. Symp. Pure Math.}
  {\bfseries 98} (2018) 259}, \href{http://arxiv.org/abs/1506.07757}{{\ttfamily
  arXiv:1506.07757 [math-ph]}}.

\end{thebibliography}\endgroup

\end{document}